\documentclass[10pt,aps,prd,twocolumn,showkeys,nofootinbib,floatfix,amsmath,amssymb,showpacs,superscriptaddress,notitlepage,preprintnumbers]{revtex4-1}

\usepackage[utf8]{inputenc}
\usepackage{rotating}
\usepackage{graphicx}
\usepackage{dcolumn}
\usepackage{bm}
\usepackage{color}
\usepackage{multirow}
\usepackage{mathtools}
\usepackage{MnSymbol}
\usepackage{tabularx}
\newcolumntype{C}{>{\centering\arraybackslash}X}
\newcolumntype{R}{>{\raggedleft\arraybackslash}X}
\newcolumntype{L}{>{\raggedright\arraybackslash}X}
\usepackage[caption=false]{subfig}
\usepackage{xr-hyper}
\usepackage[pdftex,bookmarks=true]{hyperref}
\hypersetup{
   colorlinks,
   citecolor=black,
   filecolor=black,
   linkcolor=black,
   urlcolor=black
}
\usepackage{cleveref}
\usepackage[english]{babel}
\allowdisplaybreaks

\newcommand{\bv}[1]{{\mathbf{#1}}}
\newcommand{\bs}[1]{{\boldsymbol{#1}}}
\newcommand{\kb}{\bv{k}}
\newcommand{\qb}{\bv{q}}
\newcommand{\pb}{\bv{p}}

\newcommand{\xb}{\bv{x}}
\newcommand{\yb}{\bv{y}}
\newcommand{\zb}{\bv{z}}

\def\Zlam{\mathcal{Z(\lambda)}}
\def\llangle{\left\langle}
\def\rrangle{\right\rangle}
\def\latmom{\left( \frac{2\pi}{L} \right)}
\newcommand{\ea}[1]{{\llangle #1 \rrangle}}
\newcommand{\eal}[1]{{\llangle #1 \rrangle_{\lambda}}}

\newcommand{\cosE}[2]{(e^{i \bv{#1} \cdot \bv{#2}} + e^{-i \bv{#1} \cdot \bv{#2}})}

\newcommand{\XX}{X}
\newcommand{\Dfb}{\overleftrightarrow{D}}

\usepackage[normalem]{ulem}
\renewcommand\sout{\bgroup \color{red} \ULdepth=-.5ex \ULset}

\begin{document}

	\title{Lattice QCD evaluation of the Compton amplitude employing the Feynman-Hellmann theorem}

	\author{K.~U.~Can}
	\affiliation{CSSM, Department of Physics, The University of Adelaide, Adelaide SA 5005, Australia}
	\author{A.~Hannaford-Gunn}
	\affiliation{CSSM, Department of Physics, The University of Adelaide, Adelaide SA 5005, Australia}
	\author{R.~Horsley}
	\affiliation{School of Physics and Astronomy, University of Edinburgh, Edinburgh EH9 3JZ, UK}
	\author{Y.~Nakamura}
	\affiliation{RIKEN Center for Computational Science, Kobe, Hyogo 650-0047, Japan}
	\author{H.~Perlt}
	\affiliation{Institut f\"{u}r Theoretische Physik, Universit\"{a}t Leipzig, 04103 Leipzig, Germany}
	\author{P.~E.~L.~Rakow}
	\affiliation{Theoretical Physics Division, Department of Mathematical Sciences, University of Liverpool, Liverpool L69 3BX, United Kingdom}
	\author{G.~Schierholz}
	\affiliation{Deutsches Elektronen-Synchrotron DESY, 22603 Hamburg, Germany}
	\author{K.~Y.~Somfleth}
	\affiliation{CSSM, Department of Physics, The University of Adelaide, Adelaide SA 5005, Australia}
	\author{H.~St\"{u}ben}
	\affiliation{Regionales Rechenzentrum, Universit\"{a}t Hamburg, 20146 Hamburg, Germany}
	\author{R.~D.~Young}
	\affiliation{CSSM, Department of Physics, The University of Adelaide, Adelaide SA 5005, Australia}
	\author{J.~M.~Zanotti}
	\affiliation{CSSM, Department of Physics, The University of Adelaide, Adelaide SA 5005, Australia}
	
	\collaboration{QCDSF/UKQCD/CSSM Collaborations}
	\noaffiliation

	\date{\today}

	\begin{abstract}
	    The forward Compton amplitude describes the process of virtual photon scattering from a hadron and provides an essential ingredient for the understanding of hadron structure. As a physical amplitude, the Compton tensor naturally includes all target mass corrections and higher twist effects at a fixed virtuality, $Q^2$. By making use of the second-order Feynman-Hellmann theorem, the nucleon Compton tensor is calculated in lattice QCD at an unphysical quark mass across a range of photon momenta $3 \lesssim Q^2 \lesssim 7$~GeV$^2$. This allows for the $Q^2$ dependence of the low moments of the nucleon structure functions to be studied in a lattice calculation for the first time. The results demonstrate that a systematic investigation of power corrections and the approach to parton asymptotics is now within reach.
	\end{abstract}

	\preprint{ADP-20-16/T1126, DESY 20-111, Liverpool LTH 1238}
	\maketitle

	\section{Introduction}
		Understanding the internal structure of hadrons from first principles remains one of the foremost tasks in particle and nuclear physics. It is an active field of research with important phenomenological implications in high-energy, nuclear and astroparticle physics. The static properties of hadrons, from the hybrid structure of quark and meson degrees of freedom at low energies down to the partonic structure at short distances, are encoded in structure functions. The tool for computing hadron structure functions from first principles is lattice QCD.

		The connection between nucleon structure functions and the quark structure of the nucleon is commonly rendered by the parton model. Although providing an intuitive language in which to interpret the deep-inelastic scattering data, the parton model represents an ideal case, valid if the partons are scattered elastically and incoherently by the incoming lepton. In the operator product expansion (OPE) of the Compton amplitude the operators are classified according to twist. The parton model accounts for twist-two contributions only, and does not accommodate power corrections arising from operators of higher twist. It has been known for a long time though that contributions from operators of higher twist are inseparably connected with the contributions of leading twist, as a result of operator mixing and renormalization~\cite{Martinelli:1996pk,Beneke:1998ui}. 

		So far lattice QCD calculations of nucleon structure functions have largely been limited to matrix elements of leading twist. Traditionally, that includes the calculation of a few lower moments of parton distribution functions (PDFs), building on work of~\cite{Martinelli:1988rr,Gockeler:1995wg,Gockeler:1996mu,Gockeler:1998ye}. Various strategies to overcome this limitation have appeared over the years \cite{Liu:1993cv,Capitani:1998fe,Detmold:2005gg,Braun:2007wv,Ma:2014jla,Chambers:2017dov,Bali:2017gfr,Liang:2019frk}, however the focus of recent efforts is largely directed to light-cone PDFs, which can be computed from so-called quasi-PDFs~\cite{Ji:2013dva,Lin:2014zya,Alexandrou:2015rja,Alexandrou:2017huk,Green:2017xeu,Ji:2001wha,Chen:2017mzz,Ishikawa:2017faj,Lin:2018pvv} and its extensions, pseudo-PDFs~\cite{Radyushkin:2017cyf,Orginos:2017kos,Karpie:2018zaz}. A detailed account of both approaches, including their limitations, and what has been accomplished so far, is given in~\cite{Lin:2017snn,Cichy:2018mum,Constantinou:2020pek}.

		Common to the quasi- and pseudo-PDF approaches is that the operators of interest necessarily mix with operators of higher twist under renormalization, be it in a soft renormalization scheme like $\overline{\text{MS}}$ or in a cut-off scheme like the lattice~\cite{Martinelli:1996pk,Rossi:2017muf,Braun:2018brg}. On the lattice the result is that the leading-twist Wilson coefficients diverge as $1/a^2$ ($a$ being the lattice constant). This divergence must be cancelled with that of the higher-twist operator matrix element, which demands a nonperturbative calculation of the Wilson coefficients. The usefulness of the  OPE comes from the assumption that the nonperturbative physics is contained in the operator matrix elements, known as factorization, while the Wilson coefficients are calculable in perturbation theory. This fundamental property is threatened by the presence of power divergences. Another shortcoming of present calculations is that the structure functions at medium to small Bjorken $x$ are dominated by Regge and Pomeron exchange, which are peripheral processes that proceed far off the light-cone~\cite{Brodsky:2004hh,Hautmann:2007cx}. Several attempts have been made to extend the OPE into the Regge regime~\cite{Brandt:1974nk,Brandt:1974dg} without much success~\cite{Brandt:1971ep}. The Wilson coefficients can be computed on the lattice, in principle, as presented in Refs.~\cite{Capitani:1998fe,Bietenholz:2009if,Bietenholz:2010kn}. It should be noted though that the hypercubic lattice can only accommodate operators of spin four or less, which thwarts any prediction of the  Wilson coefficients for the higher moments on the lattice.

		The structure of hadrons relevant for deep-inelastic scattering are completely characterized by the Compton amplitude. In the present work, we build upon a recent Letter \cite{Chambers:2017dov} outlining a procedure to determine nucleon structure functions from a lattice QCD calculation of the forward Compton amplitude. By working with the physical amplitude, this approach overcomes issues of operator mixing and renormalization, and the restriction to light-cone operators~\cite{Brodsky:2004hh,Hautmann:2007cx}. By working with the physical amplitude, there is no need to resort to the OPE, facing problems of factorization and renormalization, nor is the calculation bound to light-cone kinematics. However, if we were to map the OPE upon the Compton amplitude, as far as this is possible, we will find Wilson coefficients and operators being properly renormalized, including mixing effects. If the Compton amplitude is known sufficiently accurately, we can expect to obtain nucleon structure functions in closed form~\cite{Chambers:2017dov}, including power corrections. The strategy is most similar to those considered in Refs.~\cite{Detmold:2005gg,Liu:1993cv,Ji:2001wha,Fukaya:2020wpp}, and shares features with other approaches to inclusive processes \cite{Ji:2001wha,Ma:2017pxb,Hashimoto:2017wqo,Hansen:2017mnd,Feng:2020zdc,Briceno:2019opb}.

		Here we establish the theoretical foundation of the approach and present results for the Compton amplitude across a range of kinematics. The calculations are performed at the SU(3) flavor symmetric point~\cite{Bietenholz:2011qq} at an unphysical pion mass. Results are reported on the lowest four moments of the unpolarized structure functions of the nucleon for photon momenta $Q^2$ ranging from approximately $3$--$7\,{\rm GeV}^2$. The variation of $Q^2$ demonstrates the potential to provide a quantitative test of the twist expansion on the lattice for the first time. 

		In terms of the practical computation, the determination of the Compton amplitude takes advantage of the Feynman-Hellmann~\cite{Horsley:2012pz,Chambers:2014qaa,Chambers:2014pea,Chambers:2015bka,Chambers:2017tuf} approach to hadron structure---see also Refs.~\cite{Detmold:2004kw,Primer:2013pva,Davoudi:2015cba,Savage:2016kon,Bouchard:2016heu,Shanahan:2017bgi,Tiburzi:2017iux}. The use of Feynman-Hellmann provides an alternative to computing the 3- or 4-point functions. Here we also present a derivation of the second-order Feynman-Hellmann theorem necessary for the present work---a related derivation has been presented in Ref.~\cite{Agadjanov:2016cjc}.

		This paper is organized as follows: formal definitions of the Compton amplitude and the structure functions, along with the connection between the OPE and the dispersion relation are given in \Cref{sec:cs}. We explicitly derive the second order Feynman-Hellmann theorem in \Cref{sec:fh}. Our lattice setup and the implementation details are given in \Cref{sec:ls}. Results for the Compton amplitude and the moments of the structure functions are presented in \Cref{sec:rd}. We summarize our findings in \Cref{sec:sc}.

	\section{Forward Compton amplitude and the structure functions} \label{sec:cs}
		\subsection{Notation}
			At leading order in the electromagnetic interaction, the general description for the inclusive scattering of a charged lepton from a hadronic target, e.g. $eN\to e'X$, is encoded in the hadron tensor. Conventionally, the hadron tensor is expressed as a matrix element
			of the commutator of electromagnetic current operators~\cite{DevenishRobin2004Dis,Manohar:1992tz,Liu:1999ak}, \footnote{In this section, we work in Minkowski space.}
			\begin{equation} \label{eq:hadronic_tensor}
			W_{\mu\nu}(p,q) = \frac{1}{4\pi} \int d^4z\, e^{i q \cdot z} \rho_{s s^\prime} \ea{p,s^\prime \left| [\mathcal{J}_\mu(z), \mathcal{J}_\nu(0)] \right|p,s},
			\end{equation}
			for a hadron of momentum $p$ and (virtual) photon momentum $q$. For the present discussion, we will only consider
			spin-averaged observables by taking $\rho_{ss'}=\tfrac12\delta_{ss'}$. The current operator takes the familiar form as the charge-weighted sum
			of the quark vector currents, $\mathcal{J}_\mu=\sum_f \mathcal{Q}_f J_\mu^f$, with $\mathcal{Q}_f$ being the charge of quark flavor $f$. The flavor decomposition will be discussed in further detail in a later section.

			The spin-averaged nucleon tensor can be decomposed as
			\begin{align}\label{eq:comptens}
				\begin{split}
					&W_{\mu\nu}(p,q) = \left( -g_{\mu\nu} + \frac{q_\mu q_\nu}{q^2} \right) F_1(x,Q^2) \\ 
					&+ \left( p_\mu - \frac{p \cdot q}{q^2}q_\mu \right) \left( p_\nu - \frac{p \cdot q}{q^2}q_\nu \right) \frac{F_2(x,Q^2)}{p \cdot q},
				\end{split}
			\end{align}
			which is defined such that Lorentz-invariant structure functions, $F_{1,2}$, match onto their conventional partonic interpretation in the deep inelastic scaling region. These structure functions are expressed as functions of the Bjorken scaling variable ($x=Q^2/(2p\cdot q)$) and $Q^2=-q^2$.

			While the inelastic structure functions are not directly accessible within a conventional Euclidean lattice formulation, we highlight that the spacelike component of the Compton tensor can be studied within a Euclidean framework---as also discussed in Refs.~\cite{Ji:2001wha,Liu:1993cv,Detmold:2005gg}.

			The (spin-averaged) Compton tensor is defined similarly to \Cref{eq:hadronic_tensor},
			\begin{equation} \label{eq:compamp}
				T_{\mu\nu}(p,q) = i \int d^4z\, e^{i q \cdot z} \rho_{s s^\prime} \ea{p,s^\prime \left| \mathcal{T}\left\{ \mathcal{J}_\mu(z) \mathcal{J}_\nu(0) \right\} \right|p,s},
			\end{equation}
			where $\mathcal{T}$ is the time-ordering operator. This tensor can be decomposed in precisely the same way as $W_{\mu\nu}$ in
			\Cref{eq:comptens}, which defines the analogous scalar functions $\mathcal{F}_{1,2}(\omega,Q^2)$. For our purposes, it is convenient to express these in terms of the inverse Bjorken variable $\omega=2p\cdot q/Q^2$. These Compton structure functions are related to the corresponding ordinary structure functions via the optical theorem, which states:
			\begin{align} \label{eq:optical_cuts}
				\operatorname{Im}\mathcal{F}_1(\omega,Q^2) &= 2\pi F_1(x,Q^2), \\
				\operatorname{Im}\mathcal{F}_2(\omega,Q^2) &= 2\pi F_2(x,Q^2).
			\end{align}

			Analyticity and crossing symmetry means we can write a dispersion relation for $\mathcal{F}$~\cite{Drechsel:2002ar}
			\begin{align}
				\begin{split}
					\mathcal{F}_1(\omega,Q^2)-\mathcal{F}_1(0,Q^2) 
					& = \frac{2\omega^2}{\pi}\int_1^\infty d\omega'\frac{\operatorname{Im}\mathcal{F}_1(\omega',Q^2)}{\omega'\left(\omega'^2-\omega^2-i\epsilon\right)},
				\end{split} \\
				\mathcal{F}_2(\omega,Q^2)&= \frac{2\omega}{\pi}\int_{1}^\infty d\omega'\frac{\operatorname{Im}\mathcal{F}_2(\omega',Q^2)}{\omega'^2-\omega^2-i\epsilon}.
			\end{align}
			To accommodate the subtraction necessary in $\mathcal{F}_1$, we will make use of the bar notation to denote the dispersive part, $\overline{\mathcal{F}}_1(\omega,Q^2) = \mathcal{F}_1(\omega,Q^2)-\mathcal{F}_1(0,Q^2)$. The dispersion integrals can be directly connected to the hadron tensor by the optical theorem, giving:
			\begin{align}\label{eq:compomega}
				\overline{\mathcal{F}}_1(\omega,Q^2)&= 4\omega^2\int_0^1 dx \frac{x\,F_1(x,Q^2)}{1-x^2\omega^2-i\epsilon},\\
				\mathcal{F}_2(\omega,Q^2)&= 4\omega \int_{0}^1 dx\, \frac{F_2(x,Q^2)}{1-x^2\omega^2-i\epsilon}.  
			\end{align}

			The nature of the dispersion integral makes it clear that whenever $|\omega|<1$ the singularities are never encountered and the time-ordering $i\epsilon$ becomes irrelevant. Hence the current-current correlation remains spacelike and there is no distinction between the Euclidean and Minkowski amplitudes. Physically, the condition $|\omega|<1$ is simply the statement that the eigenstates which propagate between the current insertions in \Cref{eq:compamp} cannot go on-shell.

		\subsection{Operators displaced in time} \label{sec:time}
			In the Feynman-Hellmann approach employed in this work, the matrix elements calculated involve current--current correlations which are displaced in Euclidean time. It is therefore instructive to further clarify the relationship between the Compton tensor, as defined in Minkowski space, and the corresponding calculation within a Euclidean framework.

			We start by separating \Cref{eq:compamp} into two distinct time orderings by first defining the amplitude at fixed temporal separation between the currents:
			\begin{align}\label{eq:Cmink}
				\begin{split}
					\tilde{T}_{\mu\nu}^{\mathcal{M}}(p,q,t) &= i\rho_{ss'}\int d^3z \, e^{i (q_0+i\epsilon) t}e^{-i \mathbf{q} \cdot \mathbf{z}} \\ 
					&\times \ea{p,s^\prime \left| \mathcal{J}_\mu(\mathbf{z},t) \mathcal{J}_\nu(0) \right|p,s}.
				\end{split}
			\end{align}
			From this definition it is straightforward to recover the full Compton amplitude by integrating over $t$:
			\begin{align} \label{eq:comptint}
				T_{\mu\nu}(p,q)=\int_0^\infty
				dt\,\left[\tilde{T}_{\mu\nu}^{\mathcal{M}}(p,q,t)+\tilde{T}_{\nu\mu}^{\mathcal{M}}(p,-q,t)\right].
			\end{align}

			To isolate the explicit $t$ dependence in \Cref{eq:Cmink}, we insert a complete set of states and exploit translational invariance in the usual way:
			\begin{align}\label{eq:CminkB}
				\begin{split}
					\tilde{T}_{\mu\nu}^{\mathcal{M}}&(p,q,t) = i\rho_{ss'}\sumint_\XX \int d^3z\,e^{i (q_0+E_p-E_\XX+i\epsilon) t} \\ 
					&\times e^{-i(\mathbf{q}+\mathbf{p}-\mathbf{P}_\XX) \cdot \mathbf{z}}
				  	\ea{p,s^\prime \left| \mathcal{J}_\mu(0)|\XX\rangle\langle\XX |\mathcal{J}_\nu(0) \right|p,s}.
				\end{split}
			\end{align}
			The completeness integral, $I=\sumint_\XX |\XX\rangle\langle\XX |$, describes a full integral over the entire state space, implicitly including all possible momenta over all possible configurations of particles.

			Similarly to \Cref{eq:Cmink}, one can write down an expression where the current insertions are separated in Euclidean time~\cite{Liu:1993cv,Liu:1999ak}:
			\begin{align}\label{eq:Ceucl}
				\begin{split}
					\tilde{T}_{\mu\nu}^{\mathcal{E}}(p,q,\tau) &= \rho_{ss'}\int d^3z \, e^{q_0 \tau}e^{-i \mathbf{q} \cdot \mathbf{z}} \\
					&\times \ea{p,s^\prime \left| \mathcal{J}_\mu(\mathbf{z},\tau) \mathcal{J}_\nu(0) \right|p,s}.
				\end{split}
			\end{align}
			Within the Feynman-Hellmann approach, by construction, there is no external energy transfer, $q_0=0$---however, we retain this variable explicitly in our presentation for completeness. Inserting a complete set of states and using translational invariance, under Euclidean evolution, we have:
			\begin{align}\label{eq:CeuclB}
				\begin{split}
					\tilde{T}_{\mu\nu}^{\mathcal{E}}&(p,q,\tau) = \rho_{ss'}\sumint_\XX \int d^3z\, e^{(q_0+E_p-E_\XX) \tau} \\
					&\times e^{-i(\mathbf{q}+\mathbf{p}-\mathbf{P}_\XX) \cdot \mathbf{z}} \ea{p,s^\prime \left| \mathcal{J}_\mu(0)|\XX\rangle\langle\XX |\mathcal{J}_\nu(0) \right|p,s}.
				\end{split}
			\end{align}
			It is evident that~\Cref{eq:CminkB,eq:CeuclB} differ non-trivially in their dependence on the temporal coordinate---a similar point has been made in Ref.~\cite{Briceno:2017cpo}. However, upon integrating with respect to time, the Minkowski and Euclidean expressions are easily equated---subject to the caveat that we are below the elastic threshold. In particular, provided that $E_{\XX(\mathbf{p}\pm \mathbf{q})}>E_p\pm q_0$ for all non-vanishing contributions to the completeness sum $\sumint_\XX$, then the $i\epsilon$ prescription becomes irrelevant and we have:
			\begin{align}\label{eq:Tintegral}
				\int_0^\infty d\tau\,
				\tilde{T}_{\mu\nu}^{\mathcal{E}}(p,\pm q,\tau)=\int_0^\infty dt\,
				\tilde{T}_{\mu\nu}^{\mathcal{M}}(p,\pm q,t).
			\end{align}
			By summing the two terms of \Cref{eq:comptint}, this makes it clear that the Euclidean and Minkowski Compton amplitudes are identical in the unphysical region, even though the current insertions are allowed to be separated in time. We of course note that if the intermediate states can go on shell, $E_{\XX(\mathbf{p}\pm \mathbf{q})}=E_p\pm q_0$, then the Euclidean integral in \Cref{eq:Tintegral} is not well defined \cite{Liang:2019frk}. The $i\epsilon$ factor in \Cref{eq:Cmink} then becomes essential in order to define the analytic continuation required to render the integral finite. However, the fixed-$t$ Euclidean matrix elements are perfectly well defined, and there is no restriction on the kinematic thresholds---as has been studied in~\cite{Liang:2019frk,Liu:1999ak,Liu:1993cv}.
			
			Although we have just been through this careful consideration of the $t$ dependence, we note that the Feynman-Hellmann technique relies on resolving the spectrum of a perturbed Hamiltonian---which itself does not make any reference to the nature of the temporal correlations. In the present work, there is hence no need to explicitly resolve the $\tau$ dependence of \Cref{eq:Ceucl}. The connection to the Minkowski amplitude lies in \Cref{eq:Tintegral}, as the current insertions act at all times.
			
		\subsection{Moments and the OPE} \label{sec:ope}
			As will become clear in the next section, within the Feynman-Hellmann formalism, each external current momentum vector $\bv{q}$ of interest requires a unique propagator inversion. However, for each inversion, the variation of hadron Fourier momenta $\bv{p}$ allows access to multiple distinct $\omega=2\bv{p}.\bv{q}/Q^2$ values. An ensemble of $\bv{q}$ and $\bv{p}$ values therefore provides a wealth of kinematic points to resolve the Compton amplitude. It is therefore convenient to present a summary of the kinematic coverage in terms of the moments of the structure functions. We consider the Taylor series expansion of \Cref{eq:compomega} at fixed $Q^2$:
			\begin{align} \label{eq:ope_moments}
			    \overline{\mathcal{F}}_1(\omega,Q^2)&=\sum_{n=1}^\infty2\omega^{2n}M^{(1)}_{2n}(Q^2),\\
			  	\mathcal{F}_2(\omega,Q^2)&= \sum_{n=1}^\infty 4\omega^{2n-1}M^{(2)}_{2n}(Q^2),
			\end{align}
			with the moments being defined by
			\begin{align} 
			  	M^{(1)}_{2n}(Q^2)&= 2\int_0^1 dx\, x^{2n-1} F_1(x,Q^2),\\
			  	M^{(2)}_{2n}(Q^2)&= \int_{0}^1 dx\,x^{2n-2} F_2(x,Q^2).
			\end{align}

			From the perspective of the present lattice calculation, one can proceed by calculating the Compton tensor for a number of $\omega$ values and extract the moments of the structure functions. By treating $Q^2$ as an external scale, the method connects directly to the physical amplitudes of interest, and therefore circumvents the operator mixing issues discussed above. Although the described lattice calculations relate directly to the physical moments, $M_{2n}^{(1,2)}$, we note that at asymptotically large $Q^2$ the moments become dominated by their leading-twist contributions, namely the moments of the familiar parton distribution functions, $v_{2n}$,
			\begin{align} 
			    \label{eq:leading-twist}
				M^{(1)}_{2n}(Q^2)&= \sum_f C_{f,2n}^{(1)}\left(\frac{Q^2}{\mu^2},g(\mu)\right)v_{2n}^f(\mu)+\mathcal{O}\left(\frac{1}{Q^2}\right),\\
			  	M^{(2)}_{2n}(Q^2)&= \sum_f C_{f,2n}^{(2)}\left(\frac{Q^2}{\mu^2},g(\mu)\right)v_{2n}^f(\mu)+\mathcal{O}\left(\frac{1}{Q^2}\right),
			\end{align}
			where the sum runs over partonic flavors $f$. The short distance structure of the operator product in \Cref{eq:compamp} is encoded in the Wilson coefficients, $C$, and the long-distance hadronic features are encoded in the matrix elements of local operators, $v$, renormalized at some scale $\mu$. For completeness, our notation is summarized in Appendix~\ref{app:ope}. We find it necessary to reiterate that we extract the physical moments, $M_{2n}(Q^2)$, in this work, not the matrix elements of the local operators.

	\section{Second order Feynman-Hellmann theorem}\label{sec:fh}
		The purpose of applying the Feynman-Hellmann theorem to lattice QCD is to relate matrix elements of interest to energy shifts in weak external fields. In the case of a generalized Compton amplitude, described by a matrix element of two (non-local) current insertions, the conventional approach would require the evaluation of lattice 4-point functions. The application of Feynman-Hellmann then reduces the problem to a more straightforward analysis of 2-point correlation functions using spectroscopic techniques.
        We note that other related background field methods also offer alternatives to the direct evaluation of lattice 4-point functions \cite{Shanahan:2017bgi,Tiburzi:2017iux}.
        
		In order to compute the forward Compton amplitude via the Feynman-Hellmann relation, we introduce the following perturbation to the fermion action,
		\begin{equation}\label{eq:fh_perturb}
			S(\lambda) = S + \lambda \int d^4z (e^{i q \cdot z} + e^{-i q \cdot z}) \mathcal{J}_{\mu}(z) ,
		\end{equation}
		where $\lambda$ is the strength of the coupling between the quarks and the external field, $\mathcal{J}_{\mu}(x) = Z_V \bar{q}(x) \gamma_\mu q(x)$ is the electromagnetic current coupling to the quarks along the $\mu$ direction, $\bv{q}$ is the external momentum inserted by the current and $Z_V$ is the renormalization constant for the local electromagnetic current.

		The general strategy for deriving Feynman-Hellmann in a lattice QCD context is to consider the general spectral decomposition of a correlator in the presence of the background field. The differentiation of this correlation function with respect to the external field reveals a distinct temporal signature for the energy shift. By explicit evaluation of the perturbed correlator, one is able to identify this signature and hence resolve the desired relationship between the energy shift and matrix element. Our principal theoretical result here is that for the perturbed action described in \Cref{eq:fh_perturb}, the second-order energy shift of the nucleon is found to be:
		\begin{align} \label{eq:secondorder_fh}
			\left. \frac{\partial^2 E_{N_\lambda}(\bv{p})}{\partial \lambda^2} \right|_{\lambda=0} &= - \frac{T_{\mu\mu}(p,q) + T_{\mu\mu}(p,-q)}{2 E_{N}(\bv{p})}, 
		\end{align}
		where $T$ is the Compton amplitude defined in \Cref{eq:compamp}, $q=(\qb,0)$ is the external momentum encoded by \Cref{eq:fh_perturb}, and $E_{N_\lambda}(\pb)$ is the nucleon energy at momentum $\pb$ in the presence of a background field of strength $\lambda$. In the following we sketch the main steps of the derivation, and refer the interested reader to Appendix~\ref{app:deriv} for further details.

		In the presence of the external field introduced in \Cref{eq:fh_perturb}, we define the two-point correlation function projected to definite momentum as,
		\begin{equation}\label{eq:G2}
			G^{(2)}_\lambda(\bv{p};t) \equiv \int d^3 x e^{-i \bv{p} \cdot \bv{x}} \bs{\Gamma} \; \langle \Omega_\lambda | \chi(\bv{x},t) \overline{\chi}(0)|\Omega_\lambda\rangle,
		\end{equation}
		where here and in the following, a trace over Dirac indices with the spin-parity projection matrix $\bs{\Gamma}$ is understood, and $|\Omega_\lambda\rangle$ is the vacuum in the presence of the external field. The asymptotic behavior of the correlator at large Euclidean times takes the familiar form,
		\begin{equation}\label{eq:G2spec}
			G^{(2)}_\lambda(\bv{p};t) \simeq A_\lambda(\bv{p}) e^{-E_{N_\lambda}(\bv{p}) t},
		\end{equation}
		where $E_{N_\lambda}(\bv{p})$ is the energy of the ground state nucleon in the external field and $A_\lambda(\pb)$ the corresponding overlap factor.

		For the purpose of current presentation, a nucleon interpolating operator is assumed for $\chi$. However, the derivation applies to any ground-state hadron, provided the ground state in the presence of the external field is perturbatively close to the free-field state. A simple counter example could be a $\Sigma$ baryon in the presence of a strangeness-changing current, where at $\lambda=0$ the correlator behaves as $e^{-E_\Sigma t}$ but at any finite $\lambda$ this will eventually be dominated by $e^{-E_Nt}$ (kinematics permitting).

\begin{figure}[!t]
	        \centering
	        \includegraphics[width=0.485\textwidth]{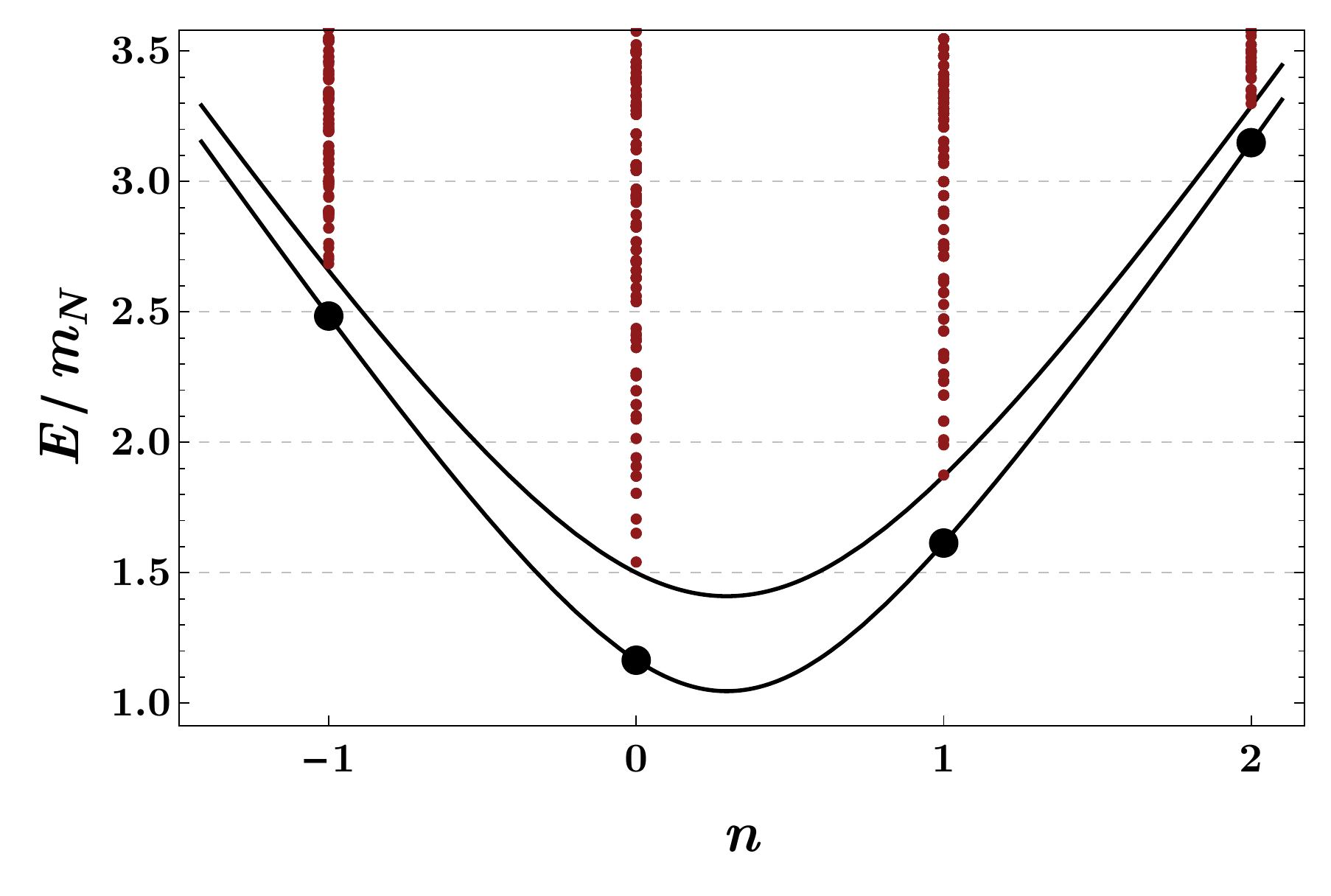}
	        \caption{\label{fig:threshold} The lower curve shows the nucleon energy for momenta along the direction of $\bv{q}$, $E=\sqrt{m_N^2+(\bv{p}+n\bv{q})^2}$. At finite external field strength, all momentum states connected by integer multiples  of $\bv{q}$ will be coupled, these are emphasized by the large dots for the ground-state nucleon. We choose an example kinematic point from the numerical results presented in the following section: $\bv{p}=2\pi/L(-1,-1,0)$ and $\bv{q}=2\pi/L(4,1,0)$. The upper curve shows the (non-interacting) two-particle $N\pi$ threshold, with the small dots representing the discrete nature of this two-body ``cut'' on the lattice.}
        \end{figure}

		It is for a similar physical reason that one must work with nucleon states that have the least possible kinetic energy among all states connected to any number of current insertions. This same condition guarantees the connection between the Euclidean and Minkowski Compton amplitudes described in the previous section. In the presence of the background field, the Hamiltonian of the system will mix momentum states connected by integer multiples of the momentum transfer $\bv{q}$.
        We hence choose the Fourier projection of our correlation
        function, \Cref{eq:G2}, such that $\bv{p}$ corresponds to the
        lowest energy of all these coupled states at finite $\lambda$.
        An example is given in Figure \ref{fig:threshold}, where we show the single nucleon energy plotted along the direction of $\bv{q}$, $E=\sqrt{m_N^2+(\bv{p}+n\bv{q})^2}$. In the example plotted, if the Fourier projection were chosen at $n=1$ (i.e.~$\bv{p}+\bv{q}$) the asymptotic behavior of the correlator would be dominated by a state near that of the free particle at $n=0$ (with an amplitude suppressed by $\lambda$ and the elastic form factor).
        
		When there is a degeneracy in the lowest energy states, this corresponds precisely to Breit-frame kinematics, where a linear response in $\lambda$ isolates the elastic form factors, see Ref.~\cite{Chambers:2017tuf}. For the purposes of the kinematics discussed here, this threshold occurs when $\bv{p}^2=(\bv{p}\pm \bv{q})^2$, or equivalently, $\omega=2\bv{p}.\bv{q}/\bv{q}^2=\pm 1$.

		Assuming that first-order perturbations of the energy vanish, as ensured by avoiding  Breit-frame kinematics---as done in this work---the second-order derivative of \Cref{eq:G2}, evaluated at $\lambda=0$, reduces to 
		\begin{align}\label{eq:G2spec_deriv}
			\begin{split}
				\frac{\partial^2 G^{(2)}_\lambda(\bv{p};t)}{\partial \lambda^2} \bigg |_{\lambda=0} =& \left( \frac{\partial^2 A_\lambda(\bv{p})}{\partial \lambda^2}  - t A(\bv{p}) \frac{\partial^2 E_{N_\lambda}(\bv{p})}{\partial \lambda^2}\right) \\ 
				&\times e^{-E_{N}(\bv{p}) t}.
			\end{split}
		\end{align}
		The derivatives of $A_\lambda(\bv{p})$ and $E_{N_\lambda}(\bv{p})$ are assumed to be evaluated at $\lambda=0$. The first term corresponds to the shift in the overlap factor
		and the second order energy shift is identified in the $t$-enhanced (or time-enhanced) term. It is this $t$ enhancement that leads to a relationship between the energy shift and matrix element. Hence to complete the derivation, we differentiate the path integral representation directly to identify the time-enhanced contributions to the correlator.

		The path integral expression for the 2-point correlator, \Cref{eq:G2}, in the background field is given by
		\begin{equation}\label{eq:G2pi}
			G^{(2)}_{\lambda}(\bv{p};t) = \int d^3 x\, e^{-i \bv{p} \cdot \bv{x}} \bv{\Gamma}\,{}_\lambda\langle \chi(\bv{x},t) \overline{\chi}(0)\rangle_\lambda,
		\end{equation}
		where ${}_\lambda\langle\cdots\rangle_\lambda$ denotes the full path integral over all fields, using the perturbed action $S(\lambda)$ given in \Cref{eq:fh_perturb}---an absence of $\lambda$ subscript is taken to imply $\lambda\to 0$. By differentiating twice with respect to $\lambda$ and evaluated at $\lambda\to 0$, one finds (see \Cref{app:deriv} for details)
		\begin{widetext}
			\begin{align} \label{eq:G2PI_deriv}
				\begin{split}
					\left.\frac{\partial^2 G^{(2)}_\lambda(\bv{p};y)}{\partial \lambda^2} \right |_{\lambda=0} = &\int d^3x\, e^{-i \bv{p} \cdot \bv{x}}\bv{\Gamma}	 \left[\ea{\chi(\bv{x},t) \overline{\chi}(0) \left(\frac{\partial S(\lambda)}{\partial \lambda}\right)^2} +\ea{\chi(\bv{x},t) \overline{\chi}(0)}\ea{\left(\frac{\partial S(\lambda)}{\partial \lambda}\right)^2}\right].
				\end{split}
			\end{align}
			To arrive at this form, it is assumed that the vacuum expectation value of a single current insertion vanishes, $\ea{\partial S(\lambda) / \partial \lambda}= 0$, such as is the case for the electromagnetic current. It is clear that the second term in \Cref{eq:G2PI_deriv} only acts to modify the unperturbed correlator, and hence cannot generate the temporal enhancement associated with the energy shift. Focusing purely on the first term, and inserting an explicit form for the electromagnetic external field, the corresponding second derivative of the correlator becomes
			\begin{align} \label{eq:4pt}
				\left. \frac{\partial^2 G^{(2)}_\lambda(\bv{p};t)}{\partial \lambda^2} \right |_{\lambda=0} = \int d^3x\, e^{-i \bv{p} \cdot \bv{x}} \bv{\Gamma}\int d^4y d^4z (e^{i \bv{q} \cdot \bv{y}} + e^{-i \bv{q} \cdot \bv{y}}) (e^{i \bv{q} \cdot \bv{z}} + e^{-i \bv{q} \cdot \bv{z}}) \ea{\chi(\bv{x},t) \mathcal{J}_{\mu}(z) \mathcal{J}_{\mu}(y) \overline{\chi}(0)}.
			\end{align}
			The correlator defined here involves a four-point correlation function with nucleon interpolating operators held at fixed temporal separation $t$, with the currents inserted across the entire four-volume. Importantly, this expression is evaluated in the absence of the external field, and hence momentum conservation is exact. It is then possible to perform a spectral decomposition of this correlator in terms of a transfer matrix that is diagonal in the momenta.

			It is a rather straightforward calculation to perform the standard procedure of inserting a complete sets of states, and then exploit translational invariance to complete the spatial integrals. Since the temporal integrals over the currents extend over all time, each distinct time ordering of the 4-point function must be treated separately. However the contribution to the energy shift can only come from the contribution where the two current operators both appear between the nucleon creation and annihilation operators. Isolating the contributions that give rise to the dominant $t e^{-{E_N(\pb)t}}$ behavior at asymptotic times gives:
			\begin{align} \label{eq:fh_pifin}
				&\left. \frac{\partial^2 G^{(2)}_\lambda(\bv{p};t)}{\partial \lambda^2} \right |_{\lambda=0} = t A(\bv{p}) \frac{e^{-E_{N}(\bv{p}) t}}{2 E_{N}(\bv{p})}
				\llangle N(\bv{p}) \left| \int d^4 z \left(e^{iq\cdot z}+e^{-iq\cdot z}\right) \mathcal{J}_{\mu}(z) \mathcal{J}_{\mu}(0) \right| N(\bv{p}) \rrangle+\ldots,
			\end{align}
		\end{widetext}
		where the subleading terms are suppressed by the ellipsis. Note that the spin indices have been suppressed here, however a detailed presentation is provided in \Cref{app:deriv}. Finally, a comparison of this form with \Cref{eq:G2spec_deriv} and the Compton amplitude, \Cref{eq:compamp}, leads to the result quoted in \Cref{eq:secondorder_fh}.
		                
		In principle the derivation presented in this section (and \Cref{app:deriv}) can be generalized to mixed currents by adding an additional perturbation (or current) to \Cref{eq:fh_perturb} with a different coupling strength, $\lambda^\prime$, and current momentum, $\bv{q}^\prime$, which can, in general, be taken to be different from $\lambda$ and $\bv{q}$. This would allow access to interference terms, allowing one to study $u$--$d$ flavor interference effects, spin-dependent amplitudes or the off-forward Compton amplitude and generalized parton distributions with $\bv{q} \ne \bv{q^\prime}$. Details of a prescription for the off-forward Compton tensor will be presented in a forthcoming paper~\cite{Alec:2020fp}.

	\section{Simulation Details}\label{sec:ls}
		\subsection{Gauge ensembles}
			\begin{table*}
				\centering
				\caption{ \label{tab:gauge_details} Details of the gauge ensembles used in this work.}
				\setlength{\extrarowheight}{2pt}
				\begin{tabularx}{\textwidth}{CCCCCCCCCCC}
					\hline\hline
					$N_f$ & $c_{SW}$ & $\kappa_l$ & $\kappa_s$ & $L^3 \times T$ & $a$ & $m_\pi$ & $m_N$ & $m_\pi L$ & $Z_V$ & $N_\text{cfg}$\\
					&&&&& [fm] &[GeV]&[GeV]&&& \\
					\hline
					$2+1$ & 2.65 & 0.1209 & 0.1209 & $32^3\times64$ & 0.074(2) & $0.467(12)$ & $1.250(39)$ & $5.6$ & 0.8611(84) & 1763 \\
					\hline\hline
				\end{tabularx}
			\end{table*}
			We use a single gauge ensemble generated by the QCDSF/UKQCD Collaborations employing a stout-smeared non-perturbatively $\mathcal{O}(a)$-improved Wilson action for the dynamical up/down and strange quarks and a tree-level Symanzik improved gauge action~\cite{Cundy:2009yy}. We work on a volume of $L^3 \times T = 32^3 \times 64$, the bare coupling parameter is $\beta = 5.5$, and the lattice spacing, $a = 0.074(2)$ fm, is set using a number of flavor-singlet quantities~\cite{PhysRevD.84.054509,Bietenholz:2010jr,Horsley:2013wqa,Bornyakov:2015eaa}. We are working on the SU$(3)$-flavor symmetric point where the masses of all three quark flavors are set to approximately the physical flavor-singlet mass, $\overline{m} = (2 m_s + m_l)/3$ and corresponds to a pion mass of $\simeq 470$ MeV and $m_\pi L = 5.6$. Further details and advantages of this choice are discussed in~\cite{Bietenholz:2011qq,Bietenholz:2010jr}. The renormalization constant for the local vector current is determined to be $Z_V = 0.8611(84)$ by imposing the charge conservation on the Sachs electric form factor calculated at $Q^2=0$. This value is in agreement within statistical precision with the value determined in the chiral limit using the RI$^\prime$-MOM scheme~\cite{Constantinou:2014fka}. We tabulate the details in \Cref{tab:gauge_details} for the Reader's convenience. 

		\subsection{Feynman-Hellmann implementation}
			We implement the second-order Feynman-Hellmann theorem through the valance quarks. It can clearly be implemented at the hybrid Monte Carlo level, which would relay the effects of the perturbations to the sea-quarks~\cite{Chambers:2015bka}, however this would lead to a significant increase in required computing resources, so in this work we focus on the quark-line connected contributions to the Compton amplitude. To this end, we add the perturbation given in \Cref{eq:fh_perturb},
			\begin{align}\label{eq:fh_perturb_quarkact}
				S(\lambda) = S &+ \lambda \int d^3 z \cosE{q}{z} \mathcal{J}_{3}(z),
			\end{align}
            to the valence quark action only, where the renormalized local vector current, $\mathcal{J}_{3}(x) = Z_V \bar{q}(x)i\gamma_3 q(x)$, is chosen to be along the $z$-direction, $\mu=3$. The second exponential term symmetrizes the Fourier transform and ensures the Hermiticity of the action. In order to evaluate the second-order energy shift with respect to $\lambda$ at $\lambda=0$, one has to compute additional quark propagators at several choices of $\lambda$. This added cost of computation is countered by optimizing the inversion of the perturbed Dirac matrix. We adopt an approach where we feed the unperturbed propagator as an initial guess to the inversion of the perturbed one, which results in roughly a factor of 10 gain in inversion time. 

            In order to improve the stability of estimating the energy-shifts at $\lambda=0$, one aims to introduce the smallest possible perturbations by choosing a suitable $\lambda$. The objective is to keep $\lambda$ sufficiently small to minimize the contamination from $\lambda^4$ effects, yet large enough to ensure that the perturbation is not lost within the numerical precision of the calculation. Our tests indicate that any choice in the range $ 10^{-1} > |\lambda| \ge 10^{-5} $ leads to meaningful results. Note that the upper bound is sensitive to the quark mass, where a too large $\lambda$ might lead to increased instabilities in the Dirac matrix inversion, particularly as one approaches the physical point.

		\subsection{Flavor decomposition}
\begin{figure}
				\includegraphics[width=.485\textwidth, trim={7cm 11cm 7cm 11cm},clip]{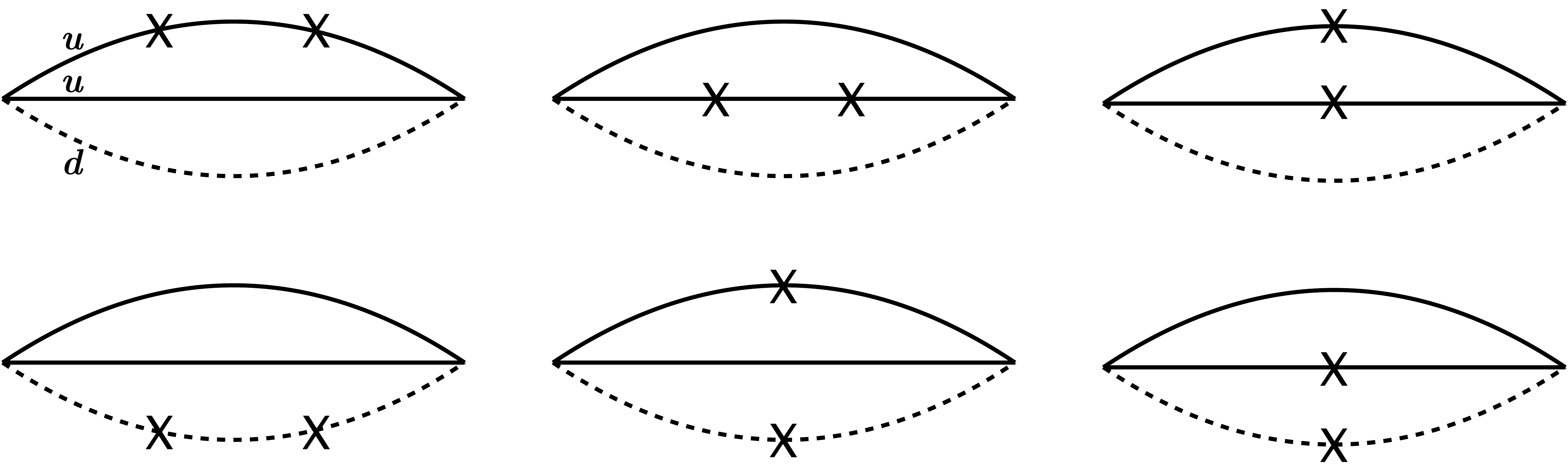}
				\caption{ \label{fig:current_inserts} The six possible ways of inserting two currents to a nucleon. Upper three correspond to the $uu$ flavor contributions while the lower leftmost is for $dd$. Remaining two are for the $ud$, which we omit in this work.}
			\end{figure}

			The implementation of the Feynman-Hellmann theorem described above effectively inserts an external current on to a quark line by computing its propagator with the perturbed quark action~\Cref{eq:fh_perturb_quarkact}. When both currents are inserted onto the $u$-quarks or the $d$-quark, we evaluate the ``$uu$'' or ``$dd$'' contributions to the Compton structure functions, respectively. By employing positive and negative pairs of $\lambda$'s (see \Cref{sec:rd}), one can form $u+d$ and $u-d$ type insertions leading to the possibility for isolating a ``$ud$'' insertion where one current hits a $u$-quark and the other the $d$ quark. The six different ways of inserting the currents are shown in \Cref{fig:current_inserts}. The $ud$ contribution is particularly interesting since it directly corresponds to a higher-twist contribution~\cite{Hannaford-Gunn:2020pvu}, i.e. the twist-4 \emph{cat's ears} diagram. An investigation of these contributions is left for future work.       

		\subsection{Isolating the energy shift}
			The energy of the ground state, $N_\lambda$, in a weakly coupled external field can be expanded as a Taylor series in $\lambda$,
			\begin{align}
				\begin{split}
					E_{N_\lambda}(\bv{p}) &= E_{N}(\bv{p}) + \lambda \left. \frac{\partial E_{N_\lambda}(\bv{p})}{\partial \lambda} \right|_{\lambda = 0} \\
					&+ \frac{\lambda^2}{2!} \left. \frac{\partial^2 E_{N_\lambda}(\bv{p})}{\partial \lambda^2} \right|_{\lambda = 0} + \mathcal{O}(\lambda^3).
				\end{split}
			\end{align}
			Collecting terms that are even and odd in $\lambda$ to all orders, we may rewrite the expansion as,
			\begin{equation} \label{eq:enshift_evenodd}
			 	E_{N_\lambda}(\bv{p}) = E_{N}(\bv{p}) + \Delta E_{N_\lambda}^e(\bv{p}) + \Delta E_{N_\lambda}^o(\bv{p}),
			\end{equation}
			where $E_N(\bv{p})$ in the above two expressions corresponds to the unperturbed ($\lambda=0$) energy. In order to extract the second order energy shift from the lattice correlation functions, we construct a ratio which isolates the even-$\lambda$ energy shift, $\Delta E_{N_\lambda}^e(\bv{p})$,
			\begin{align} 
				\label{eq:ratio}
				\mathcal{R}^e_\lambda (\bv{p},t) \equiv& \frac{G^{(2)}_{+\lambda}(\bv{p},t) G^{(2)}_{-\lambda}(\bv{p},t)}{\left( G^{(2)}(\bv{p},t) \right)^2} \\ 
				\label{eq:ratio_largetime}
				\xrightarrow{t \gg 0}& A_\lambda(\bv{p}) e^{-2\Delta E^e_{N_\lambda}(\bv{p}) t},
			\end{align}
			where the perturbed two-point functions, $G^{(2)}_{\pm\lambda}(\bv{p},t)$, are defined in \Cref{eq:G2,eq:G2spec} and $G^{(2)}(\bv{p},t)$ is the unperturbed one. The large $t$ behavior given in \Cref{eq:ratio_largetime} is arrived at by combining \Cref{eq:G2spec,eq:enshift_evenodd}. Note that the ratio in \Cref{eq:ratio} further eliminates any remaining (highly suppressed) contributions odd in $\lambda$ (i.e. $\mathcal{O}(\lambda)$, $\mathcal{O}(\lambda^3)$, $\cdots$). Therefore, any contribution at the order of $\lambda$ is eliminated by construction and higher-order contributions (i.e. $ \mathcal{O}(\lambda^4)$, $\cdots$) are heavily suppressed due to the weakness of the perturbations. While not necessary for our discussion, for completeness we note that the overlap factor is 
			\begin{align}
				\begin{split}
					&A_\lambda(\bv{p}) = \frac{|\langle \Omega_\lambda | \chi(0) | N_\lambda(\bv{p}) \rangle|^2}{2 E_{N_\lambda}(\bv{p})} \times \\ 
					&\frac{|\langle \Omega_\lambda | \chi(0) | N_{-\lambda}(\bv{p}) \rangle|^2}{2E_{N_{-\lambda}}(\bv{p})} 
					\left(\frac{ |\langle \Omega | \chi(0) | N(\bv{p}) \rangle|^2 }{2 E_N(\bv{p})}\right)^{-2}.
				\end{split}
			\end{align}
			Extraction of the even-$\lambda$ energy shift $\Delta E_{N_\lambda}^e$ then follows standard spectroscopy methods by fitting $\mathcal{R}^e_\lambda (\bv{p},t)$ defined in \Cref{eq:ratio} with a single exponential at sufficiently large times. Details follow in the next section.

	\section{Results And Discussion}\label{sec:rd}
		\subsection{Extracting the structure functions and the moments}
			In order to illustrate the feasibility and the versatility of the method, we carry out simulations with several values of current momentum, $Q^2$, in the range $3 \lesssim Q^2 \lesssim 7$ GeV$^2$. Utilizing up to six randomly placed quark sources per configuration, we perform up to $\mathcal{O}(10^4)$ measurements for each pair of $\lambda$ and $\bv{q}$. Quark fields are smeared in a gauge-invariant manner by Jacobi smearing~\cite{Allton:1993wc}, where the smearing parameters are tuned to produce a rms radius of $\simeq 0.5$ fm. We bin the measurements to account for the autocorrelations. In order to estimate the statistical errors, we pull a set of bootstrap samples from the binned dataset and perform all steps of the analysis on each sample. We access multiple $\omega$ values at each simulated value of $\bv{q}$ by varying the nucleon momentum $\bv{p}$ as shown in \Cref{tab:pmom}. 
			\begin{table}
				\centering
				\caption{\label{tab:pmom} Multiple $\omega$ values that we can access with several combinations of $\bv{p} = (p_x,p_y,p_z)$ and $\bv{q} = (q_x,q_y,q_z)$ in lattice units. Note that the $\omega \ge \emph{1}$ values are omitted in the analysis---values of $\omega$ outside the allowed range are indicated by italics. }
				\setlength{\extrarowheight}{2pt}
				\begin{tabularx}{.485\textwidth}{C|CCCCC}
					\hline\hline
					& \multicolumn{5}{c}{$\omega = 2 \bv{p}\cdot\bv{q}/Q^2$} \\
					\hline
					\multirow{2}{*}{$\bv{p} / (2\pi/L)$}  & \multicolumn{5}{c}{$\bv{q} / (2\pi/L)$} \\
					& $(3,1,0)$ & $(3,2,0)$ & $(4,1,0)$ & $(4,2,0)$ & $(5,1,0)$ \\
					\hline
					$(0,0,0)$	& $0.00$ & $0.00$ & $0.00$ & $0.00$ & $0.00$ \\
					$(0,1,0)$	& $0.20$ & $0.31$ & $0.12$ & $0.20$ & $0.08$ \\
					$(0,2,0)$	& $0.40$ & $0.62$ & $0.24$ & $0.40$ & $0.15$ \\
					$(1,-2,0)$	& $0.20$ & $0.16$ & $0.24$ & $0.00$ & $0.23$ \\
					$(1,-1,0)$	& $0.40$ & $0.16$ & $0.35$ & $0.20$ & $0.31$ \\
					$(1,0,0)$	& $0.60$ & $0.46$ & $0.47$ & $0.40$ & $0.39$ \\
					$(1,1,0)$	& $0.80$ & $0.77$ & $0.59$ & $0.60$ & $0.46$ \\
					$(1,2,0)$	& $\emph{1.00}$ & $\emph{1.08}$ & $0.71$ & $0.80$ & $0.54$ \\
					$(2,-1,0)$	& $\emph{1.00}$ & $0.62$ & $0.82$ & $0.60$ & $0.69$ \\
					$(2,0,0)$	& $\emph{1.20}$ & $0.92$ & $0.94$ & $0.80$ & $0.77$ \\
					$(2,1,0)$	& $\emph{1.40}$ & $\emph{1.23}$ & $\emph{1.06}$ & $\emph{1.00}$ & $0.85$ \\
					\hline\hline
				\end{tabularx}
			\end{table} 

			Setting $p_z = 0$ and $q_z = 0$, along with our choice of $\mu=3$, simplifies the Compton tensor such that the second order energy shift in \Cref{eq:secondorder_fh} corresponds to the $\mathcal{F}_1$ Compton structure function directly, 
			\begin{align} \label{eq:comptostruct}
				\begin{split}
					\left. \frac{\partial^2 E_{N_\lambda}(\bv{p})}{\partial^2 \lambda} \right|_{\lambda = \bv{0}} &= -\frac{T_{33}(p,q) + T_{33}(p,-q)}{2E_N(\bv{p})} \\
					&= -\frac{\mathcal{F}_1(\omega, Q^2)}{E_N(\bv{p})},
				\end{split}
			\end{align}
			where the energy of the nucleon is calculated via the continuum dispersion relation, $E_N(\bv{p}) = \sqrt{m_N^2 + \bv{p}^2}$.

\begin{figure}
				\centering
				\includegraphics[width=.485\textwidth]{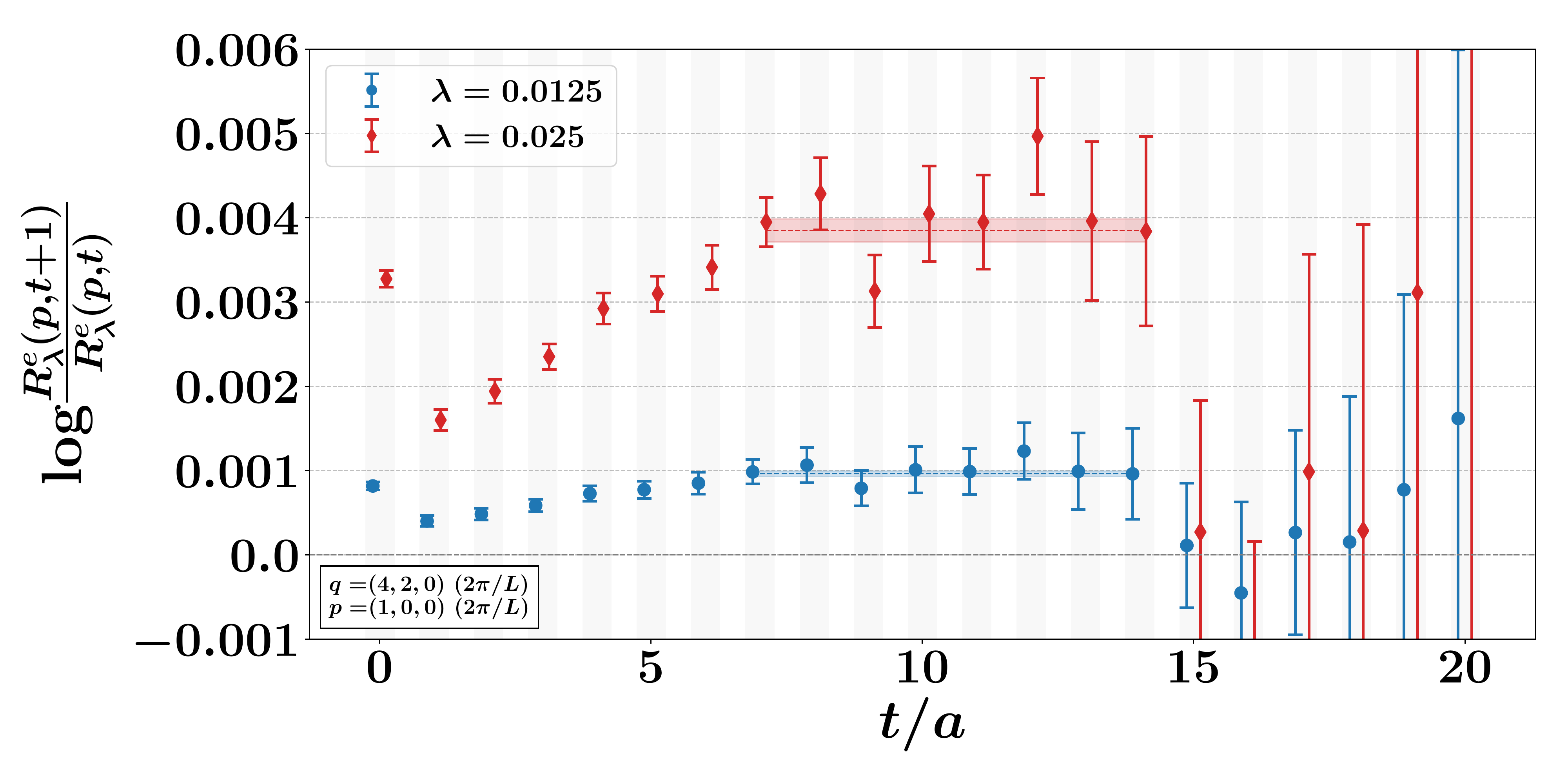}
				\caption{\label{fig:effmass}Effective mass plot of the ratio given in \Cref{eq:ratio} for $\bv{q} = (4,2,0) \latmom$ and $\bv{p} = (1,0,0) \latmom$. Shaded horizontal regions indicate the fit windows along with the extracted $\lambda$ values with their $1\sigma$ error margins. Data points are shifted for clarity.}
			\end{figure}

\begin{figure}
				\centering
				\includegraphics[width=0.465\textwidth]{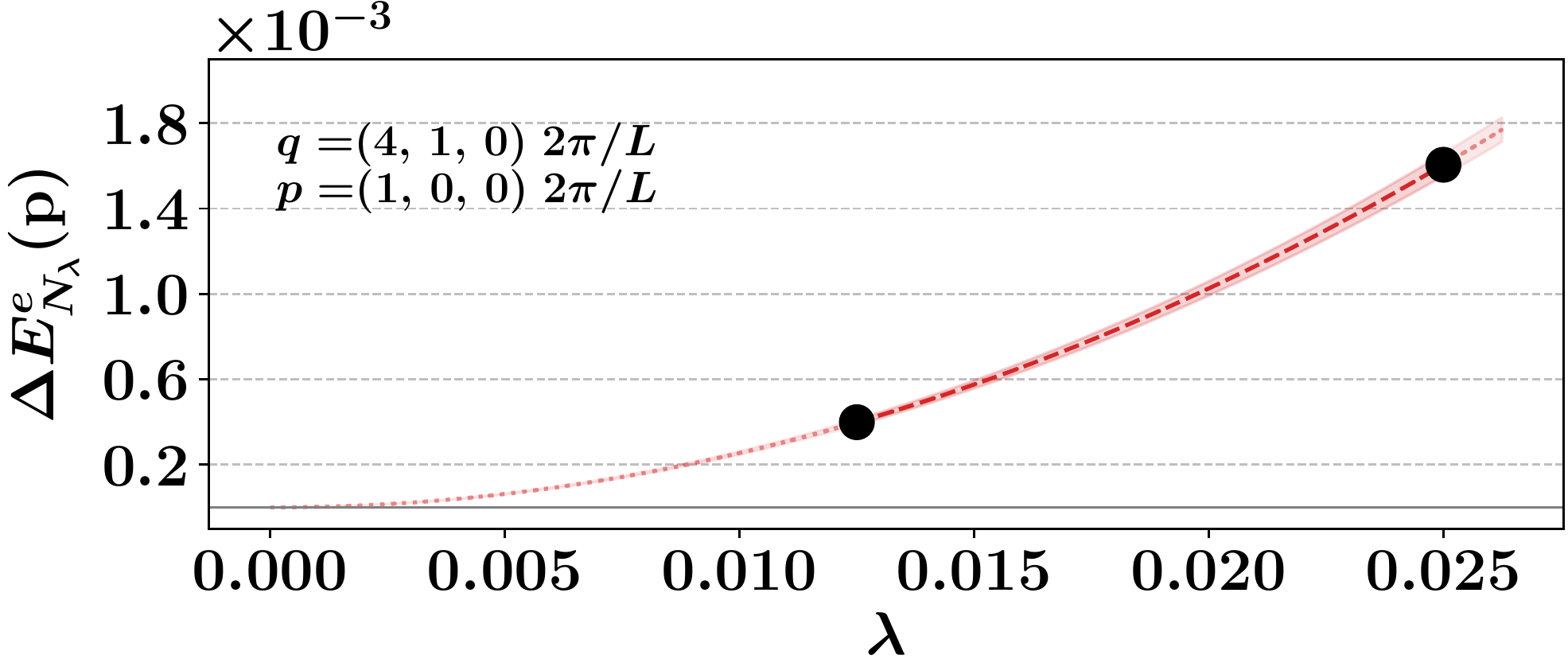}
				\caption{\label{fig:lambda_dep_q410}$\lambda$ dependence of $\Delta E^e_{N_\lambda}(\bv{p})$ given in \Cref{eq:even_Eshift}. Fit form is $f(\lambda) = b \lambda^2$. Error bars are smaller than the symbols.}
			\end{figure}

\begin{figure}
			 	\centering
				\includegraphics[width=.475\textwidth]{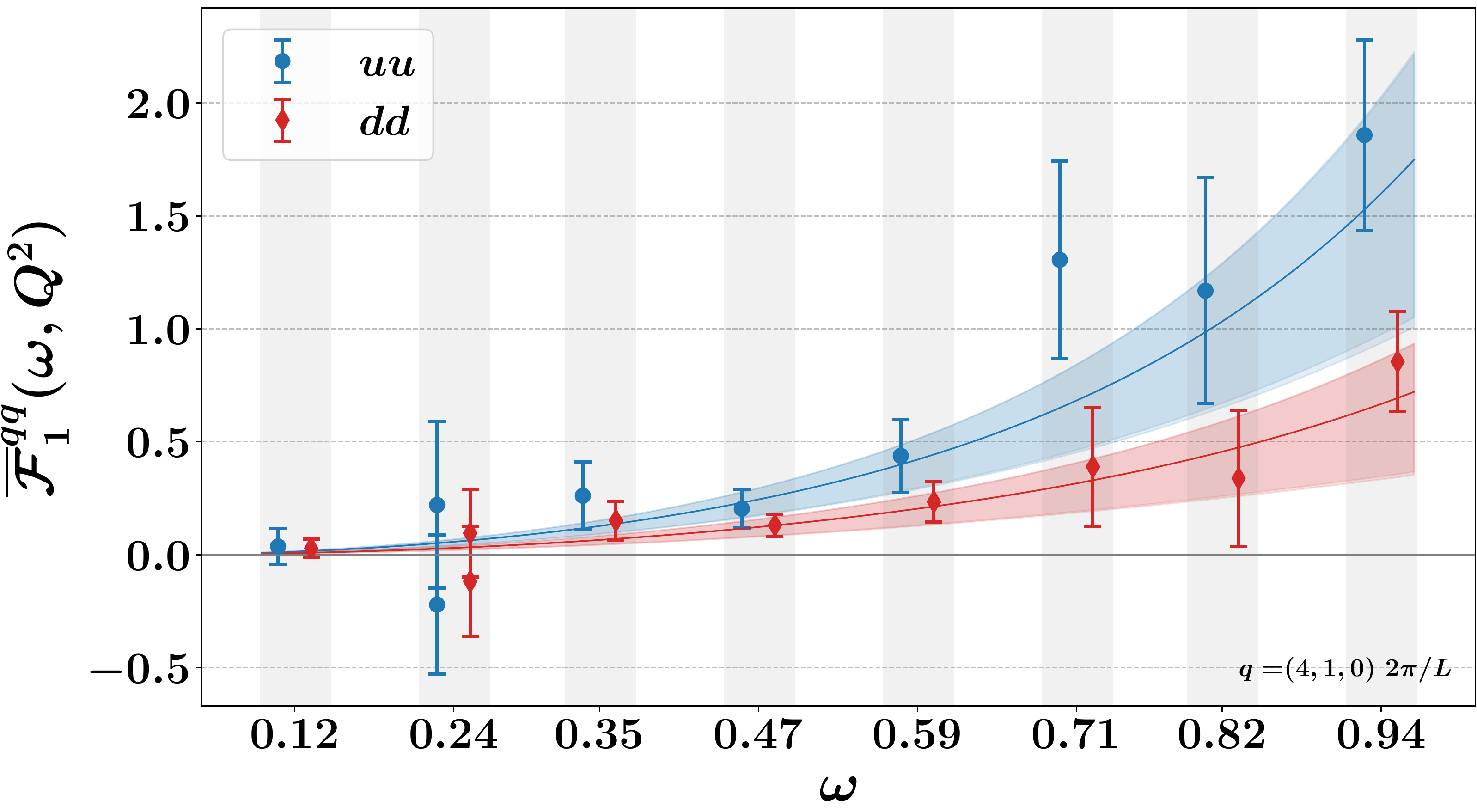}
				\caption{\label{fig:q410_moments} $\omega$ dependence of the subtracted Compton structure function $\overline{\mathcal{F}}_1^{qq}(\omega, Q^2)$ of nucleon at $Q^2 = 4.66$ GeV$^2$. Fits depicting the extraction of the moments via \Cref{eq:moments_taylor} with $n=6$ are shown as well. Shaded curves correspond to the $68\%$ credible region of the highest posterior density. Points are shifted for clarity. Corresponding moments are given in \Cref{tab:moments}.}
			\end{figure}

			We compute the perturbed two-point correlation functions (\Cref{eq:G2}) with four values of $\lambda = [\pm 0.0125, \pm 0.025]$. Even-$\lambda$ energy shifts are extracted from the ratio of correlation functions given in \Cref{eq:ratio} following a covariance-matrix based $\chi^2$ analysis to pick the best available range for each ratio. We show a representative case for $\bv{q} = (4,2,0) \latmom$ and $\bv{p} = (1,0,0) \latmom$ in \Cref{fig:effmass}. 
			Having even-$\lambda$ energy shifts for two $\lambda$ values, we perform polynomial fits of the form,
			\begin{equation} \label{eq:even_Eshift}
			 	\Delta E^e_{N_\lambda}(\bv{p}) = \frac{\lambda^2}{2} \left. \frac{\partial^2 E_{N_\lambda}(\bv{p})}{\partial \lambda^2} \right|_{\lambda = \bv{0}} + \mathcal{O}(\lambda^4),
			\end{equation} 
			to determine the second order energy shift. Given the smallness of our $\lambda$ values, higher order $\mathcal{O}(\lambda^4)$ terms are heavily suppressed, hence the fit form reduces to a simple one parameter polynomial. A representative fit of \Cref{eq:even_Eshift} to the energy shifts for $\bv{q} = (4,1,0) \latmom$ and $\bv{p} = (1,0,0) \latmom$ is shown in \Cref{fig:lambda_dep_q410}. 
			In \Cref{fig:q410_moments} we show the subtracted Compton structure function for $uu$ and $dd$ insertions as obtained from the energy shifts via \Cref{eq:comptostruct} for each value of $\omega$ for $\bv{q} = (4,1,0) \latmom$. 

			With our particular choice of Lorentz indices and momenta, we can connect the lattice Compton amplitude (\Cref{eq:comptostruct}) to the moments of the structure functions (\Cref{eq:ope_moments}) as,
			\begin{align} \label{eq:moments_taylor}
				\begin{split}
					\overline{\mathcal{F}}_1(\omega,Q^2) = 4 (\omega^2 M^{(1)}_{2}(Q^2) + \omega^4 M^{(1)}_{4}(Q^2) &\\
					 + \cdot \cdot \cdot + \omega^{2n} M^{(1)}_{2n}(Q^2) + \cdot \cdot \cdot)&.
				\end{split}
			\end{align}
			Since the Compton amplitude is directly related to the experimental cross-section, it must be positive definitive for the entire kinematic region. Consequently, this holds for the $uu$ and $dd$ moments as well. Hence, the moments $M^{(1)}_{2n}$ are constrained to be monotonically decreasing,
			\begin{equation} \label{eq:monotonic_moments}
				M^{(1)}_2(Q^2) \ge M^{(1)}_4(Q^2) \ge \cdot \cdot \cdot \ge M^{(1)}_{2n}(Q^2) \ge \cdot \cdot \cdot \ge 0.
			\end{equation}
			More generally, the sequence of moments satisfy the Hausdorff moment criteria \cite{Hausdorff}, yet the simple monotonic decreasing form of \Cref{eq:monotonic_moments} allows an assessment of the constraint on the moments provided by the data. This series is rapidly converging and stable with respect to the truncation order. However, imposing the above condition in a least-squares analysis is not so straightforward, but it can be easily implemented in a Bayesian approach. The particular Bayesian inference implementation we are employing~\cite{Salvatier:2016swf} has the advantage of using adaptive algorithms to optimize the hybrid Monte Carlo parameters~\cite{JMLR:v15:hoffman14a}, which removes the extra effort of fine-tuning such parameters and returns the sampling results within mere minutes with convergence checks implemented.  

			In the present analysis, we sample the moments from uniform distributions with bounds $M^{(1)}_{2}(Q^2) \in [0,1]$ and $M^{(1)}_{2n}(Q^2) \in [0, M^{(1)}_{2n-2}(Q^2)]$, for $n > 1$, to enforce the monotonic decreasing nature of the moments. Uniform prior distributions are chosen since these are uninformative distributions and remove a source of bias. The sequences of individual $uu$ or $dd$ moments are selected according to a multivariate probability distribution, $\operatorname{exp}(-\chi^2/2)$, where,
			\begin{equation}\label{eq:chi2}
			 	\chi^2 = \sum_{i,j} \left[ \overline{\mathcal{F}}_{1,i} - \overline{\mathcal{F}}_1^\text{obs}(\omega_i) \right] C^{-1}_{ij} \left[ \overline{\mathcal{F}}_{1,j} - \overline{\mathcal{F}}_1^\text{obs}(\omega_j) \right],
			\end{equation} 
			is the $\chi^2$ function with the covariance matrix $C_{ij}$, ensuring the correlations between the data points are taken into account. Fits depicting the extraction of the moments are shown in \Cref{fig:q410_moments}. Note that, \Cref{eq:monotonic_moments} is not necessarily true for the isovector, $uu-dd$, moments. Therefore, the Bayesian priors for the $uu$ and $dd$ are treated independently. However, by sampling the $uu$ and the $dd$ datasets within the same trajectory, we ensure underlying correlations between those datasets are accounted for. Hence, the indices $i$, $j$ in \Cref{eq:chi2} run through all the $\omega$ values and both flavors.

\begin{figure}
			 	\centering
				\includegraphics[width=.485\textwidth]{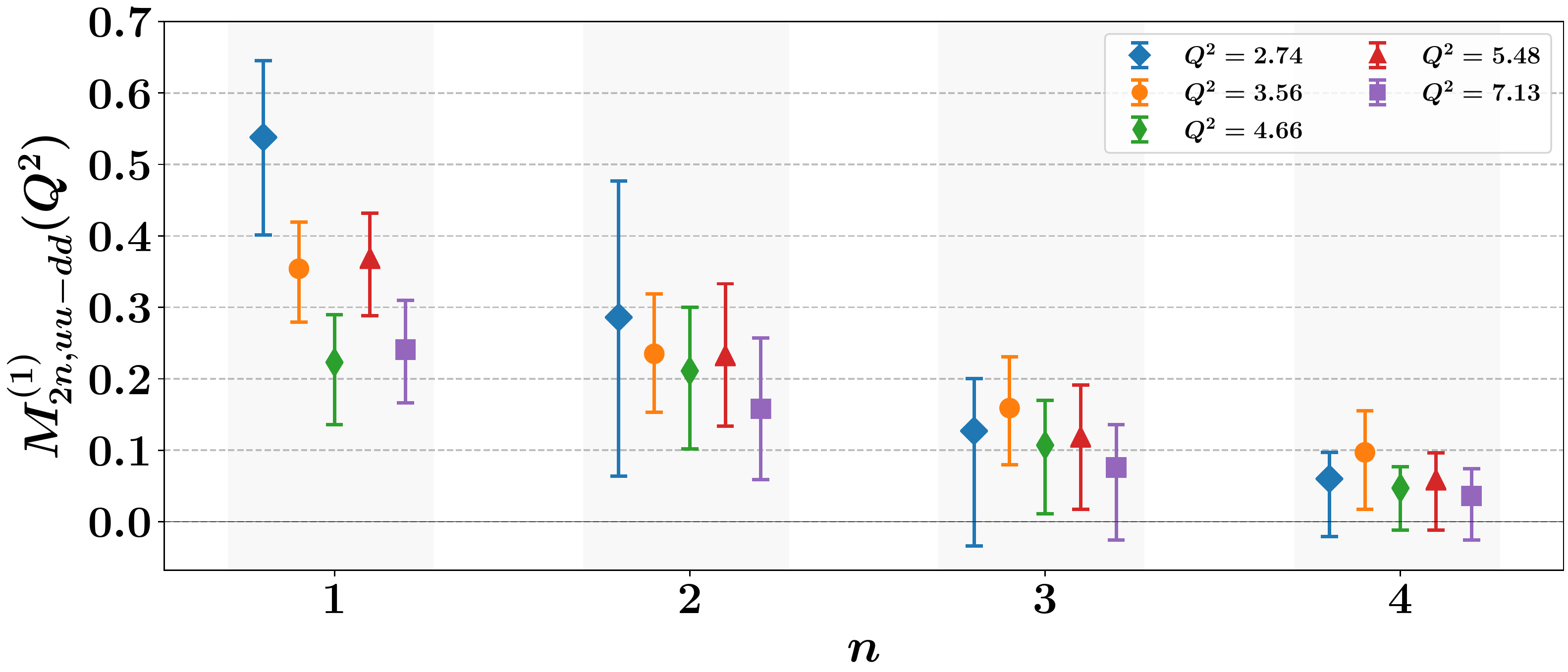}
				\caption{\label{fig:moments}Isovector moments given in \Cref{tab:moments}. $Q^2$ are given in GeV$^2$.}
			\end{figure}

			The first few moments extracted are given in \Cref{tab:moments} and the isovector moments are plotted in \Cref{fig:moments} for each choice of $Q^2$. Error margins correspond to the highest posterior density interval with a $68\%$ credible region and the asymmetric intervals reflect the shape of the posterior distributions. We find that the lower moments have a negligible dependence on the truncation order of the series in \Cref{eq:moments_taylor} for $n \ge 3$.
			
			It is evident from \Cref{fig:moments} that the lowest $Q^2$ point has considerably larger uncertainties than the others. The increased uncertainty of the moments is largely caused by the fact that as $Q^2$ is decreased, there are less kinematically-available $\omega$ points (see \Cref{tab:pmom}). As a consequence, the series expansion coefficients in \Cref{eq:moments_taylor} are less well constrained,
			despite the statistical precision of the point-wise Compton amplitude being similar across all $Q^2$ values.
			\begin{table}
				\centering
				\caption{\label{tab:moments} First few moments of the structure function $\mathcal{\overline{F}}_1$ for several values of $Q^2$. Contributions of each flavor are given along with the isovector quantity. Errors are statistical uncertainties only. }
				\setlength{\extrarowheight}{5pt}
				\resizebox{.485\textwidth}{!}{
				\begin{tabular}{cccccc}
					\hline\hline
					$\bv{q} / (2\pi/L)$ & $Q^2$ [GeV$^2$] 	
						& $M^{(1)}_{2n}(Q^2)$ & $uu$ 	& $dd$ 		& $uu-dd$	\\
					\hline
					\multirow{4}{*}{$(3,1,0)$} & \multirow{4}{*}{$2.74$}	
						&  $M^{(1)}_{2}$ 	& $0.797^{+0.152}_{-0.167}$ 
													& $0.259^{+0.088}_{-0.096}$	
													& $0.538^{+0.107}_{-0.137}$ \\
						&& $M^{(1)}_{4}$ 	& $0.374^{+0.204}_{-0.226}$
													& $0.088^{+0.023}_{-0.088}$
													& $0.286^{+0.191}_{-0.222}$ \\
						&& $M^{(1)}_{6}$ 	& $0.166^{+0.043}_{-0.166}$
													& $0.039^{+0.008}_{-0.039}$
													& $0.127^{+0.073}_{-0.161}$ \\
						&& $M^{(1)}_{8}$ 	& $0.078^{+0.013}_{-0.078}$
													& $0.019^{+0.001}_{-0.019}$
													& $0.060^{+0.037}_{-0.081}$ \\												
					\hline
					\multirow{4}{*}{$(3,2,0)$} & \multirow{4}{*}{$3.56$}	
						&  $M^{(1)}_{2}$ 	& $0.509^{+0.088}_{-0.095}$
													& $0.155^{+0.046}_{-0.052}$
													& $0.354^{+0.065}_{-0.075}$ \\
						&& $M^{(1)}_{4}$ 	& $0.321^{+0.082}_{-0.097}$
													& $0.085^{+0.045}_{-0.046}$
													& $0.235^{+0.084}_{-0.082}$ \\
						&& $M^{(1)}_{6}$ 	& $0.199^{+0.079}_{-0.084}$
													& $0.040^{+0.011}_{-0.040}$
													& $0.159^{+0.072}_{-0.079}$ \\
						&& $M^{(1)}_{8}$ 	& $0.115^{+0.050}_{-0.094}$
													& $0.018^{+0.004}_{-0.018}$
													& $0.097^{+0.058}_{-0.080}$ \\
					\hline
					\multirow{4}{*}{$(4,1,0)$} & \multirow{4}{*}{$4.66$}
						&  $M^{(1)}_{2}$ 	& $0.479^{+0.089}_{-0.125}$
													& $0.257^{+0.067}_{-0.075}$
													& $0.223^{+0.067}_{-0.087}$	\\
						&& $M^{(1)}_{4}$ 	& $0.300^{+0.099}_{-0.109}$
													& $0.090^{+0.036}_{-0.073}$
													& $0.211^{+0.089}_{-0.109}$	\\
						&& $M^{(1)}_{6}$ 	& $0.145^{+0.066}_{-0.108}$
													& $0.039^{+0.009}_{-0.039}$
													& $0.107^{+0.063}_{-0.096}$	\\
						&& $M^{(1)}_{8}$ 	& $0.066^{+0.015}_{-0.066}$
													& $0.018^{+0.003}_{-0.018}$
													& $0.047^{+0.030}_{-0.059}$ \\
					\hline
					\multirow{4}{*}{$(4,2,0)$} & \multirow{4}{*}{$5.48$}			
						&  $M^{(1)}_{2}$ 	& $0.576^{+0.095}_{-0.099}$
													& $0.208^{+0.051}_{-0.060}$
													& $0.368^{+0.064}_{-0.080}$	\\
						&& $M^{(1)}_{4}$ 	& $0.329^{+0.114}_{-0.112}$
													& $0.097^{+0.050}_{-0.065}$
													& $0.232^{+0.101}_{-0.098}$ \\	
						&& $M^{(1)}_{6}$ 	& $0.160^{+0.072}_{-0.117}$
													& $0.043^{+0.010}_{-0.043}$
													& $0.118^{+0.073}_{-0.101}$ \\
						&& $M^{(1)}_{8}$ 	& $0.078^{+0.020}_{-0.078}$	
													& $0.020^{+0.002}_{-0.020}$
													& $0.058^{+0.038}_{-0.070}$ \\
					\hline
					\multirow{4}{*}{$(5,1,0)$} & \multirow{4}{*}{$7.13$}				
						&  $M^{(1)}_{2}$ 	& $0.429^{+0.102}_{-0.114}$
													& $0.188^{+0.064}_{-0.077}$
													& $0.241^{+0.069}_{-0.075}$ \\
						&& $M^{(1)}_{4}$ 	& $0.261^{+0.119}_{-0.123}$
													& $0.103^{+0.050}_{-0.073}$
													& $0.158^{+0.099}_{-0.099}$ \\
						&& $M^{(1)}_{6}$ 	& $0.132^{+0.035}_{-0.132}$
													& $0.055^{+0.015}_{-0.055}$
													& $0.076^{+0.060}_{-0.102}$ \\
						&& $M^{(1)}_{8}$ 	& $0.064^{+0.014}_{-0.064}$
													& $0.029^{+0.004}_{-0.029}$
													& $0.036^{+0.038}_{-0.062}$ \\
					\hline\hline
				\end{tabular}
				}
			\end{table} 
			
			We note that although the fall-off of the moments is quite evident, the second moments do not decrease as rapidly as one would expect from DIS data. Combined with the interplay between the $u$ and the $d$ moments, this leads to rather large second moments for the isovector $u-d$ combination, which are even comparable to that of the first moment in some cases. While this is likely due to the limited statistics of the current simulations, it may also be a signal of significant power corrections which we discuss in the next section. 

        \subsection{Power corrections and scaling}
	        The moments $M^{(1)}_{2n}(Q^2)$ in Table III appear to indicate that power corrections are present throughout our range of photon momenta, $3 \lesssim Q^2 \lesssim 7\,\mbox{GeV}^2$. The leading power corrections to moments of structure functions have essentially two sources, target mass (together with possible threshold effects) and mixing with operators of higher twist. Target mass and threshold effects can be accounted for, to a certain
	        extent, by replacing the Bjorken $x$ scaling variable by a generalized scaling variable, $\xi$ (e.g. \cite{Bloom:1970xb,Nachtmann:1973mr,Kuroda:1978jx}). For example, a commonly used form proposed by Nachtmann
	        \cite{Nachtmann:1973mr} is
	        \begin{equation}
	            \xi = \frac{2x}{1+\sqrt{1+4m_N^2x^2/Q^2}},   
	        \end{equation}
	        where $m_N$ is the nucleon mass. Besides being power corrected, the various moments defined in terms of these new $\xi$ scaling variables are mixtures of the (Cornwall-Norton~\cite{Cornwall:1968cx}) moments defined in terms of $x$, with the mixings suppressed by powers of $1/Q^2$. For a generalized scaling variable incorporating the analytic structure of the forward Compton amplitude see Ref.~\cite{Bloom:1970xb}. 

	        The second source of power corrections in the structure function moments are due to contributions from operators with twist-4 and above, represented by the $\mathcal{O}(1/Q^2)$ terms in \Cref{eq:leading-twist}. We note that while in principle it is possible to compute the Wilson coefficient for the twist-4 contribution to $M_2^{(1)}(Q^2)$ non-perturbatively on the lattice \cite{Capitani:1998fe,CAPITANI1999173,Bietenholz:2009rb}, the hyper-cubic nature of the lattice can only accommodate operators of spin four or less, which thwarts any direct prediction of the Wilson coefficients for higher moments.

\begin{figure}
	            \centering
	            \includegraphics[width=.485\textwidth]{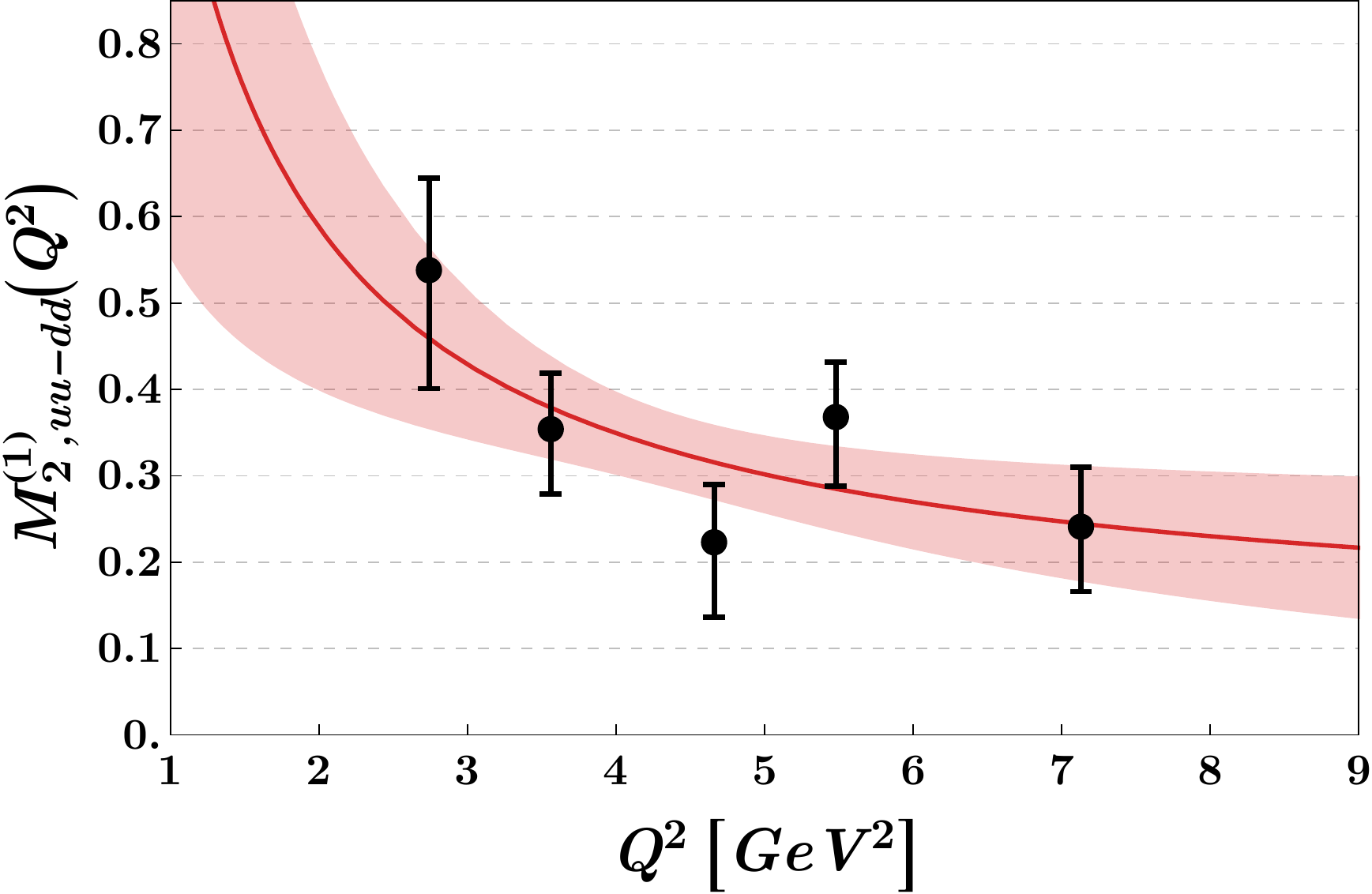}
	            \caption{\label{fig:moments_qDep}$Q^2$ dependence of the isovector $M^{(1)}_{2,uu-dd}(Q^2)$ moment. Black data points (tabulated in \Cref{tab:moments}) are obtained from independent fits to the $\omega$-dependence of $\overline{\mathcal{F}}_1(\omega,Q^2)$ at fixed $Q^2$. Curve shows the fit to \Cref{eq:moment_q2_correc_fit}.
	            }
	        \end{figure}

	        Since the moments computed in this work are determined from a fit to the full Compton amplitude, they naturally include all possible power corrections. While the present data do not have the precision or range of $Q^2$ to isolate individual power corrections, we are able to account for the observed $Q^2$-dependence of each moment by fitting with the functional form
	        \begin{equation} \label{eq:moment_q2_correc_fit}
	            M^{(1)}_{2n}(Q^2) = M^{(1)}_{2n} + C_{2n}/Q^2 + \mathcal{O}(1/Q^4)\,.
	        \end{equation}
	        Of particular interest is a fit to the lowest isovector moment $M^{(1)}_{2,uu-dd}(Q^2)$, which we show in \Cref{fig:moments_qDep} as a function of $Q^2$. Here we clearly see that the current data is well described by \Cref{eq:moment_q2_correc_fit}, with the fit form suggesting large power corrections may be present at low $Q^2$. However we add a cautionary note that the behavior at small $Q^2$ is heavily influenced by the presence of the large value for the moment at $Q^2=2.74$~GeV$^2$ and neglecting this point would result in a much softer, although still non-trivial, $Q^2$-dependence in the small-$Q^2$ region.

	        The phenomenological values of the moments in the language of the parton model, namely $v_{2n}(\mu)$ in \Cref{eq:leading-twist}, are commonly quoted at the scale $\mu=2$ GeV. The results presented in \Cref{fig:moments_qDep} indicate that in our approach this number should be obtained from first taking the asymptotic value of $M^{(1)}_{2,uu-dd}(Q^2)$, say at $Q^2 \gtrsim 16$ GeV$^2$, then performing a perturbative rescaling down to $\mu=2 \text{ GeV}$. In practice, to achieve a reliable prediction would require an extension of the current simulations to larger values of $Q^2$ and a further increase in statistics.

	        At this stage we refrain from extending the present analysis to the higher moments, however an immediate avenue of study will be to investigate if the observed enhancement of the $M_4^{(1)}$ isovector moment persists with higher statistics over a larger $Q^2$ range. Should this be the case, it will be interesting to compare with the enhancement observed in the empirical results at small-$Q^2$ (see e.g. \cite{Melnitchouk:2005zr}).

	        In addition to studying the $Q^2$-dependence of the $uu$, $dd$ and $uu-dd$ moments of $\overline{\mathcal{F}}_1(\omega,Q^2)$, the techniques described in this work can be easily extended to investigations of additional quantities such as the higher-twist $ud$ moments (see the final two
	        diagrams in \Cref{fig:current_inserts}) or a test of the Callen-Gross relation at small $Q^2$. For a first attempt see~\cite{Hannaford-Gunn:2020pvu}. We emphasize that this work clearly demonstrates that a study of the $Q^2$-dependence of such observable on the lattice is now possible.

    \subsection{Comments on systematics}
        The focus of the present manuscript is to expand on the proposal introduced in Ref.~\cite{Chambers:2017dov}, where we here provide the theoretical foundation and present first detailed numerical results. However, as is common to all lattice QCD calculations, systematic uncertainties must be addressed in order to make a direct comparison with experimental and/or empirical results. The method presented here is also subject to the usual systematics encountered in conventional lattice observables: finite volume, finite lattice spacing and (possibly) unphysical quark masses. However, it is worth highlighting a couple of the challenges regarding the particular systematics in the present formulation.
        
        The rotational symmetry is always broken on the lattice. Hence, while the formulation avoids the need to invoke the OPE, the lattice Compton amplitude will necessarily be subject to discretisation artefacts that reflect the loss of rotational symmetry. From this perspective, it would be desirable to extrapolate the Compton amplitude, at fixed $\omega$ and $Q^2$, to the continuum limit {\em before} the determination of the corresponding moments. A detailed investigation and exploration of implementation strategies remain beyond the scope of the present work.
        
        Given that the energy eigenstates extracted all correspond to the ground-state hadron in that channel, the finite volume corrections must scale with $e^{-m_\pi L}$ in the same way as ordinary masses (or energies) \cite{Luscher:1985dn}. In future precision studies, it may be essential to resolve these effects, motivating a variant of the work presented in Ref.~\cite{Briceno:2018lfj}.
        
        The Feynman-Hellmann technique simplifies the study of excited-state contamination, since only one time variable is required. For most of the kinematics presented in this work, the additional thresholds introduced at finite external field strength $\lambda$ are energetically suppressed compared to the free-field excitations---such as seen in \Cref{fig:threshold}. At the relatively-large $Q^2$ values considered here, the low-lying nucleon poles will be further suppressed by the square of the elastic form factors. In this case, ground-state saturation times will be comparable to the free-field correlator. However, at lower $Q^2$ values, and especially as $\omega$ approaches 1, the elastic nucleon poles could lie relatively low compared to the free-field excitations. This could give rise to a distinct excited-state contamination in the correlator at ${\cal O}(\lambda)$---hence requiring some form of variational-like approach across multiple Fourier momenta.

	\section{Summary and Conclusions}\label{sec:sc}
		We have presented a derivation of the second-order Feynman-Hellmann theorem and its relationship to the forward Compton amplitude. In particular, the Compton amplitude can be computed directly on the lattice with a simple extension of the already established lattice QCD implementations of the Feynman-Hellmann theorem, devoid of operator mixing and related complications of the conventional approach. In order to illustrate the feasibility and the versatility of this method in directly probing nucleon structure functions, we have performed high-statistics simulations for several photon momenta, $Q^2$, on the $2+1$-flavor, $32^3 \times 64$ QCDSF/UKQCD lattices at the $SU(3)$ flavor symmetric point corresponding to a pion mass of $\simeq 470$ MeV. By studying the Compton amplitude across a range of kinematics, we have presented non-trivial signals for the first few moments of the nucleon structure functions. By revealing the $Q^2$ dependence of the low moments, there is a clear opportunity to directly study the evolution to the partonic regime  — more detailed investigations of the power corrections will be pursued in future work. Beyond studying the approach to the partonic regime, the method could also be applied at smaller $Q^2$ and probe the dynamics of low-energy Compton scattering processes \cite{Lensky:2017bwi,Seng:2019plg}. While moments of structure functions are relatively straightforward, there is also a prospect to invert the Compton amplitude to extract the $x$-dependence of the structure functions directly — see Ref.~\cite{Horsley:2020ltc} for some first attempts.

	\acknowledgments
		The numerical configuration generation (using the BQCD lattice QCD program~\cite{Haar:2017ubh})) and data analysis (using the Chroma software library~\cite{Edwards:2004sx}) was carried out on the IBM BlueGene/Q and HP Tesseract using DIRAC 2 resources (EPCC, Edinburgh, UK), the IBM BlueGene/Q (NIC, J\"{u}lich, Germany) and the Cray XC40 at HLRN (The North-German Supercomputer Alliance), the NCI National Facility in Canberra, Australia (supported by the Australian Commonwealth Government) and Phoenix (University of Adelaide). RH is supported by STFC through grant ST/P000630/1. HP is supported by DFG Grant No. PE 2792/2-1. PELR is supported in part by the STFC under contract ST/G00062X/1. GS is supported by DFG Grant No. SCHI 179/8-1. KUC, RDY and JMZ are supported by the Australian Research Council grant DP190100297.

	\appendix
		\section{Operator product expansion} \label{app:ope}
			For completeness, we summarize the relevant notation to complement the OPE discussion in Section~\ref{sec:ope}. The (leading-order) Wilson coefficients are given by:
			\begin{equation}
				C_{f,2n}^{(j)}=\mathcal{Q}_f^2+\mathcal{O}(g^2), \quad j=1,2\\
			\end{equation}
			and the hadron matrix elements are defined by
			\begin{equation}
				\langle p, s|\left[\mathcal{O}_f^{\{\mu_1\ldots \mu_n\}}-\operatorname{Tr}\right] |p,s\rangle =
				2v_n^f\left[p^{\mu_1}\ldots p^{\mu_n}-\operatorname{Tr} \right],
			\end{equation}
			in terms of the traceless and symmetric parts of the local quark bilinears:
			\begin{equation}
				\mathcal{O}^{\{\mu_1\ldots\mu_n\}}_q=i^{n-1}\overline{\psi}_q\gamma^{\mu_1}\Dfb^{\mu_2}\cdots \Dfb^{\mu_n}\psi_q,
			\end{equation}
			and similarly for the gluons
			\begin{equation}
				\mathcal{O}^{\{\mu_1\ldots\mu_n\}}_g=i^{n-2}\operatorname{Tr} F^{\mu_1\nu}\Dfb^{\mu_2}\cdots \Dfb^{\mu_n} F^{\mu_n}_{\phantom{\mu_n}\nu},
			\end{equation}
			where $\Dfb=\tfrac12(\overrightarrow{D}-\overleftarrow{D})$.
			  
		\section{Second order Feynman-Hellmann Theorem} \label{app:deriv}
			For the interested reader, we provide some of the key intermediate steps to produce the principal derivation presented in the main text, \Cref{eq:secondorder_fh}.

			The 2-point nucleon correlator in an external field, \Cref{eq:fh_perturb}, is given by:
			\begin{equation}\label{eq:2pt}
				G^{(2)}_\lambda(\bv{p};t) \equiv \int d^3 x e^{-i \bv{p} \cdot \bv{x}} \bs{\Gamma} \; \langle \Omega_\lambda | \chi(\bv{x},t) \overline{\chi}(0)|\Omega_\lambda\rangle,
			\end{equation}
			where $\bs{\Gamma}$ the spin-parity projection matrix, with trace implied, and $|\Omega_\lambda\rangle$ is the vacuum in the presence of the external field. A nucleon interpolating operator is assumed for $\chi$.
			
			The strategy to derive the second-order Feynman-Hellmann relation is to consider the general form of the spectral decomposition of the Euclidean correlator, and match the energy shift against the explicit decomposition of the correlator in the presence of a weak external field. Following the usual procedure of
			inserting a complete set of states in between the operators, $\sumint_\XX \frac{d^3k}{(2\pi)^3} \frac{1}{2E_{X_\lambda}(\bv{k})} |X_\lambda(\bv{k})\rangle \langle X_\lambda(\bv{k})|$, and carrying out the momentum integral, the spectral decomposition of \Cref{eq:2pt} in the large (Euclidean) time limit, where the ground state dominance is realized, is given as,
			\begin{equation}\label{eq:2pt_spec}
				G^{(2)}_\lambda(\bv{p};t) \simeq A_\lambda(\bv{p}) e^{-E_{N_\lambda}(\bv{p}) t},
			\end{equation}
			where $E_{N_\lambda}(\bv{p})$ and $A_\lambda(\pb)$ are the energy of the ground state nucleon and overlap factor, respectively, in the background field. We note that in the presence of the background field, the Hamiltonian of the system will mix momentum states $\bv{p}\pm n \bv{q}$---with that $\bv{p}$ chosen to correspond to the lowest kinetic energy, $|\bv{p}|<|\bv{p}+n\bv{q}|\ \forall n\in\mathbb{Z}$. 

			The second order derivative of \Cref{eq:2pt_spec} with respect to $\lambda$, evaluated at $\lambda=0$, is given by:
			\begin{widetext}
				\begin{align}
					 \left. \frac{\partial^2 G^{(2)}_\lambda(\bv{p};t)}{\partial \lambda^2} \right |_{\lambda=0} &= e^{-E_{N}(\bv{p}) t} \left[ \frac{\partial^2 A_\lambda(\bv{p})}{\partial \lambda^2} - t \left( 2\frac{\partial A_\lambda(\bv{p})}{\partial \lambda} \frac{\partial E_{N_\lambda}(\bv{p})}{\partial \lambda} + A(\bv{p}) \frac{\partial^2 E_{N_\lambda}}{\partial \lambda^2} \right) + t^2 A(\bv{p}) \left(\frac{\partial E_{N_\lambda}(\bv{p})}{\partial \lambda}\right)^2 \right].
				\end{align} 
				The derivatives of $A_\lambda(\bv{p})$ and $E_{N_\lambda}(\bv{p})$ are understood to be evaluated at $\lambda=0$. The first-order energy shifts vanish, $\partial E_N / \partial \lambda = 0$, provided we restrict ourselves to the non-Breit-frame kinematics, i.e. $|\bv{p}| \ne |\bv{p} \pm \bv{q}|$ \cite{Chambers:2014qaa,Chambers:2017tuf}. In this case, the above equation thus reduces to
				\begin{align}\label{eq:fh_spec}
					\left. \frac{\partial^2 G^{(2)}_\lambda(\bv{p};t)}{\partial \lambda^2} \right |_{\lambda=0} = e^{-E_{N}(\bv{p}) t}\left[ \frac{\partial^2 A_\lambda(\bv{p})}{\partial \lambda^2}  - t A(\bv{p}) \frac{\partial^2 E_{N_\lambda}(\bv{p})}{\partial \lambda^2} \right].
				\end{align}
				where the first term corresponds to the shift in the overlap factor and the second order energy shift is identified in the $t$-enhanced or the time-enhanced term. The familiar overlap factor is given by:
				\begin{equation}
				    A(\pb) = \sum_s \frac{1}{2E_{N}(\bv{p})} \bv{\Gamma}\ea{\Omega | \chi(0) |N(\pb,s)}\ea{N(\pb,s)|\overline{\chi}(0)|\Omega}.
				\end{equation}

				We now directly evaluate the second-order derivative within the path integral formalism. The 2-point correlation function takes the form:
				\begin{align}
				  {}_\lambda\langle\chi(\xb,t)\overline{\chi}(0)\rangle_\lambda =\frac{1}{\mathcal{Z}(\lambda)}\int \mathcal{D}\psi\mathcal{D}\overline{\psi}\mathcal{D}U\,\chi(\xb,t)\overline{\chi}(0)\,e^{-S(\lambda)},
				\end{align}
				where $S(\lambda)$ is the perturbed action given in \Cref{eq:fh_perturb}, and $\Zlam$ is the corresponding partition function. Projecting the 2-point function to definite momenta and spin gives the standard correlator,
				\begin{equation}\label{eq:2pt_path}
					G^{(2)}_{\lambda}(\bv{p};t) = \int d^3 x\, e^{-i \bv{p} \cdot \bv{x}}\bv{\Gamma}\,   {}_\lambda\langle\chi(\xb,t)\overline{\chi}(0)\rangle_\lambda
				\end{equation}

				To simplify the following expressions, we use the shorthand notation to describe the product of interpolating operators,
				\begin{align}
				  \mathcal{G}=\int d^3 x\, e^{-i \bv{p} \cdot \bv{x}} \bv{\Gamma}\chi(\bv{x},t) \overline{\chi}(0).
				\end{align}
				The first-order derivative of the correlator is then given by
				\begin{align} \label{eq:fh_firstorder}
				\frac{\partial \eal{\mathcal{G}}}{\partial \lambda} = \ea{\mathcal{G}}_\lambda  \ea{\frac{\partial S(\lambda)}{\partial \lambda}}_\lambda 
					- \ea{\mathcal{G} \frac{\partial S(\lambda)}{\partial \lambda}}_\lambda.
				\end{align}
				The first term corresponds to a vacuum shift and the second term encodes a the three-point correlation function that is related to the first-order energy shift. This term has been discussed in detail and applied to the calculation of forward matrix elements \cite{Chambers:2014qaa} and form factors \cite{Chambers:2017tuf}. For the Compton amplitude, the second order derivative is required, which is straightforward to evaluate,
				\begin{align} \label{eq:pi_secondorder}
					\begin{split}
						\frac{\partial^2 \eal{\mathcal{G}}}{\partial \lambda^2} &= \ea{\mathcal{G}}_\lambda \ea{\frac{\partial^2 S(\lambda)}{\partial \lambda^2}}_\lambda + \ea{\mathcal{G} \frac{\partial^2 S(\lambda)}{\partial \lambda^2}}_\lambda + \ea{\mathcal{G}}_\lambda \ea{\left(\frac{\partial S(\lambda)}{\partial \lambda}\right)^2}_\lambda + 2\ea{\mathcal{G}}_\lambda \ea{\frac{\partial S(\lambda)}{\partial \lambda}}_\lambda \ea{\frac{\partial S(\lambda)}{\partial \lambda}}_\lambda \\
						&\quad- 2\ea{\mathcal{G} \frac{\partial S(\lambda)}{\partial \lambda}}_\lambda \ea{\frac{\partial S(\lambda)}{\partial \lambda}}_\lambda + \ea{\mathcal{G} \left(\frac{\partial S(\lambda)}{\partial \lambda}\right)^2}_\lambda.
					\end{split}
				\end{align}
				The first two terms vanish when the external perturbation is purely linear in $\lambda$. In the limit $\lambda\to0$, vacuum matrix elements of the external
				fields vanish, $\ea{\partial S(\lambda) / \partial \lambda}=0$, assuming the operator doesn't carry vacuum quantum numbers, such as the electromagnetic current---the scalar current would be an obvious counter example. The term involving $\ea{\left(\partial S(\lambda)/\partial \lambda\right)^2}$ will not in general vanish, however this can only act as a multiplicative factor on the free-field correlator and hence cannot contribute to the time-enhanced term in \Cref{eq:fh_spec}. The second-order energy shift can therefore only arise from the final term in \Cref{eq:pi_secondorder},
				\begin{equation}\label{eq:fh_pigen} 
					\left. \frac{\partial^2 \ea{\mathcal{G}}_\lambda}{\partial \lambda^2} \right |_{\lambda=0} = \ea{\mathcal{G} \left(\frac{\partial S(\lambda)}{\partial \lambda}\right)^2}+\ldots,
				\end{equation}
				where the ellipsis denotes terms that are not time-enhanced. By restoring the explicit form for $\mathcal{G}$, we have
				\begin{align} \label{eq:fh_pi}
					\begin{split}
						\left.\frac{\partial^2 G^{(2)}_\lambda(\bv{p};y)}{\partial \lambda^2} \right |_{\lambda=0} = &\int d^3x\, e^{-i \bv{p} \cdot \bv{x}}\bv{\Gamma} \ea{\chi(\bv{x},t) \overline{\chi}(0) \left(\frac{\partial S(\lambda)}{\partial \lambda}\right)^2},
					\end{split}
				\end{align}

				Using our explicit form for the electromagnetic external field, the corresponding second derivative of the correlator is given by
				\begin{align} \label{Aeq:4pt}
					\left. \frac{\partial^2 G^{(2)}_\lambda(\bv{p};t)}{\partial \lambda^2} \right |_{\lambda=0} = \int d^3x e^{-i \bv{p} \cdot \bv{x}} \bv{\Gamma}\int d^4y d^4z (e^{i \bv{q} \cdot \bv{y}} + e^{-i \bv{q} \cdot \bv{y}}) (e^{i \bv{q} \cdot \bv{z}} + e^{-i \bv{q} \cdot \bv{z}}) \ea{\chi(\bv{x},t) \mathcal{J}_{\mu}(z) \mathcal{J}_{\mu}(y) \overline{\chi}(0)}.
				\end{align}
				The correlator defined here involves a four-point correlation function with nucleon interpolating operators held at fixed temporal separation $t$, with the currents inserted across the entire four-volume. Importantly, this expression is evaluated in the absence of the external field, and hence momentum conservation is exact. It is then possible to perform a spectral decomposition of this
				correlator in terms of a transfer matrix that is diagonal in the momenta. Given that the Fourier projection of the nucleon sink is at definite momentum $\bv{p}$, and $|\bv{p}|<|\bv{p}\pm\bv{q}|$ (as discussed above \Cref{eq:fh_spec}), the leading asymptotic behavior of the correlator must have an exponential behavior given by $e^{-E_N(\bv{p})t}$. By resolving the corresponding $t$-enhanced coefficient of this exponential, we can identify the second-order energy shift, as given in \Cref{eq:fh_spec}.

				Assuming that the temporal length is sufficiently large that we can neglect the temporal boundary conditions, there are six distinct time orderings of where the current insertions can act relative to the nucleon interpolating fields. They are shown in \Cref{fig:time}.
				\begin{figure}
							\centering
							\includegraphics[width=0.465\textwidth]{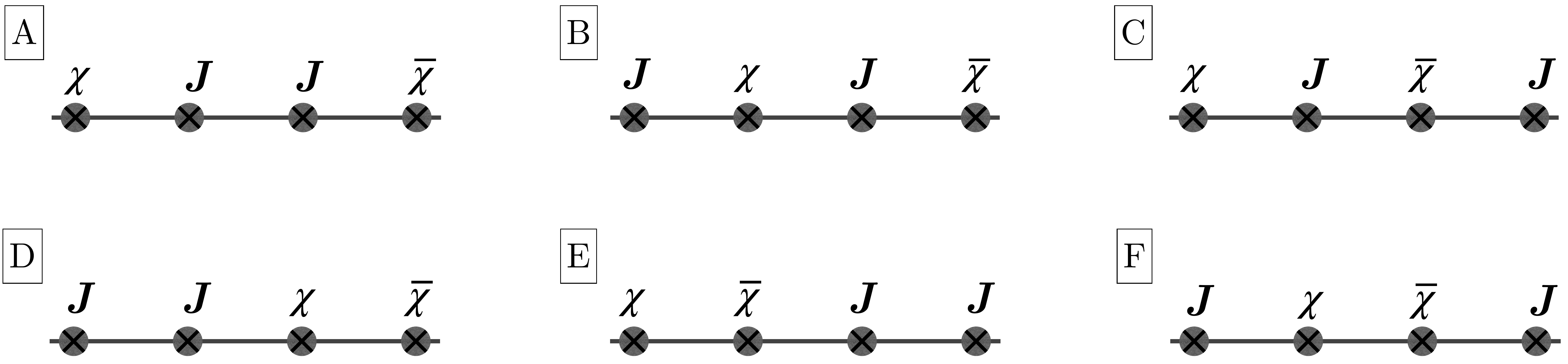}
							\caption{Distinct time orderings of the current insertions, with increasing time assumed from left to right. \label{fig:time}}
				\end{figure}
				Configuration A is the obvious ordering that contains the desired Compton amplitude. This corresponds to ground-state saturation of the nucleon on either side of the current insertions. The $t$ dependence of this particular contribution, including explicit integrals over the current insertion times, will take the form:
				\begin{align}\label{eq:timeA1}
					\int_0^t d\tau'\int_0^{\tau'} d\tau \langle\chi(t)J(\tau')J(\tau)\overline{\chi}(0)\rangle \sim \int_0^t d\tau'\int_0^{\tau'} d\tau \, e^{-E_N(\pb)(t-\tau')}e^{-E_X(\pb+\qb)(\tau'-\tau)}e^{-E_N(\pb)(\tau)}
				\end{align}
				It is convenient to isolate the current separation time by transforming the coordinates to:
				\begin{align}
					\Delta&=\tau'-\tau\\ 
					\bar{\tau}&=(\tau+\tau')/2,
				\end{align}
				and hence
				\begin{align} \label{eq:timeintegralA}
					\int_0^t d\tau'\int_0^{\tau'} d\tau\, e^{-E_N(\pb)(t-\tau')}e^{-E_X(\pb+\qb)(\tau'-\tau)}e^{-E_N(\pb)(\tau)} &=\int_0^td\Delta\int_{\Delta/2}^{t-\Delta/2}d\bar{\tau}\, e^{-E_N(\pb)t}e^{-(E_X(\pb+\qb)-E_N(\pb))\Delta},\\
					&=e^{-E_N(\pb)t}\int_0^td\Delta\, e^{-(E_X(\pb+\qb)-E_N(\pb))\Delta}(t-\Delta).
				\end{align}
				The term linear in $t$ corresponds to the anticipated time enhancement of \Cref{eq:fh_spec}---details of the connection to the Compton amplitude are given below. Given the condition that $E_X>E_N$, the damping ensures that the term proportional to $\Delta$ is independent of $t$ for large times. It is this damping which ensures the current separation remains localized in time, and allows the nucleon to saturate to the ground state on either side of the current.

				Having selected the term of interest, it is necessary to confirm that none of the other possible configurations can scale as $te^{-E_N(\pb)t}$ at large times. One potential example would be to consider the nucleon at the source to carry momentum $\pb+2\qb$. This case gives a temporal behavior according to
				\begin{align}
				&\int_0^t d\tau'\int_0^{\tau'} d\tau \langle\chi(t)J(\tau')J(\tau)\overline{\chi}(0)\rangle \sim \int_0^t d\tau'\int_0^{\tau'} d\tau\, e^{-E_N(\pb)(t-\tau')}e^{-E_X(\pb+\qb)(\tau'-\tau)}e^{-E_N(\pb+2\qb)(\tau)},\\
				\label{eq:timeTransition}
				&\quad=\int_0^td\Delta\, \left(\frac{e^{-E_N(\pb+2\qb)t}e^{-(E_X(\pb+\qb)-E_N(\pb+2\qb))\Delta} -e^{-E_N(\pb)t}e^{-(E_X(\pb+\qb)-E_N(\pb))\Delta}}{E_N(\pb+2\qb)-E_N(\pb)}\right).
				\end{align}
				The second term clearly contains a damped exponential and hence the integral over $\Delta$ converges for large $t$. In the first term, the ordering of the levels $E_X(\pb+\qb)$ and $E_N(\pb+2\qb)$ will govern which contribution dominates at large $t$. However in either case, this term is exponentially-suppressed relative to $e^{-E_N(\pb)t}$. This example, that does not exhibit the desired $te^{-E_N(\pb)t}$ behavior, makes it clear that in order to generate the coefficient linear in $t$, one must have 2 intermediate propagators of the lowest energy nucleon, such as in \Cref{eq:timeA1}. With three available time windows and the momentum transfer through the current insertion, it is only possible to achieve this with the lowest-energy nucleons separated by an intermediate, energetic state.

				It is then straightforward to conclude that it is not possible for any of the temporal configurations B to F to generate a contribution $te^{-E_N(\pb)t}$. To highlight how these other terms contribute, we consider the behavior of the B-type ordering. One of the contributions would take the form
				\begin{align}
				  \int_t^\infty d\tau'\int_0^t d\tau \langle J(\tau')\chi(t)J(\tau)\overline{\chi}(0)\rangle \sim \int_t^\infty d\tau'\int_0^t d\tau\, e^{-E_V(\qb)(\tau'-t)}e^{-E_X(\pb+\qb)(t-\tau)}e^{-E_N(\pb)\tau}.
				\end{align}
				Although a ``light'' vector meson propagates outside the nucleon interpolators, the $\tau'$ integral is convergent and no remnant of this mass scale can appear in the $t$-dependent exponent. And even though the momentum states were chosen to highlight a $e^{-E_N(p)t}$ contribution, there cannot be a temporal enhancement since the kinematics are chosen to ensure $E_X(\pb+\qb)>E_N(\pb)$.

				Given that the contribution to the second-order energy shift must come from the temporal orientation of type A, we demonstrate how this relates to the Compton amplitude. Explicitly written out, configuration A gives rise to the 4-point function:
				\begin{align}
					\begin{split}
						  \left. \frac{\partial^2 G^{(2)}_\lambda(\bv{p};t)}{\partial \lambda^2} \right |_{\lambda=0}^A &= 2\int d^3x\, e^{-i \bv{p} \cdot \bv{x}} \int d^3{y} d^3{z} \int_{0}^{t} d\tau' \int_{0}^{\tau'} d\tau (e^{i \bv{q} \cdot \bv{y}} + e^{-i \bv{q} \cdot \bv{y}}) (e^{i \bv{q} \cdot \bv{z}} + e^{-i \bv{q} \cdot \bv{z}}) \\
						  &\quad\times\bv{\Gamma}\ea{\chi(x) | \mathcal{J}_{\mu}(\zb,\tau') \mathcal{J}_{\mu}(\yb,\tau) | \overline{\chi}(0)}.
					\end{split}
				\end{align}
				We insert complete sets of states next to the nucleon interpolating operators, and translate the operator expressions according to the standard form, $\chi(x) = e^{-i \hat{P}.x} \chi(0) e^{i \hat{P}.x}$ and $\mathcal{J}_{\mu}(z)\mathcal{J}_{\mu}(y) = e^{-i \hat{P}.y} \mathcal{J}_{\mu}(z-y)\mathcal{J}_{\mu}(0) e^{i \hat{P}.y}$, which leads to
				\begin{align}
					\begin{split}
						  \left. \frac{\partial^2 G^{(2)}_\lambda(\bv{p};t)}{\partial \lambda^2} \right |_{\lambda=0}^A &= 2 \int d^3{y} d^3{z}
						  \int_{0}^{t} d\tau' \int_{0}^{\tau'} d\tau\sum_{X,Y}\int\frac{d^3k}{(2\pi)^3}\frac{e^{-E_X(\pb)t}e^{-(E_Y(\kb)-E_X(\pb))\tau}}{4E_X(\pb)E_Y(\kb)}\\
						  &\quad\times e^{i(\kb-\pb)\cdot\yb}(e^{i \bv{q} \cdot \bv{y}} + e^{-i \bv{q} \cdot \bv{y}}) (e^{i \bv{q} \cdot \bv{z}} + e^{-i \bv{q} \cdot \bv{z}})\\
						  &\quad\times \bv{\Gamma} \ea{\Omega|\chi(0) |X(\pb)}\ea{X(\pb)| \mathcal{J}_{\mu}(\zb-\yb,\tau'-\tau) \mathcal{J}_{\mu}(\bv{0},0) | Y(\kb)}\ea{Y(\kb)|\overline{\chi}(0)|\Omega}.
					\end{split}
				\end{align}

				By adopting the transformation, $\zb'=\zb-\yb$,  $\yb'=\yb$, the Fourier integral over $\yb'$ can be eliminated, and hence eliminate the $\kb$ integral:
				\begin{align}
					\begin{split}
						\left. \frac{\partial^2 G^{(2)}_\lambda(\bv{p};t)}{\partial \lambda^2} \right |_{\lambda=0}^A &= 2 \int d^3{z'} \int_{0}^{t} d\tau' \int_{0}^{\tau'} d\tau\sum_{X,Y}e^{i\qb\cdot\zb'}\frac{e^{-E_X(\pb)(t-\tau)}}{2E_X(\pb)}\\
						&\quad\times\left[\frac{e^{-E_Y(\pb-2\qb)\tau}}{2E_Y(\pb-2\qb)} \bv{\Gamma}\ea{\Omega|\chi(0) |X(\pb)}\ea{X(\pb)| \mathcal{J}_{\mu}(\zb',\tau'-\tau) \mathcal{J}_{\mu}(\bv{0},0) | Y(\pb-2\qb)}\ea{Y(\pb-2\qb)|\overline{\chi}(0)|\Omega}\right.\\
					    &\quad\quad\left.+\frac{e^{-E_Y(\pb)\tau}}{2E_Y(\pb)}\bv{\Gamma}\ea{\Omega|\chi(0) |X(\pb)}\ea{X(\pb)| \mathcal{J}_{\mu}(\zb',\tau'-\tau) \mathcal{J}_{\mu}(\bv{0},0) | Y(\pb)}\ea{Y(\pb)|\overline{\chi}(0)|\Omega}
					    \right]+(\qb\to-\qb)
					\end{split}
				\end{align}

				As described above in \Cref{eq:timeTransition}, the term involving the momentum transfer between in and out states cannot contribute to the energy shift, it is only the term involving a $\pb\to\pb$ matrix element that is of interest. By applying the result of \Cref{eq:timeintegralA}, and noting that
				at large $t$, the correlator must be dominated by the state $E_X=E_Y=E_N$:
				\begin{align}
					\begin{split}
						\left. \frac{\partial^2 G^{(2)}_\lambda(\bv{p};t)}{\partial \lambda^2} \right |_{\lambda=0}^A &= 2 \int d^3{z}\, e^{i\qb\cdot\zb}\frac{e^{-E_N(\pb)t}}{(2E_N(\pb))^2} \sum_{s,s'}\int_0^td\Delta(t-\Delta)\\
						&\quad\times \bv{\Gamma}\ea{\Omega|\chi(0) |N(\pb,s)}\ea{N(\pb,s)| \mathcal{J}_{\mu}(\zb,\Delta) \mathcal{J}_{\mu}(\bv{0},0) | N(\pb,s')}\ea{N(\pb,s')|\overline{\chi}(0)|\Omega} +(\qb\to-\qb)+\ldots,
					\end{split}
				\end{align}
				where the ellipsis represents terms that are suppressed at large $t$, relative to $te^{-E_N(\pb)t}$. Here, in identifying the ground-state nucleon, the spin sums implied by the $\sum_{X,Y}$ have been restored. Because the matrix element $\ea{N|J(\Delta)J(0)|N}$ is exponentially damped at large $\Delta$, the term in the integrand proportional to $\Delta$ also cannot generate a contribution to the second-order energy shift. Hence the only remaining term contributing to the energy shift is:
				\begin{align}\label{eq:appResult}
					\begin{split}
						\left. \frac{\partial^2 G^{(2)}_\lambda(\bv{p};t)}{\partial \lambda^2} \right |_{\lambda=0}^A &= \sum_{ss'} \frac{A_{ss'}(\pb)t\,e^{-E_N(\pb)t}}{2E_N(\pb)} 2\left[\int_0^td\Delta \int d^3{z}\, e^{i\qb\cdot\zb} \ea{N(\pb,s)| \mathcal{J}_{\mu}(\zb,\Delta) \mathcal{J}_{\mu}(\bv{0},0) | N(\pb,s')} +(\qb\to-\qb)\right]+\ldots,
					\end{split}
				\end{align}
				where a spin-density overlap is used:
				\begin{align}
					A_{ss'}(\pb)=\frac{1}{2E_N(\pb)}\bv{\Gamma}\ea{\Omega|\chi(0)|N(\pb,s)}\ea{N(\pb,s')|\overline{\chi}(0)|\Omega}=\frac12\delta_{ss'}A(\pb).
				\end{align}
				A comparison of the form presented in \Cref{eq:appResult} with \Cref{eq:fh_spec}, together with the Compton amplitude in Equations.~(\ref{eq:Tintegral}) and (\ref{eq:comptint}) with $q_0=0$, yields our result quoted in \Cref{eq:secondorder_fh}. 
			\end{widetext}


\begin{thebibliography}{89}%
\makeatletter
\providecommand \@ifxundefined [1]{%
 \@ifx{#1\undefined}
}%
\providecommand \@ifnum [1]{%
 \ifnum #1\expandafter \@firstoftwo
 \else \expandafter \@secondoftwo
 \fi
}%
\providecommand \@ifx [1]{%
 \ifx #1\expandafter \@firstoftwo
 \else \expandafter \@secondoftwo
 \fi
}%
\providecommand \natexlab [1]{#1}%
\providecommand \enquote  [1]{``#1''}%
\providecommand \bibnamefont  [1]{#1}%
\providecommand \bibfnamefont [1]{#1}%
\providecommand \citenamefont [1]{#1}%
\providecommand \href@noop [0]{\@secondoftwo}%
\providecommand \href [0]{\begingroup \@sanitize@url \@href}%
\providecommand \@href[1]{\@@startlink{#1}\@@href}%
\providecommand \@@href[1]{\endgroup#1\@@endlink}%
\providecommand \@sanitize@url [0]{\catcode `\\12\catcode `\$12\catcode
  `\&12\catcode `\#12\catcode `\^12\catcode `\_12\catcode `\%12\relax}%
\providecommand \@@startlink[1]{}%
\providecommand \@@endlink[0]{}%
\providecommand \url  [0]{\begingroup\@sanitize@url \@url }%
\providecommand \@url [1]{\endgroup\@href {#1}{\urlprefix }}%
\providecommand \urlprefix  [0]{URL }%
\providecommand \Eprint [0]{\href }%
\providecommand \doibase [0]{http://dx.doi.org/}%
\providecommand \selectlanguage [0]{\@gobble}%
\providecommand \bibinfo  [0]{\@secondoftwo}%
\providecommand \bibfield  [0]{\@secondoftwo}%
\providecommand \translation [1]{[#1]}%
\providecommand \BibitemOpen [0]{}%
\providecommand \bibitemStop [0]{}%
\providecommand \bibitemNoStop [0]{.\EOS\space}%
\providecommand \EOS [0]{\spacefactor3000\relax}%
\providecommand \BibitemShut  [1]{\csname bibitem#1\endcsname}%
\let\auto@bib@innerbib\@empty
\bibitem [{\citenamefont {Martinelli}\ and\ \citenamefont
  {Sachrajda}(1996)}]{Martinelli:1996pk}%
  \BibitemOpen
  \bibfield  {author} {\bibinfo {author} {\bibfnamefont {G.}~\bibnamefont
  {Martinelli}}\ and\ \bibinfo {author} {\bibfnamefont {C.~T.}\ \bibnamefont
  {Sachrajda}},\ }\href {\doibase 10.1016/0550-3213(96)00415-4} {\bibfield
  {journal} {\bibinfo  {journal} {Nucl. Phys. B}\ }\textbf {\bibinfo {volume}
  {478}},\ \bibinfo {pages} {660} (\bibinfo {year} {1996})},\ \Eprint
  {http://arxiv.org/abs/hep-ph/9605336} {arXiv:hep-ph/9605336} \BibitemShut
  {NoStop}%
\bibitem [{\citenamefont {Beneke}(1999)}]{Beneke:1998ui}%
  \BibitemOpen
  \bibfield  {author} {\bibinfo {author} {\bibfnamefont {M.}~\bibnamefont
  {Beneke}},\ }\href {\doibase 10.1016/S0370-1573(98)00130-6} {\bibfield
  {journal} {\bibinfo  {journal} {Phys. Rept.}\ }\textbf {\bibinfo {volume}
  {317}},\ \bibinfo {pages} {1} (\bibinfo {year} {1999})},\ \Eprint
  {http://arxiv.org/abs/hep-ph/9807443} {arXiv:hep-ph/9807443} \BibitemShut
  {NoStop}%
\bibitem [{\citenamefont {Martinelli}\ and\ \citenamefont
  {Sachrajda}(1989)}]{Martinelli:1988rr}%
  \BibitemOpen
  \bibfield  {author} {\bibinfo {author} {\bibfnamefont {G.}~\bibnamefont
  {Martinelli}}\ and\ \bibinfo {author} {\bibfnamefont {C.~T.}\ \bibnamefont
  {Sachrajda}},\ }\href {\doibase 10.1016/0550-3213(89)90035-7} {\bibfield
  {journal} {\bibinfo  {journal} {Nucl. Phys. B}\ }\textbf {\bibinfo {volume}
  {316}},\ \bibinfo {pages} {355} (\bibinfo {year} {1989})}\BibitemShut
  {NoStop}%
\bibitem [{\citenamefont {G{\"o}ckeler}\ \emph
  {et~al.}(1996{\natexlab{a}})\citenamefont {G{\"o}ckeler}, \citenamefont
  {Horsley}, \citenamefont {Ilgenfritz}, \citenamefont {Perlt}, \citenamefont
  {Rakow}, \citenamefont {Schierholz},\ and\ \citenamefont
  {Schiller}}]{Gockeler:1995wg}%
  \BibitemOpen
  \bibfield  {author} {\bibinfo {author} {\bibfnamefont {M.}~\bibnamefont
  {G{\"o}ckeler}}, \bibinfo {author} {\bibfnamefont {R.}~\bibnamefont
  {Horsley}}, \bibinfo {author} {\bibfnamefont {E.-M.}\ \bibnamefont
  {Ilgenfritz}}, \bibinfo {author} {\bibfnamefont {H.}~\bibnamefont {Perlt}},
  \bibinfo {author} {\bibfnamefont {P.~E.~L.}\ \bibnamefont {Rakow}}, \bibinfo
  {author} {\bibfnamefont {G.}~\bibnamefont {Schierholz}}, \ and\ \bibinfo
  {author} {\bibfnamefont {A.}~\bibnamefont {Schiller}},\ }\href {\doibase
  10.1103/PhysRevD.53.2317} {\bibfield  {journal} {\bibinfo  {journal} {Phys.
  Rev. D}\ }\textbf {\bibinfo {volume} {53}},\ \bibinfo {pages} {2317}
  (\bibinfo {year} {1996}{\natexlab{a}})},\ \Eprint
  {http://arxiv.org/abs/hep-lat/9508004} {arXiv:hep-lat/9508004} \BibitemShut
  {NoStop}%
\bibitem [{\citenamefont {G{\"o}ckeler}\ \emph
  {et~al.}(1996{\natexlab{b}})\citenamefont {G{\"o}ckeler}, \citenamefont
  {Horsley}, \citenamefont {Ilgenfritz}, \citenamefont {Perlt}, \citenamefont
  {Rakow}, \citenamefont {Schierholz},\ and\ \citenamefont
  {Schiller}}]{Gockeler:1996mu}%
  \BibitemOpen
  \bibfield  {author} {\bibinfo {author} {\bibfnamefont {M.}~\bibnamefont
  {G{\"o}ckeler}}, \bibinfo {author} {\bibfnamefont {R.}~\bibnamefont
  {Horsley}}, \bibinfo {author} {\bibfnamefont {E.-M.}\ \bibnamefont
  {Ilgenfritz}}, \bibinfo {author} {\bibfnamefont {H.}~\bibnamefont {Perlt}},
  \bibinfo {author} {\bibfnamefont {P.~E.~L.}\ \bibnamefont {Rakow}}, \bibinfo
  {author} {\bibfnamefont {G.}~\bibnamefont {Schierholz}}, \ and\ \bibinfo
  {author} {\bibfnamefont {A.}~\bibnamefont {Schiller}},\ }\href {\doibase
  10.1103/PhysRevD.54.5705} {\bibfield  {journal} {\bibinfo  {journal} {Phys.
  Rev. D}\ }\textbf {\bibinfo {volume} {54}},\ \bibinfo {pages} {5705}
  (\bibinfo {year} {1996}{\natexlab{b}})},\ \Eprint
  {http://arxiv.org/abs/hep-lat/9602029} {arXiv:hep-lat/9602029} \BibitemShut
  {NoStop}%
\bibitem [{\citenamefont {G{\"o}ckeler}\ \emph {et~al.}(1999)\citenamefont
  {G{\"o}ckeler}, \citenamefont {Horsley}, \citenamefont {Oelrich},
  \citenamefont {Perlt}, \citenamefont {Petters}, \citenamefont {Rakow},
  \citenamefont {Schafer}, \citenamefont {Schierholz},\ and\ \citenamefont
  {Schiller}}]{Gockeler:1998ye}%
  \BibitemOpen
  \bibfield  {author} {\bibinfo {author} {\bibfnamefont {M.}~\bibnamefont
  {G{\"o}ckeler}}, \bibinfo {author} {\bibfnamefont {R.}~\bibnamefont
  {Horsley}}, \bibinfo {author} {\bibfnamefont {H.}~\bibnamefont {Oelrich}},
  \bibinfo {author} {\bibfnamefont {H.}~\bibnamefont {Perlt}}, \bibinfo
  {author} {\bibfnamefont {D.}~\bibnamefont {Petters}}, \bibinfo {author}
  {\bibfnamefont {P.~E.~L.}\ \bibnamefont {Rakow}}, \bibinfo {author}
  {\bibfnamefont {A.}~\bibnamefont {Schafer}}, \bibinfo {author} {\bibfnamefont
  {G.}~\bibnamefont {Schierholz}}, \ and\ \bibinfo {author} {\bibfnamefont
  {A.}~\bibnamefont {Schiller}},\ }\href {\doibase
  10.1016/S0550-3213(99)00036-X} {\bibfield  {journal} {\bibinfo  {journal}
  {Nucl. Phys. B}\ }\textbf {\bibinfo {volume} {544}},\ \bibinfo {pages} {699}
  (\bibinfo {year} {1999})},\ \Eprint {http://arxiv.org/abs/hep-lat/9807044}
  {arXiv:hep-lat/9807044} \BibitemShut {NoStop}%
\bibitem [{\citenamefont {Liu}\ and\ \citenamefont {Dong}(1994)}]{Liu:1993cv}%
  \BibitemOpen
  \bibfield  {author} {\bibinfo {author} {\bibfnamefont {K.-F.}\ \bibnamefont
  {Liu}}\ and\ \bibinfo {author} {\bibfnamefont {S.-J.}\ \bibnamefont {Dong}},\
  }\href {\doibase 10.1103/PhysRevLett.72.1790} {\bibfield  {journal} {\bibinfo
   {journal} {Phys. Rev. Lett.}\ }\textbf {\bibinfo {volume} {72}},\ \bibinfo
  {pages} {1790} (\bibinfo {year} {1994})},\ \Eprint
  {http://arxiv.org/abs/hep-ph/9306299} {arXiv:hep-ph/9306299} \BibitemShut
  {NoStop}%
\bibitem [{\citenamefont {Capitani}\ \emph
  {et~al.}(1999{\natexlab{a}})\citenamefont {Capitani}, \citenamefont
  {G{\"o}ckeler}, \citenamefont {Horsley}, \citenamefont {Oelrich},
  \citenamefont {Petters}, \citenamefont {Rakow},\ and\ \citenamefont
  {Schierholz}}]{Capitani:1998fe}%
  \BibitemOpen
  \bibfield  {author} {\bibinfo {author} {\bibfnamefont {S.}~\bibnamefont
  {Capitani}}, \bibinfo {author} {\bibfnamefont {M.}~\bibnamefont
  {G{\"o}ckeler}}, \bibinfo {author} {\bibfnamefont {R.}~\bibnamefont
  {Horsley}}, \bibinfo {author} {\bibfnamefont {H.}~\bibnamefont {Oelrich}},
  \bibinfo {author} {\bibfnamefont {D.}~\bibnamefont {Petters}}, \bibinfo
  {author} {\bibfnamefont {P.~E.~L.}\ \bibnamefont {Rakow}}, \ and\ \bibinfo
  {author} {\bibfnamefont {G.}~\bibnamefont {Schierholz}},\ }\href {\doibase
  10.1016/S0920-5632(99)85050-6} {\bibfield  {journal} {\bibinfo  {journal}
  {Nucl. Phys. B Proc. Suppl.}\ }\textbf {\bibinfo {volume} {73}},\ \bibinfo
  {pages} {288} (\bibinfo {year} {1999}{\natexlab{a}})},\ \Eprint
  {http://arxiv.org/abs/hep-lat/9809171} {arXiv:hep-lat/9809171} \BibitemShut
  {NoStop}%
\bibitem [{\citenamefont {Detmold}\ and\ \citenamefont
  {Lin}(2006)}]{Detmold:2005gg}%
  \BibitemOpen
  \bibfield  {author} {\bibinfo {author} {\bibfnamefont {W.}~\bibnamefont
  {Detmold}}\ and\ \bibinfo {author} {\bibfnamefont {C.~J.~D.}\ \bibnamefont
  {Lin}},\ }\href {\doibase 10.1103/PhysRevD.73.014501} {\bibfield  {journal}
  {\bibinfo  {journal} {Phys. Rev.}\ }\textbf {\bibinfo {volume} {D73}},\
  \bibinfo {pages} {014501} (\bibinfo {year} {2006})},\ \Eprint
  {http://arxiv.org/abs/hep-lat/0507007} {arXiv:hep-lat/0507007 [hep-lat]}
  \BibitemShut {NoStop}%
\bibitem [{\citenamefont {Braun}\ and\ \citenamefont
  {M{\"u}ller}(2008)}]{Braun:2007wv}%
  \BibitemOpen
  \bibfield  {author} {\bibinfo {author} {\bibfnamefont {V.}~\bibnamefont
  {Braun}}\ and\ \bibinfo {author} {\bibfnamefont {D.}~\bibnamefont
  {M{\"u}ller}},\ }\href {\doibase 10.1140/epjc/s10052-008-0608-4} {\bibfield
  {journal} {\bibinfo  {journal} {Eur. Phys. J. C}\ }\textbf {\bibinfo {volume}
  {55}},\ \bibinfo {pages} {349} (\bibinfo {year} {2008})},\ \Eprint
  {http://arxiv.org/abs/0709.1348} {arXiv:0709.1348 [hep-ph]} \BibitemShut
  {NoStop}%
\bibitem [{\citenamefont {Ma}\ and\ \citenamefont
  {Qiu}(2018{\natexlab{a}})}]{Ma:2014jla}%
  \BibitemOpen
  \bibfield  {author} {\bibinfo {author} {\bibfnamefont {Y.-Q.}\ \bibnamefont
  {Ma}}\ and\ \bibinfo {author} {\bibfnamefont {J.-W.}\ \bibnamefont {Qiu}},\
  }\href {\doibase 10.1103/PhysRevD.98.074021} {\bibfield  {journal} {\bibinfo
  {journal} {Phys. Rev. D}\ }\textbf {\bibinfo {volume} {98}},\ \bibinfo
  {pages} {074021} (\bibinfo {year} {2018}{\natexlab{a}})},\ \Eprint
  {http://arxiv.org/abs/1404.6860} {arXiv:1404.6860 [hep-ph]} \BibitemShut
  {NoStop}%
\bibitem [{\citenamefont {Chambers}\ \emph
  {et~al.}(2017{\natexlab{a}})\citenamefont {Chambers}, \citenamefont
  {Horsley}, \citenamefont {Nakamura}, \citenamefont {Perlt}, \citenamefont
  {Rakow}, \citenamefont {Schierholz}, \citenamefont {Schiller}, \citenamefont
  {Somfleth}, \citenamefont {Young},\ and\ \citenamefont
  {Zanotti}}]{Chambers:2017dov}%
  \BibitemOpen
  \bibfield  {author} {\bibinfo {author} {\bibfnamefont {A.~J.}\ \bibnamefont
  {Chambers}}, \bibinfo {author} {\bibfnamefont {R.}~\bibnamefont {Horsley}},
  \bibinfo {author} {\bibfnamefont {Y.}~\bibnamefont {Nakamura}}, \bibinfo
  {author} {\bibfnamefont {H.}~\bibnamefont {Perlt}}, \bibinfo {author}
  {\bibfnamefont {P.~E.~L.}\ \bibnamefont {Rakow}}, \bibinfo {author}
  {\bibfnamefont {G.}~\bibnamefont {Schierholz}}, \bibinfo {author}
  {\bibfnamefont {A.}~\bibnamefont {Schiller}}, \bibinfo {author}
  {\bibfnamefont {K.}~\bibnamefont {Somfleth}}, \bibinfo {author}
  {\bibfnamefont {R.~D.}\ \bibnamefont {Young}}, \ and\ \bibinfo {author}
  {\bibfnamefont {J.~M.}\ \bibnamefont {Zanotti}},\ }\href {\doibase
  10.1103/PhysRevLett.118.242001} {\bibfield  {journal} {\bibinfo  {journal}
  {Phys. Rev. Lett.}\ }\textbf {\bibinfo {volume} {118}},\ \bibinfo {pages}
  {242001} (\bibinfo {year} {2017}{\natexlab{a}})},\ \Eprint
  {http://arxiv.org/abs/1703.01153} {arXiv:1703.01153 [hep-lat]} \BibitemShut
  {NoStop}%
\bibitem [{\citenamefont {Bali}\ \emph {et~al.}(2018)\citenamefont {Bali},
  \citenamefont {Braun}, \citenamefont {Gl{\"a}{\ss}le}, \citenamefont
  {G{\"o}ckeler}, \citenamefont {Gruber}, \citenamefont {Hutzler},
  \citenamefont {Korcyl}, \citenamefont {Lang}, \citenamefont {Sch{\"a}fer},
  \citenamefont {Wein},\ and\ \citenamefont {Zhang}}]{Bali:2017gfr}%
  \BibitemOpen
  \bibfield  {author} {\bibinfo {author} {\bibfnamefont {G.~S.}\ \bibnamefont
  {Bali}}, \bibinfo {author} {\bibfnamefont {V.~M.}\ \bibnamefont {Braun}},
  \bibinfo {author} {\bibfnamefont {B.}~\bibnamefont {Gl{\"a}{\ss}le}},
  \bibinfo {author} {\bibfnamefont {M.}~\bibnamefont {G{\"o}ckeler}}, \bibinfo
  {author} {\bibfnamefont {M.}~\bibnamefont {Gruber}}, \bibinfo {author}
  {\bibfnamefont {F.}~\bibnamefont {Hutzler}}, \bibinfo {author} {\bibfnamefont
  {P.}~\bibnamefont {Korcyl}}, \bibinfo {author} {\bibfnamefont
  {B.}~\bibnamefont {Lang}}, \bibinfo {author} {\bibfnamefont {A.}~\bibnamefont
  {Sch{\"a}fer}}, \bibinfo {author} {\bibfnamefont {P.}~\bibnamefont {Wein}}, \
  and\ \bibinfo {author} {\bibfnamefont {J.-H.}\ \bibnamefont {Zhang}},\ }\href
  {\doibase 10.1140/epjc/s10052-018-5700-9} {\bibfield  {journal} {\bibinfo
  {journal} {Eur. Phys. J. C}\ }\textbf {\bibinfo {volume} {78}},\ \bibinfo
  {pages} {217} (\bibinfo {year} {2018})},\ \Eprint
  {http://arxiv.org/abs/1709.04325} {arXiv:1709.04325 [hep-lat]} \BibitemShut
  {NoStop}%
\bibitem [{\citenamefont {Liang}\ \emph {et~al.}(2020)\citenamefont {Liang},
  \citenamefont {Draper}, \citenamefont {Liu}, \citenamefont {Rothkopf},\ and\
  \citenamefont {Yang}}]{Liang:2019frk}%
  \BibitemOpen
  \bibfield  {author} {\bibinfo {author} {\bibfnamefont {J.}~\bibnamefont
  {Liang}}, \bibinfo {author} {\bibfnamefont {T.}~\bibnamefont {Draper}},
  \bibinfo {author} {\bibfnamefont {K.-F.}\ \bibnamefont {Liu}}, \bibinfo
  {author} {\bibfnamefont {A.}~\bibnamefont {Rothkopf}}, \ and\ \bibinfo
  {author} {\bibfnamefont {Y.-B.}\ \bibnamefont {Yang}} (\bibinfo
  {collaboration} {{XQCD Collaboration}}),\ }\href {\doibase
  10.1103/PhysRevD.101.114503} {\bibfield  {journal} {\bibinfo  {journal}
  {Phys. Rev. D}\ }\textbf {\bibinfo {volume} {101}},\ \bibinfo {pages}
  {114503} (\bibinfo {year} {2020})},\ \Eprint
  {http://arxiv.org/abs/1906.05312} {arXiv:1906.05312 [hep-ph]} \BibitemShut
  {NoStop}%
\bibitem [{\citenamefont {Ji}(2013)}]{Ji:2013dva}%
  \BibitemOpen
  \bibfield  {author} {\bibinfo {author} {\bibfnamefont {X.}~\bibnamefont
  {Ji}},\ }\href {\doibase 10.1103/PhysRevLett.110.262002} {\bibfield
  {journal} {\bibinfo  {journal} {Phys. Rev. Lett.}\ }\textbf {\bibinfo
  {volume} {110}},\ \bibinfo {pages} {262002} (\bibinfo {year} {2013})},\
  \Eprint {http://arxiv.org/abs/1305.1539} {arXiv:1305.1539 [hep-ph]}
  \BibitemShut {NoStop}%
\bibitem [{\citenamefont {Lin}\ \emph {et~al.}(2015)\citenamefont {Lin},
  \citenamefont {Chen}, \citenamefont {Cohen},\ and\ \citenamefont
  {Ji}}]{Lin:2014zya}%
  \BibitemOpen
  \bibfield  {author} {\bibinfo {author} {\bibfnamefont {H.-W.}\ \bibnamefont
  {Lin}}, \bibinfo {author} {\bibfnamefont {J.-W.}\ \bibnamefont {Chen}},
  \bibinfo {author} {\bibfnamefont {S.~D.}\ \bibnamefont {Cohen}}, \ and\
  \bibinfo {author} {\bibfnamefont {X.}~\bibnamefont {Ji}},\ }\href {\doibase
  10.1103/PhysRevD.91.054510} {\bibfield  {journal} {\bibinfo  {journal} {Phys.
  Rev. D}\ }\textbf {\bibinfo {volume} {91}},\ \bibinfo {pages} {054510}
  (\bibinfo {year} {2015})},\ \Eprint {http://arxiv.org/abs/1402.1462}
  {arXiv:1402.1462 [hep-ph]} \BibitemShut {NoStop}%
\bibitem [{\citenamefont {Alexandrou}\ \emph {et~al.}(2015)\citenamefont
  {Alexandrou}, \citenamefont {Cichy}, \citenamefont {Drach}, \citenamefont
  {Garcia-Ramos}, \citenamefont {Hadjiyiannakou}, \citenamefont {Jansen},
  \citenamefont {Steffens},\ and\ \citenamefont {Wiese}}]{Alexandrou:2015rja}%
  \BibitemOpen
  \bibfield  {author} {\bibinfo {author} {\bibfnamefont {C.}~\bibnamefont
  {Alexandrou}}, \bibinfo {author} {\bibfnamefont {K.}~\bibnamefont {Cichy}},
  \bibinfo {author} {\bibfnamefont {V.}~\bibnamefont {Drach}}, \bibinfo
  {author} {\bibfnamefont {E.}~\bibnamefont {Garcia-Ramos}}, \bibinfo {author}
  {\bibfnamefont {K.}~\bibnamefont {Hadjiyiannakou}}, \bibinfo {author}
  {\bibfnamefont {K.}~\bibnamefont {Jansen}}, \bibinfo {author} {\bibfnamefont
  {F.}~\bibnamefont {Steffens}}, \ and\ \bibinfo {author} {\bibfnamefont
  {C.}~\bibnamefont {Wiese}},\ }\href {\doibase 10.1103/PhysRevD.92.014502}
  {\bibfield  {journal} {\bibinfo  {journal} {Phys. Rev. D}\ }\textbf {\bibinfo
  {volume} {92}},\ \bibinfo {pages} {014502} (\bibinfo {year} {2015})},\
  \Eprint {http://arxiv.org/abs/1504.07455} {arXiv:1504.07455 [hep-lat]}
  \BibitemShut {NoStop}%
\bibitem [{\citenamefont {Alexandrou}\ \emph {et~al.}(2017)\citenamefont
  {Alexandrou}, \citenamefont {Cichy}, \citenamefont {Constantinou},
  \citenamefont {Hadjiyiannakou}, \citenamefont {Jansen}, \citenamefont
  {Panagopoulos},\ and\ \citenamefont {Steffens}}]{Alexandrou:2017huk}%
  \BibitemOpen
  \bibfield  {author} {\bibinfo {author} {\bibfnamefont {C.}~\bibnamefont
  {Alexandrou}}, \bibinfo {author} {\bibfnamefont {K.}~\bibnamefont {Cichy}},
  \bibinfo {author} {\bibfnamefont {M.}~\bibnamefont {Constantinou}}, \bibinfo
  {author} {\bibfnamefont {K.}~\bibnamefont {Hadjiyiannakou}}, \bibinfo
  {author} {\bibfnamefont {K.}~\bibnamefont {Jansen}}, \bibinfo {author}
  {\bibfnamefont {H.}~\bibnamefont {Panagopoulos}}, \ and\ \bibinfo {author}
  {\bibfnamefont {F.}~\bibnamefont {Steffens}},\ }\href {\doibase
  10.1016/j.nuclphysb.2017.08.012} {\bibfield  {journal} {\bibinfo  {journal}
  {Nucl. Phys. B}\ }\textbf {\bibinfo {volume} {923}},\ \bibinfo {pages} {394}
  (\bibinfo {year} {2017})},\ \Eprint {http://arxiv.org/abs/1706.00265}
  {arXiv:1706.00265 [hep-lat]} \BibitemShut {NoStop}%
\bibitem [{\citenamefont {Green}\ \emph {et~al.}(2018)\citenamefont {Green},
  \citenamefont {Jansen},\ and\ \citenamefont {Steffens}}]{Green:2017xeu}%
  \BibitemOpen
  \bibfield  {author} {\bibinfo {author} {\bibfnamefont {J.}~\bibnamefont
  {Green}}, \bibinfo {author} {\bibfnamefont {K.}~\bibnamefont {Jansen}}, \
  and\ \bibinfo {author} {\bibfnamefont {F.}~\bibnamefont {Steffens}},\ }\href
  {\doibase 10.1103/PhysRevLett.121.022004} {\bibfield  {journal} {\bibinfo
  {journal} {Phys. Rev. Lett.}\ }\textbf {\bibinfo {volume} {121}},\ \bibinfo
  {pages} {022004} (\bibinfo {year} {2018})},\ \Eprint
  {http://arxiv.org/abs/1707.07152} {arXiv:1707.07152 [hep-lat]} \BibitemShut
  {NoStop}%
\bibitem [{\citenamefont {Ji}\ and\ \citenamefont {Jung}(2001)}]{Ji:2001wha}%
  \BibitemOpen
  \bibfield  {author} {\bibinfo {author} {\bibfnamefont {X.}~\bibnamefont
  {Ji}}\ and\ \bibinfo {author} {\bibfnamefont {C.}~\bibnamefont {Jung}},\
  }\href {\doibase 10.1103/PhysRevLett.86.208} {\bibfield  {journal} {\bibinfo
  {journal} {Phys. Rev. Lett.}\ }\textbf {\bibinfo {volume} {86}},\ \bibinfo
  {pages} {208} (\bibinfo {year} {2001})},\ \Eprint
  {http://arxiv.org/abs/hep-lat/0101014} {arXiv:hep-lat/0101014} \BibitemShut
  {NoStop}%
\bibitem [{\citenamefont {Chen}\ \emph {et~al.}(2018)\citenamefont {Chen},
  \citenamefont {Ishikawa}, \citenamefont {Jin}, \citenamefont {Lin},
  \citenamefont {Yang}, \citenamefont {Zhang},\ and\ \citenamefont
  {Zhao}}]{Chen:2017mzz}%
  \BibitemOpen
  \bibfield  {author} {\bibinfo {author} {\bibfnamefont {J.-W.}\ \bibnamefont
  {Chen}}, \bibinfo {author} {\bibfnamefont {T.}~\bibnamefont {Ishikawa}},
  \bibinfo {author} {\bibfnamefont {L.}~\bibnamefont {Jin}}, \bibinfo {author}
  {\bibfnamefont {H.-W.}\ \bibnamefont {Lin}}, \bibinfo {author} {\bibfnamefont
  {Y.-B.}\ \bibnamefont {Yang}}, \bibinfo {author} {\bibfnamefont {J.-H.}\
  \bibnamefont {Zhang}}, \ and\ \bibinfo {author} {\bibfnamefont
  {Y.}~\bibnamefont {Zhao}},\ }\href {\doibase 10.1103/PhysRevD.97.014505}
  {\bibfield  {journal} {\bibinfo  {journal} {Phys. Rev. D}\ }\textbf {\bibinfo
  {volume} {97}},\ \bibinfo {pages} {014505} (\bibinfo {year} {2018})},\
  \Eprint {http://arxiv.org/abs/1706.01295} {arXiv:1706.01295 [hep-lat]}
  \BibitemShut {NoStop}%
\bibitem [{\citenamefont {Ishikawa}\ \emph {et~al.}(2017)\citenamefont
  {Ishikawa}, \citenamefont {Ma}, \citenamefont {Qiu},\ and\ \citenamefont
  {Yoshida}}]{Ishikawa:2017faj}%
  \BibitemOpen
  \bibfield  {author} {\bibinfo {author} {\bibfnamefont {T.}~\bibnamefont
  {Ishikawa}}, \bibinfo {author} {\bibfnamefont {Y.-Q.}\ \bibnamefont {Ma}},
  \bibinfo {author} {\bibfnamefont {J.-W.}\ \bibnamefont {Qiu}}, \ and\
  \bibinfo {author} {\bibfnamefont {S.}~\bibnamefont {Yoshida}},\ }\href
  {\doibase 10.1103/PhysRevD.96.094019} {\bibfield  {journal} {\bibinfo
  {journal} {Phys. Rev. D}\ }\textbf {\bibinfo {volume} {96}},\ \bibinfo
  {pages} {094019} (\bibinfo {year} {2017})},\ \Eprint
  {http://arxiv.org/abs/1707.03107} {arXiv:1707.03107 [hep-ph]} \BibitemShut
  {NoStop}%
\bibitem [{\citenamefont {Lin}\ \emph {et~al.}(2018{\natexlab{a}})\citenamefont
  {Lin}, \citenamefont {Chen}, \citenamefont {Ji}, \citenamefont {Jin},
  \citenamefont {Li}, \citenamefont {Liu}, \citenamefont {Yang}, \citenamefont
  {Zhang},\ and\ \citenamefont {Zhao}}]{Lin:2018pvv}%
  \BibitemOpen
  \bibfield  {author} {\bibinfo {author} {\bibfnamefont {H.-W.}\ \bibnamefont
  {Lin}}, \bibinfo {author} {\bibfnamefont {J.-W.}\ \bibnamefont {Chen}},
  \bibinfo {author} {\bibfnamefont {X.}~\bibnamefont {Ji}}, \bibinfo {author}
  {\bibfnamefont {L.}~\bibnamefont {Jin}}, \bibinfo {author} {\bibfnamefont
  {R.}~\bibnamefont {Li}}, \bibinfo {author} {\bibfnamefont {Y.-S.}\
  \bibnamefont {Liu}}, \bibinfo {author} {\bibfnamefont {Y.-B.}\ \bibnamefont
  {Yang}}, \bibinfo {author} {\bibfnamefont {J.-H.}\ \bibnamefont {Zhang}}, \
  and\ \bibinfo {author} {\bibfnamefont {Y.}~\bibnamefont {Zhao}},\ }\href
  {\doibase 10.1103/PhysRevLett.121.242003} {\bibfield  {journal} {\bibinfo
  {journal} {Phys. Rev. Lett.}\ }\textbf {\bibinfo {volume} {121}},\ \bibinfo
  {pages} {242003} (\bibinfo {year} {2018}{\natexlab{a}})},\ \Eprint
  {http://arxiv.org/abs/1807.07431} {arXiv:1807.07431 [hep-lat]} \BibitemShut
  {NoStop}%
\bibitem [{\citenamefont {Radyushkin}(2017)}]{Radyushkin:2017cyf}%
  \BibitemOpen
  \bibfield  {author} {\bibinfo {author} {\bibfnamefont {A. V.}~\bibnamefont
  {Radyushkin}},\ }\href {\doibase 10.1103/PhysRevD.96.034025} {\bibfield
  {journal} {\bibinfo  {journal} {Phys. Rev. D}\ }\textbf {\bibinfo {volume}
  {96}},\ \bibinfo {pages} {034025} (\bibinfo {year} {2017})},\ \Eprint
  {http://arxiv.org/abs/1705.01488} {arXiv:1705.01488 [hep-ph]} \BibitemShut
  {NoStop}%
\bibitem [{\citenamefont {Orginos}\ \emph {et~al.}(2017)\citenamefont
  {Orginos}, \citenamefont {Radyushkin}, \citenamefont {Karpie},\ and\
  \citenamefont {Zafeiropoulos}}]{Orginos:2017kos}%
  \BibitemOpen
  \bibfield  {author} {\bibinfo {author} {\bibfnamefont {K.}~\bibnamefont
  {Orginos}}, \bibinfo {author} {\bibfnamefont {A.}~\bibnamefont {Radyushkin}},
  \bibinfo {author} {\bibfnamefont {J.}~\bibnamefont {Karpie}}, \ and\ \bibinfo
  {author} {\bibfnamefont {S.}~\bibnamefont {Zafeiropoulos}},\ }\href {\doibase
  10.1103/PhysRevD.96.094503} {\bibfield  {journal} {\bibinfo  {journal} {Phys.
  Rev. D}\ }\textbf {\bibinfo {volume} {96}},\ \bibinfo {pages} {094503}
  (\bibinfo {year} {2017})},\ \Eprint {http://arxiv.org/abs/1706.05373}
  {arXiv:1706.05373 [hep-ph]} \BibitemShut {NoStop}%
\bibitem [{\citenamefont {Karpie}\ \emph {et~al.}(2018)\citenamefont {Karpie},
  \citenamefont {Orginos},\ and\ \citenamefont
  {Zafeiropoulos}}]{Karpie:2018zaz}%
  \BibitemOpen
  \bibfield  {author} {\bibinfo {author} {\bibfnamefont {J.}~\bibnamefont
  {Karpie}}, \bibinfo {author} {\bibfnamefont {K.}~\bibnamefont {Orginos}}, \
  and\ \bibinfo {author} {\bibfnamefont {S.}~\bibnamefont {Zafeiropoulos}},\
  }\href {\doibase 10.1007/JHEP11(2018)178} {\bibfield  {journal} {\bibinfo
  {journal} {JHEP}\ }\textbf {\bibinfo {volume} {11}},\ \bibinfo {pages} {178}
  (\bibinfo {year} {2018})},\ \Eprint {http://arxiv.org/abs/1807.10933}
  {arXiv:1807.10933 [hep-lat]} \BibitemShut {NoStop}%
\bibitem [{\citenamefont {Lin}\ \emph {et~al.}(2018{\natexlab{b}})\citenamefont
  {Lin}, \citenamefont {Nocera}, \citenamefont {Olness}, \citenamefont
  {Orginos}, \citenamefont {Rojo}, \citenamefont {Accardi}, \citenamefont
  {Alexandrou}, \citenamefont {Bacchetta}, \citenamefont {Bozzi}, \citenamefont
  {Chen}, \citenamefont {Collins}, \citenamefont {Cooper-Sarkar}, \citenamefont
  {Constantinou}, \citenamefont {Debbio}, \citenamefont {Engelhardt},
  \citenamefont {Green}, \citenamefont {Gupta}, \citenamefont {Harland-Lang},
  \citenamefont {Ishikawa}, \citenamefont {Kusina}, \citenamefont {Liu},
  \citenamefont {Liuti}, \citenamefont {Monahan}, \citenamefont {Nadolsky},
  \citenamefont {Qiu}, \citenamefont {Schienbein}, \citenamefont {Schierholz},
  \citenamefont {Thorne}, \citenamefont {Vogelsang}, \citenamefont {Wittig},
  \citenamefont {Yuan},\ and\ \citenamefont {Zanotti}}]{Lin:2017snn}%
  \BibitemOpen
  \bibfield  {author} {\bibinfo {author} {\bibfnamefont {H.-W.}\ \bibnamefont
  {Lin}}, \bibinfo {author} {\bibfnamefont {E.~R.}\ \bibnamefont {Nocera}},
  \bibinfo {author} {\bibfnamefont {F.}~\bibnamefont {Olness}}, \bibinfo
  {author} {\bibfnamefont {K.}~\bibnamefont {Orginos}}, \bibinfo {author}
  {\bibfnamefont {J.}~\bibnamefont {Rojo}}, \bibinfo {author} {\bibfnamefont
  {A.}~\bibnamefont {Accardi}}, \bibinfo {author} {\bibfnamefont
  {C.}~\bibnamefont {Alexandrou}}, \bibinfo {author} {\bibfnamefont
  {A.}~\bibnamefont {Bacchetta}}, \bibinfo {author} {\bibfnamefont
  {G.}~\bibnamefont {Bozzi}}, \bibinfo {author} {\bibfnamefont {J.-W.}\
  \bibnamefont {Chen}}, \bibinfo {author} {\bibfnamefont {S.}~\bibnamefont
  {Collins}}, \bibinfo {author} {\bibfnamefont {A.}~\bibnamefont
  {Cooper-Sarkar}}, \bibinfo {author} {\bibfnamefont {M.}~\bibnamefont
  {Constantinou}}, \bibinfo {author} {\bibfnamefont {L.~D.}\ \bibnamefont
  {Debbio}}, \bibinfo {author} {\bibfnamefont {M.}~\bibnamefont {Engelhardt}},
  \bibinfo {author} {\bibfnamefont {J.}~\bibnamefont {Green}}, \bibinfo
  {author} {\bibfnamefont {R.}~\bibnamefont {Gupta}}, \bibinfo {author}
  {\bibfnamefont {L.~A.}\ \bibnamefont {Harland-Lang}}, \bibinfo {author}
  {\bibfnamefont {T.}~\bibnamefont {Ishikawa}}, \bibinfo {author}
  {\bibfnamefont {A.}~\bibnamefont {Kusina}}, \bibinfo {author} {\bibfnamefont
  {K.-F.}\ \bibnamefont {Liu}}, \bibinfo {author} {\bibfnamefont
  {S.}~\bibnamefont {Liuti}}, \bibinfo {author} {\bibfnamefont
  {C.}~\bibnamefont {Monahan}}, \bibinfo {author} {\bibfnamefont
  {P.}~\bibnamefont {Nadolsky}}, \bibinfo {author} {\bibfnamefont {J.-W.}\
  \bibnamefont {Qiu}}, \bibinfo {author} {\bibfnamefont {I.}~\bibnamefont
  {Schienbein}}, \bibinfo {author} {\bibfnamefont {G.}~\bibnamefont
  {Schierholz}}, \bibinfo {author} {\bibfnamefont {R.~S.}\ \bibnamefont
  {Thorne}}, \bibinfo {author} {\bibfnamefont {W.}~\bibnamefont {Vogelsang}},
  \bibinfo {author} {\bibfnamefont {H.}~\bibnamefont {Wittig}}, \bibinfo
  {author} {\bibfnamefont {C.-P.}\ \bibnamefont {Yuan}}, \ and\ \bibinfo
  {author} {\bibfnamefont {J.~M.}\ \bibnamefont {Zanotti}},\ }\href {\doibase
  10.1016/j.ppnp.2018.01.007} {\bibfield  {journal} {\bibinfo  {journal} {Prog.
  Part. Nucl. Phys.}\ }\textbf {\bibinfo {volume} {100}},\ \bibinfo {pages}
  {107} (\bibinfo {year} {2018}{\natexlab{b}})},\ \Eprint
  {http://arxiv.org/abs/1711.07916} {arXiv:1711.07916 [hep-ph]} \BibitemShut
  {NoStop}%
\bibitem [{\citenamefont {Cichy}\ and\ \citenamefont
  {Constantinou}(2019)}]{Cichy:2018mum}%
  \BibitemOpen
  \bibfield  {author} {\bibinfo {author} {\bibfnamefont {K.}~\bibnamefont
  {Cichy}}\ and\ \bibinfo {author} {\bibfnamefont {M.}~\bibnamefont
  {Constantinou}},\ }\href {\doibase 10.1155/2019/3036904} {\bibfield
  {journal} {\bibinfo  {journal} {Adv. High Energy Phys.}\ }\textbf {\bibinfo
  {volume} {2019}},\ \bibinfo {pages} {3036904} (\bibinfo {year} {2019})},\
  \Eprint {http://arxiv.org/abs/1811.07248} {arXiv:1811.07248 [hep-lat]}
  \BibitemShut {NoStop}%
\bibitem [{\citenamefont {Constantinou}(2020)}]{Constantinou:2020pek}%
  \BibitemOpen
  \bibfield  {author} {\bibinfo {author} {\bibfnamefont {M.}~\bibnamefont
  {Constantinou}},\ }in\ \href@noop {} {\emph {\bibinfo {booktitle} {{38th
  International Symposium on Lattice Field Theory}}}}\ (\bibinfo {year}
  {2020})\ \Eprint {http://arxiv.org/abs/2010.02445} {arXiv:2010.02445
  [hep-lat]} \BibitemShut {NoStop}%
\bibitem [{\citenamefont {Rossi}\ and\ \citenamefont
  {Testa}(2017)}]{Rossi:2017muf}%
  \BibitemOpen
  \bibfield  {author} {\bibinfo {author} {\bibfnamefont {G. C.}~\bibnamefont
  {Rossi}}\ and\ \bibinfo {author} {\bibfnamefont {M.}~\bibnamefont {Testa}},\
  }\href {\doibase 10.1103/PhysRevD.96.014507} {\bibfield  {journal} {\bibinfo
  {journal} {Phys. Rev. D}\ }\textbf {\bibinfo {volume} {96}},\ \bibinfo
  {pages} {014507} (\bibinfo {year} {2017})},\ \Eprint
  {http://arxiv.org/abs/1706.04428} {arXiv:1706.04428 [hep-lat]} \BibitemShut
  {NoStop}%
\bibitem [{\citenamefont {Braun}\ \emph {et~al.}(2019)\citenamefont {Braun},
  \citenamefont {Vladimirov},\ and\ \citenamefont {Zhang}}]{Braun:2018brg}%
  \BibitemOpen
  \bibfield  {author} {\bibinfo {author} {\bibfnamefont {V.~M.}\ \bibnamefont
  {Braun}}, \bibinfo {author} {\bibfnamefont {A.}~\bibnamefont {Vladimirov}}, \
  and\ \bibinfo {author} {\bibfnamefont {J.-H.}\ \bibnamefont {Zhang}},\ }\href
  {\doibase 10.1103/PhysRevD.99.014013} {\bibfield  {journal} {\bibinfo
  {journal} {Phys. Rev. D}\ }\textbf {\bibinfo {volume} {99}},\ \bibinfo
  {pages} {014013} (\bibinfo {year} {2019})},\ \Eprint
  {http://arxiv.org/abs/1810.00048} {arXiv:1810.00048 [hep-ph]} \BibitemShut
  {NoStop}%
\bibitem [{\citenamefont {Brodsky}(2005)}]{Brodsky:2004hh}%
  \BibitemOpen
  \bibfield  {author} {\bibinfo {author} {\bibfnamefont {S.~J.}\ \bibnamefont
  {Brodsky}},\ }\href@noop {} {\bibfield  {journal} {\bibinfo  {journal} {Acta
  Phys. Polon. B}\ }\textbf {\bibinfo {volume} {36}},\ \bibinfo {pages} {635}
  (\bibinfo {year} {2005})},\ \Eprint {http://arxiv.org/abs/hep-ph/0411028}
  {arXiv:hep-ph/0411028} \BibitemShut {NoStop}%
\bibitem [{\citenamefont {Hautmann}\ and\ \citenamefont
  {Soper}(2007)}]{Hautmann:2007cx}%
  \BibitemOpen
  \bibfield  {author} {\bibinfo {author} {\bibfnamefont {F.}~\bibnamefont
  {Hautmann}}\ and\ \bibinfo {author} {\bibfnamefont {D.~E.}\ \bibnamefont
  {Soper}},\ }\href {\doibase 10.1103/PhysRevD.75.074020} {\bibfield  {journal}
  {\bibinfo  {journal} {Phys. Rev. D}\ }\textbf {\bibinfo {volume} {75}},\
  \bibinfo {pages} {074020} (\bibinfo {year} {2007})},\ \Eprint
  {http://arxiv.org/abs/hep-ph/0702077} {arXiv:hep-ph/0702077} \BibitemShut
  {NoStop}%
\bibitem [{\citenamefont {Brandt}(1974{\natexlab{a}})}]{Brandt:1974nk}%
  \BibitemOpen
  \bibfield  {author} {\bibinfo {author} {\bibfnamefont {R.~A.}\ \bibnamefont
  {Brandt}},\ }\href {\doibase 10.1016/0550-3213(74)90225-9} {\bibfield
  {journal} {\bibinfo  {journal} {Nucl. Phys. B}\ }\textbf {\bibinfo {volume}
  {72}},\ \bibinfo {pages} {125} (\bibinfo {year}
  {1974}{\natexlab{a}})}\BibitemShut {NoStop}%
\bibitem [{\citenamefont {Brandt}(1974{\natexlab{b}})}]{Brandt:1974dg}%
  \BibitemOpen
  \bibfield  {author} {\bibinfo {author} {\bibfnamefont {R.~A.}\ \bibnamefont
  {Brandt}},\ }\href {\doibase 10.1016/0550-3213(74)90074-1} {\bibfield
  {journal} {\bibinfo  {journal} {Nucl. Phys. B}\ }\textbf {\bibinfo {volume}
  {83}},\ \bibinfo {pages} {60} (\bibinfo {year}
  {1974}{\natexlab{b}})}\BibitemShut {NoStop}%
\bibitem [{\citenamefont {Brandt}\ and\ \citenamefont
  {Orzalesi}(1971)}]{Brandt:1971ep}%
  \BibitemOpen
  \bibfield  {author} {\bibinfo {author} {\bibfnamefont {R.~A.}\ \bibnamefont
  {Brandt}}\ and\ \bibinfo {author} {\bibfnamefont {C.}~\bibnamefont
  {Orzalesi}},\ }\href {\doibase 10.1016/0370-2693(71)90161-4} {\bibfield
  {journal} {\bibinfo  {journal} {Phys. Lett. B}\ }\textbf {\bibinfo {volume}
  {34}},\ \bibinfo {pages} {641} (\bibinfo {year} {1971})}\BibitemShut
  {NoStop}%
\bibitem [{\citenamefont {Bietenholz}\ \emph
  {et~al.}(2009{\natexlab{a}})\citenamefont {Bietenholz}, \citenamefont
  {Cundy}, \citenamefont {G{\"o}ckeler}, \citenamefont {Horsley}, \citenamefont
  {Perlt}, \citenamefont {Pleiter}, \citenamefont {Rakow}, \citenamefont
  {Schierholz}, \citenamefont {Schiller}, \citenamefont {Streuer},\ and\
  \citenamefont {Zanotti}}]{Bietenholz:2009if}%
  \BibitemOpen
  \bibfield  {author} {\bibinfo {author} {\bibfnamefont {W.}~\bibnamefont
  {Bietenholz}}, \bibinfo {author} {\bibfnamefont {N.}~\bibnamefont {Cundy}},
  \bibinfo {author} {\bibfnamefont {M.}~\bibnamefont {G{\"o}ckeler}}, \bibinfo
  {author} {\bibfnamefont {R.}~\bibnamefont {Horsley}}, \bibinfo {author}
  {\bibfnamefont {H.}~\bibnamefont {Perlt}}, \bibinfo {author} {\bibfnamefont
  {D.}~\bibnamefont {Pleiter}}, \bibinfo {author} {\bibfnamefont {P.~E.~L.}\
  \bibnamefont {Rakow}}, \bibinfo {author} {\bibfnamefont {G.}~\bibnamefont
  {Schierholz}}, \bibinfo {author} {\bibfnamefont {A.}~\bibnamefont
  {Schiller}}, \bibinfo {author} {\bibfnamefont {T.}~\bibnamefont {Streuer}}, \
  and\ \bibinfo {author} {\bibfnamefont {J.~M.}\ \bibnamefont {Zanotti}}
  (\bibinfo {collaboration} {QCDSF Collaboration}),\ }\href {\doibase
  10.22323/1.091.0139} {\bibfield  {journal} {\bibinfo  {journal} {PoS}\
  }\textbf {\bibinfo {volume} {LAT2009}},\ \bibinfo {pages} {139} (\bibinfo
  {year} {2009}{\natexlab{a}})},\ \Eprint {http://arxiv.org/abs/0911.4892}
  {arXiv:0911.4892 [hep-lat]} \BibitemShut {NoStop}%
\bibitem [{\citenamefont {Bietenholz}\ \emph
  {et~al.}(2010{\natexlab{a}})\citenamefont {Bietenholz}, \citenamefont
  {Cundy}, \citenamefont {G{\"o}ckeler}, \citenamefont {Horsley}, \citenamefont
  {Perlt}, \citenamefont {Pleiter}, \citenamefont {Rakow}, \citenamefont
  {Schierholz}, \citenamefont {Schiller}, \citenamefont {Streuer},\ and\
  \citenamefont {Zanotti}}]{Bietenholz:2010kn}%
  \BibitemOpen
  \bibfield  {author} {\bibinfo {author} {\bibfnamefont {W.}~\bibnamefont
  {Bietenholz}}, \bibinfo {author} {\bibfnamefont {N.}~\bibnamefont {Cundy}},
  \bibinfo {author} {\bibfnamefont {M.}~\bibnamefont {G{\"o}ckeler}}, \bibinfo
  {author} {\bibfnamefont {R.}~\bibnamefont {Horsley}}, \bibinfo {author}
  {\bibfnamefont {H.}~\bibnamefont {Perlt}}, \bibinfo {author} {\bibfnamefont
  {D.}~\bibnamefont {Pleiter}}, \bibinfo {author} {\bibfnamefont {P.~E.~L.}\
  \bibnamefont {Rakow}}, \bibinfo {author} {\bibfnamefont {G.}~\bibnamefont
  {Schierholz}}, \bibinfo {author} {\bibfnamefont {A.}~\bibnamefont
  {Schiller}}, \bibinfo {author} {\bibfnamefont {T.}~\bibnamefont {Streuer}}, \
  and\ \bibinfo {author} {\bibfnamefont {J.~M.}\ \bibnamefont {Zanotti}}
  (\bibinfo {collaboration} {QCDSF Collaboration}),\ }\href {\doibase
  10.1088/1742-6596/239/1/012011} {\bibfield  {journal} {\bibinfo  {journal}
  {J. Phys. Conf. Ser.}\ }\textbf {\bibinfo {volume} {239}},\ \bibinfo {pages}
  {012011} (\bibinfo {year} {2010}{\natexlab{a}})},\ \Eprint
  {http://arxiv.org/abs/1004.2100} {arXiv:1004.2100 [hep-lat]} \BibitemShut
  {NoStop}%
\bibitem [{\citenamefont {Fukaya}\ \emph {et~al.}(2020)\citenamefont {Fukaya},
  \citenamefont {Hashimoto}, \citenamefont {Kaneko},\ and\ \citenamefont
  {Ohki}}]{Fukaya:2020wpp}%
  \BibitemOpen
  \bibfield  {author} {\bibinfo {author} {\bibfnamefont {H.}~\bibnamefont
  {Fukaya}}, \bibinfo {author} {\bibfnamefont {S.}~\bibnamefont {Hashimoto}},
  \bibinfo {author} {\bibfnamefont {T.}~\bibnamefont {Kaneko}}, \ and\ \bibinfo
  {author} {\bibfnamefont {H.}~\bibnamefont {Ohki}},\ }\href@noop {} {\
  (\bibinfo {year} {2020})},\ \Eprint {http://arxiv.org/abs/2010.01253}
  {arXiv:2010.01253 [hep-lat]} \BibitemShut {NoStop}%
\bibitem [{\citenamefont {Ma}\ and\ \citenamefont
  {Qiu}(2018{\natexlab{b}})}]{Ma:2017pxb}%
  \BibitemOpen
  \bibfield  {author} {\bibinfo {author} {\bibfnamefont {Y.-Q.}\ \bibnamefont
  {Ma}}\ and\ \bibinfo {author} {\bibfnamefont {J.-W.}\ \bibnamefont {Qiu}},\
  }\href {\doibase 10.1103/PhysRevLett.120.022003} {\bibfield  {journal}
  {\bibinfo  {journal} {Phys. Rev. Lett.}\ }\textbf {\bibinfo {volume} {120}},\
  \bibinfo {pages} {022003} (\bibinfo {year} {2018}{\natexlab{b}})},\ \Eprint
  {http://arxiv.org/abs/1709.03018} {arXiv:1709.03018 [hep-ph]} \BibitemShut
  {NoStop}%
\bibitem [{\citenamefont {Hashimoto}(2017)}]{Hashimoto:2017wqo}%
  \BibitemOpen
  \bibfield  {author} {\bibinfo {author} {\bibfnamefont {S.}~\bibnamefont
  {Hashimoto}},\ }\href {\doibase 10.1093/ptep/ptx052} {\bibfield  {journal}
  {\bibinfo  {journal} {PTEP}\ }\textbf {\bibinfo {volume} {2017}},\ \bibinfo
  {pages} {053B03} (\bibinfo {year} {2017})},\ \Eprint
  {http://arxiv.org/abs/1703.01881} {arXiv:1703.01881 [hep-lat]} \BibitemShut
  {NoStop}%
\bibitem [{\citenamefont {Hansen}\ \emph {et~al.}(2017)\citenamefont {Hansen},
  \citenamefont {Meyer},\ and\ \citenamefont {Robaina}}]{Hansen:2017mnd}%
  \BibitemOpen
  \bibfield  {author} {\bibinfo {author} {\bibfnamefont {M.~T.}\ \bibnamefont
  {Hansen}}, \bibinfo {author} {\bibfnamefont {H.~B.}\ \bibnamefont {Meyer}}, \
  and\ \bibinfo {author} {\bibfnamefont {D.}~\bibnamefont {Robaina}},\ }\href
  {\doibase 10.1103/PhysRevD.96.094513} {\bibfield  {journal} {\bibinfo
  {journal} {Phys. Rev. D}\ }\textbf {\bibinfo {volume} {96}},\ \bibinfo
  {pages} {094513} (\bibinfo {year} {2017})},\ \Eprint
  {http://arxiv.org/abs/1704.08993} {arXiv:1704.08993 [hep-lat]} \BibitemShut
  {NoStop}%
\bibitem [{\citenamefont {Feng}\ \emph {et~al.}(2020)\citenamefont {Feng},
  \citenamefont {Gorchtein}, \citenamefont {Jin}, \citenamefont {Ma},\ and\
  \citenamefont {Seng}}]{Feng:2020zdc}%
  \BibitemOpen
  \bibfield  {author} {\bibinfo {author} {\bibfnamefont {X.}~\bibnamefont
  {Feng}}, \bibinfo {author} {\bibfnamefont {M.}~\bibnamefont {Gorchtein}},
  \bibinfo {author} {\bibfnamefont {L.-C.}\ \bibnamefont {Jin}}, \bibinfo
  {author} {\bibfnamefont {P.-X.}\ \bibnamefont {Ma}}, \ and\ \bibinfo {author}
  {\bibfnamefont {C.-Y.}\ \bibnamefont {Seng}},\ }\href {\doibase
  10.1103/PhysRevLett.124.192002} {\bibfield  {journal} {\bibinfo  {journal}
  {Phys. Rev. Lett.}\ }\textbf {\bibinfo {volume} {124}},\ \bibinfo {pages}
  {192002} (\bibinfo {year} {2020})},\ \Eprint
  {http://arxiv.org/abs/2003.09798} {arXiv:2003.09798 [hep-lat]} \BibitemShut
  {NoStop}%
\bibitem [{\citenamefont {Brice{\~n}o}\ \emph {et~al.}(2020)\citenamefont
  {Brice{\~n}o}, \citenamefont {Davoudi}, \citenamefont {Hansen}, \citenamefont
  {Schindler},\ and\ \citenamefont {Baroni}}]{Briceno:2019opb}%
  \BibitemOpen
  \bibfield  {author} {\bibinfo {author} {\bibfnamefont {R.~A.}\ \bibnamefont
  {Brice{\~n}o}}, \bibinfo {author} {\bibfnamefont {Z.}~\bibnamefont
  {Davoudi}}, \bibinfo {author} {\bibfnamefont {M.~T.}\ \bibnamefont {Hansen}},
  \bibinfo {author} {\bibfnamefont {M.~R.}\ \bibnamefont {Schindler}}, \ and\
  \bibinfo {author} {\bibfnamefont {A.}~\bibnamefont {Baroni}},\ }\href
  {\doibase 10.1103/PhysRevD.101.014509} {\bibfield  {journal} {\bibinfo
  {journal} {Phys. Rev. D}\ }\textbf {\bibinfo {volume} {101}},\ \bibinfo
  {pages} {014509} (\bibinfo {year} {2020})},\ \Eprint
  {http://arxiv.org/abs/1911.04036} {arXiv:1911.04036 [hep-lat]} \BibitemShut
  {NoStop}%
\bibitem [{\citenamefont {Bietenholz}\ \emph
  {et~al.}(2011{\natexlab{a}})\citenamefont {Bietenholz}, \citenamefont
  {Bornyakov}, \citenamefont {G{\"o}ckeler}, \citenamefont {Horsley},
  \citenamefont {Lockhart}, \citenamefont {Nakamura}, \citenamefont {Perlt},
  \citenamefont {Pleiter}, \citenamefont {Rakow}, \citenamefont {Schierholz},
  \citenamefont {Schiller}, \citenamefont {Streuer}, \citenamefont
  {St{\"u}ben}, \citenamefont {Winter},\ and\ \citenamefont
  {Zanotti}}]{Bietenholz:2011qq}%
  \BibitemOpen
  \bibfield  {author} {\bibinfo {author} {\bibfnamefont {W.}~\bibnamefont
  {Bietenholz}}, \bibinfo {author} {\bibfnamefont {V.}~\bibnamefont
  {Bornyakov}}, \bibinfo {author} {\bibfnamefont {M.}~\bibnamefont
  {G{\"o}ckeler}}, \bibinfo {author} {\bibfnamefont {R.}~\bibnamefont
  {Horsley}}, \bibinfo {author} {\bibfnamefont {W.~G.}\ \bibnamefont
  {Lockhart}}, \bibinfo {author} {\bibfnamefont {Y.}~\bibnamefont {Nakamura}},
  \bibinfo {author} {\bibfnamefont {H.}~\bibnamefont {Perlt}}, \bibinfo
  {author} {\bibfnamefont {D.}~\bibnamefont {Pleiter}}, \bibinfo {author}
  {\bibfnamefont {P.~E.~L.}\ \bibnamefont {Rakow}}, \bibinfo {author}
  {\bibfnamefont {G.}~\bibnamefont {Schierholz}}, \bibinfo {author}
  {\bibfnamefont {A.}~\bibnamefont {Schiller}}, \bibinfo {author}
  {\bibfnamefont {T.}~\bibnamefont {Streuer}}, \bibinfo {author} {\bibfnamefont
  {H.}~\bibnamefont {St{\"u}ben}}, \bibinfo {author} {\bibfnamefont
  {F.}~\bibnamefont {Winter}}, \ and\ \bibinfo {author} {\bibfnamefont {J.~M.}\
  \bibnamefont {Zanotti}},\ }\href {\doibase 10.1103/PhysRevD.84.054509}
  {\bibfield  {journal} {\bibinfo  {journal} {Phys. Rev. D}\ }\textbf {\bibinfo
  {volume} {84}},\ \bibinfo {pages} {054509} (\bibinfo {year}
  {2011}{\natexlab{a}})},\ \Eprint {http://arxiv.org/abs/1102.5300}
  {arXiv:1102.5300 [hep-lat]} \BibitemShut {NoStop}%
\bibitem [{\citenamefont {Horsley}\ \emph {et~al.}(2012)\citenamefont
  {Horsley}, \citenamefont {Millo}, \citenamefont {Nakamura}, \citenamefont
  {Perlt}, \citenamefont {Pleiter}, \citenamefont {Rakow}, \citenamefont
  {Schierholz}, \citenamefont {Schiller}, \citenamefont {Winter},\ and\
  \citenamefont {Zanotti}}]{Horsley:2012pz}%
  \BibitemOpen
  \bibfield  {author} {\bibinfo {author} {\bibfnamefont {R.}~\bibnamefont
  {Horsley}}, \bibinfo {author} {\bibfnamefont {R.}~\bibnamefont {Millo}},
  \bibinfo {author} {\bibfnamefont {Y.}~\bibnamefont {Nakamura}}, \bibinfo
  {author} {\bibfnamefont {H.}~\bibnamefont {Perlt}}, \bibinfo {author}
  {\bibfnamefont {D.}~\bibnamefont {Pleiter}}, \bibinfo {author} {\bibfnamefont
  {P.~E.~L.}\ \bibnamefont {Rakow}}, \bibinfo {author} {\bibfnamefont
  {G.}~\bibnamefont {Schierholz}}, \bibinfo {author} {\bibfnamefont
  {A.}~\bibnamefont {Schiller}}, \bibinfo {author} {\bibfnamefont
  {F.}~\bibnamefont {Winter}}, \ and\ \bibinfo {author} {\bibfnamefont {J.~M.}\
  \bibnamefont {Zanotti}} (\bibinfo {collaboration} {QCDSF Collaboration, UKQCD
  Collaboration}),\ }\href {\doibase 10.1016/j.physletb.2012.07.004} {\bibfield
   {journal} {\bibinfo  {journal} {Phys. Lett. B}\ }\textbf {\bibinfo {volume}
  {714}},\ \bibinfo {pages} {312} (\bibinfo {year} {2012})},\ \Eprint
  {http://arxiv.org/abs/1205.6410} {arXiv:1205.6410 [hep-lat]} \BibitemShut
  {NoStop}%
\bibitem [{\citenamefont {Chambers}\ \emph {et~al.}(2014)\citenamefont
  {Chambers}, \citenamefont {Horsley}, \citenamefont {Nakamura}, \citenamefont
  {Perlt}, \citenamefont {Pleiter}, \citenamefont {Rakow}, \citenamefont
  {Schierholz}, \citenamefont {Schiller}, \citenamefont {St{\"u}ben},
  \citenamefont {Young},\ and\ \citenamefont {Zanotti}}]{Chambers:2014qaa}%
  \BibitemOpen
  \bibfield  {author} {\bibinfo {author} {\bibfnamefont {A.~J.}\ \bibnamefont
  {Chambers}}, \bibinfo {author} {\bibfnamefont {R.}~\bibnamefont {Horsley}},
  \bibinfo {author} {\bibfnamefont {Y.}~\bibnamefont {Nakamura}}, \bibinfo
  {author} {\bibfnamefont {H.}~\bibnamefont {Perlt}}, \bibinfo {author}
  {\bibfnamefont {D.}~\bibnamefont {Pleiter}}, \bibinfo {author} {\bibfnamefont
  {P.~E.~L.}\ \bibnamefont {Rakow}}, \bibinfo {author} {\bibfnamefont
  {G.}~\bibnamefont {Schierholz}}, \bibinfo {author} {\bibfnamefont
  {A.}~\bibnamefont {Schiller}}, \bibinfo {author} {\bibfnamefont
  {H.}~\bibnamefont {St{\"u}ben}}, \bibinfo {author} {\bibfnamefont {R.~D.}\
  \bibnamefont {Young}}, \ and\ \bibinfo {author} {\bibfnamefont {J.~M.}\
  \bibnamefont {Zanotti}} (\bibinfo {collaboration} {CSSM Collaboration, QCDSF
  Collaboration, UKQCD Collaboration}),\ }\href {\doibase
  10.1103/PhysRevD.90.014510} {\bibfield  {journal} {\bibinfo  {journal} {Phys.
  Rev. D}\ }\textbf {\bibinfo {volume} {90}},\ \bibinfo {pages} {014510}
  (\bibinfo {year} {2014})},\ \Eprint {http://arxiv.org/abs/1405.3019}
  {arXiv:1405.3019 [hep-lat]} \BibitemShut {NoStop}%
\bibitem [{\citenamefont {Chambers}\ \emph
  {et~al.}(2015{\natexlab{a}})\citenamefont {Chambers}, \citenamefont
  {Horsley}, \citenamefont {Nakamura}, \citenamefont {Perlt}, \citenamefont
  {Rakow}, \citenamefont {Schierholz}, \citenamefont {Schiller},\ and\
  \citenamefont {Zanotti}}]{Chambers:2014pea}%
  \BibitemOpen
  \bibfield  {author} {\bibinfo {author} {\bibfnamefont {A.}~\bibnamefont
  {Chambers}}, \bibinfo {author} {\bibfnamefont {R.}~\bibnamefont {Horsley}},
  \bibinfo {author} {\bibfnamefont {Y.}~\bibnamefont {Nakamura}}, \bibinfo
  {author} {\bibfnamefont {H.}~\bibnamefont {Perlt}}, \bibinfo {author}
  {\bibfnamefont {P.~E.~L.}\ \bibnamefont {Rakow}}, \bibinfo {author}
  {\bibfnamefont {G.}~\bibnamefont {Schierholz}}, \bibinfo {author}
  {\bibfnamefont {A.}~\bibnamefont {Schiller}}, \ and\ \bibinfo {author}
  {\bibfnamefont {J.~M.}\ \bibnamefont {Zanotti}} (\bibinfo {collaboration}
  {QCDSF Collaboration}),\ }\href {\doibase 10.1016/j.physletb.2014.11.033}
  {\bibfield  {journal} {\bibinfo  {journal} {Phys. Lett. B}\ }\textbf
  {\bibinfo {volume} {740}},\ \bibinfo {pages} {30} (\bibinfo {year}
  {2015}{\natexlab{a}})},\ \Eprint {http://arxiv.org/abs/1410.3078}
  {arXiv:1410.3078 [hep-lat]} \BibitemShut {NoStop}%
\bibitem [{\citenamefont {Chambers}\ \emph
  {et~al.}(2015{\natexlab{b}})\citenamefont {Chambers}, \citenamefont
  {Horsley}, \citenamefont {Nakamura}, \citenamefont {Perlt}, \citenamefont
  {Pleiter}, \citenamefont {Rakow}, \citenamefont {Schierholz}, \citenamefont
  {Schiller}, \citenamefont {St{\"u}ben}, \citenamefont {Young}, ,\ and\
  \citenamefont {Zanotti}}]{Chambers:2015bka}%
  \BibitemOpen
  \bibfield  {author} {\bibinfo {author} {\bibfnamefont {A.~J.}\ \bibnamefont
  {Chambers}}, \bibinfo {author} {\bibfnamefont {R.}~\bibnamefont {Horsley}},
  \bibinfo {author} {\bibfnamefont {Y.}~\bibnamefont {Nakamura}}, \bibinfo
  {author} {\bibfnamefont {H.}~\bibnamefont {Perlt}}, \bibinfo {author}
  {\bibfnamefont {D.}~\bibnamefont {Pleiter}}, \bibinfo {author} {\bibfnamefont
  {P.~E.~L.}\ \bibnamefont {Rakow}}, \bibinfo {author} {\bibfnamefont
  {G.}~\bibnamefont {Schierholz}}, \bibinfo {author} {\bibfnamefont
  {A.}~\bibnamefont {Schiller}}, \bibinfo {author} {\bibfnamefont
  {H.}~\bibnamefont {St{\"u}ben}}, \bibinfo {author} {\bibfnamefont {R.~D.}\
  \bibnamefont {Young}}, , \ and\ \bibinfo {author} {\bibfnamefont {J.~M.}\
  \bibnamefont {Zanotti}},\ }\href {\doibase 10.1103/PhysRevD.92.114517}
  {\bibfield  {journal} {\bibinfo  {journal} {Phys. Rev. D}\ }\textbf {\bibinfo
  {volume} {92}},\ \bibinfo {pages} {114517} (\bibinfo {year}
  {2015}{\natexlab{b}})},\ \Eprint {http://arxiv.org/abs/1508.06856}
  {arXiv:1508.06856 [hep-lat]} \BibitemShut {NoStop}%
\bibitem [{\citenamefont {Chambers}\ \emph
  {et~al.}(2017{\natexlab{b}})\citenamefont {Chambers}, \citenamefont {Dragos},
  \citenamefont {Horsley}, \citenamefont {Nakamura}, \citenamefont {Perlt},
  \citenamefont {Pleiter}, \citenamefont {Rakow}, \citenamefont {Schierholz},
  \citenamefont {Schiller}, \citenamefont {Somfleth}, \citenamefont
  {St{\"u}ben}, \citenamefont {Young},\ and\ \citenamefont
  {Zanotti}}]{Chambers:2017tuf}%
  \BibitemOpen
  \bibfield  {author} {\bibinfo {author} {\bibfnamefont {A.~J.}\ \bibnamefont
  {Chambers}}, \bibinfo {author} {\bibfnamefont {J.}~\bibnamefont {Dragos}},
  \bibinfo {author} {\bibfnamefont {R.}~\bibnamefont {Horsley}}, \bibinfo
  {author} {\bibfnamefont {Y.}~\bibnamefont {Nakamura}}, \bibinfo {author}
  {\bibfnamefont {H.}~\bibnamefont {Perlt}}, \bibinfo {author} {\bibfnamefont
  {D.}~\bibnamefont {Pleiter}}, \bibinfo {author} {\bibfnamefont {P.~E.~L.}\
  \bibnamefont {Rakow}}, \bibinfo {author} {\bibfnamefont {G.}~\bibnamefont
  {Schierholz}}, \bibinfo {author} {\bibfnamefont {A.}~\bibnamefont
  {Schiller}}, \bibinfo {author} {\bibfnamefont {K.}~\bibnamefont {Somfleth}},
  \bibinfo {author} {\bibfnamefont {H.}~\bibnamefont {St{\"u}ben}}, \bibinfo
  {author} {\bibfnamefont {R.~D.}\ \bibnamefont {Young}}, \ and\ \bibinfo
  {author} {\bibfnamefont {J.~M.}\ \bibnamefont {Zanotti}} (\bibinfo
  {collaboration} {CSSM Collaboration, QCDSF Collaboration, UKQCD
  Collaboration}),\ }\href {\doibase 10.1103/PhysRevD.96.114509} {\bibfield
  {journal} {\bibinfo  {journal} {Phys. Rev. D}\ }\textbf {\bibinfo {volume}
  {96}},\ \bibinfo {pages} {114509} (\bibinfo {year} {2017}{\natexlab{b}})},\
  \Eprint {http://arxiv.org/abs/1702.01513} {arXiv:1702.01513 [hep-lat]}
  \BibitemShut {NoStop}%
\bibitem [{\citenamefont {Detmold}(2005)}]{Detmold:2004kw}%
  \BibitemOpen
  \bibfield  {author} {\bibinfo {author} {\bibfnamefont {W.}~\bibnamefont
  {Detmold}},\ }\href {\doibase 10.1103/PhysRevD.71.054506} {\bibfield
  {journal} {\bibinfo  {journal} {Phys. Rev. D}\ }\textbf {\bibinfo {volume}
  {71}},\ \bibinfo {pages} {054506} (\bibinfo {year} {2005})},\ \Eprint
  {http://arxiv.org/abs/hep-lat/0410011} {arXiv:hep-lat/0410011} \BibitemShut
  {NoStop}%
\bibitem [{\citenamefont {Primer}\ \emph {et~al.}(2014)\citenamefont {Primer},
  \citenamefont {Kamleh}, \citenamefont {Leinweber},\ and\ \citenamefont
  {Burkardt}}]{Primer:2013pva}%
  \BibitemOpen
  \bibfield  {author} {\bibinfo {author} {\bibfnamefont {T.}~\bibnamefont
  {Primer}}, \bibinfo {author} {\bibfnamefont {W.}~\bibnamefont {Kamleh}},
  \bibinfo {author} {\bibfnamefont {D.}~\bibnamefont {Leinweber}}, \ and\
  \bibinfo {author} {\bibfnamefont {M.}~\bibnamefont {Burkardt}},\ }\href
  {\doibase 10.1103/PhysRevD.89.034508} {\bibfield  {journal} {\bibinfo
  {journal} {Phys. Rev. D}\ }\textbf {\bibinfo {volume} {89}},\ \bibinfo
  {pages} {034508} (\bibinfo {year} {2014})},\ \Eprint
  {http://arxiv.org/abs/1307.1509} {arXiv:1307.1509 [hep-lat]} \BibitemShut
  {NoStop}%
\bibitem [{\citenamefont {Davoudi}\ and\ \citenamefont
  {Detmold}(2015)}]{Davoudi:2015cba}%
  \BibitemOpen
  \bibfield  {author} {\bibinfo {author} {\bibfnamefont {Z.}~\bibnamefont
  {Davoudi}}\ and\ \bibinfo {author} {\bibfnamefont {W.}~\bibnamefont
  {Detmold}},\ }\href {\doibase 10.1103/PhysRevD.92.074506} {\bibfield
  {journal} {\bibinfo  {journal} {Phys. Rev. D}\ }\textbf {\bibinfo {volume}
  {92}},\ \bibinfo {pages} {074506} (\bibinfo {year} {2015})},\ \Eprint
  {http://arxiv.org/abs/1507.01908} {arXiv:1507.01908 [hep-lat]} \BibitemShut
  {NoStop}%
\bibitem [{\citenamefont {Savage}\ \emph {et~al.}(2017)\citenamefont {Savage},
  \citenamefont {Shanahan}, \citenamefont {Tiburzi}, \citenamefont {Wagman},
  \citenamefont {Winter}, \citenamefont {Beane}, \citenamefont {Chang},
  \citenamefont {Davoudi}, \citenamefont {Detmold},\ and\ \citenamefont
  {Orginos}}]{Savage:2016kon}%
  \BibitemOpen
  \bibfield  {author} {\bibinfo {author} {\bibfnamefont {M.~J.}\ \bibnamefont
  {Savage}}, \bibinfo {author} {\bibfnamefont {P.~E.}\ \bibnamefont
  {Shanahan}}, \bibinfo {author} {\bibfnamefont {B.~C.}\ \bibnamefont
  {Tiburzi}}, \bibinfo {author} {\bibfnamefont {M.~L.}\ \bibnamefont {Wagman}},
  \bibinfo {author} {\bibfnamefont {F.}~\bibnamefont {Winter}}, \bibinfo
  {author} {\bibfnamefont {S.~R.}\ \bibnamefont {Beane}}, \bibinfo {author}
  {\bibfnamefont {E.}~\bibnamefont {Chang}}, \bibinfo {author} {\bibfnamefont
  {Z.}~\bibnamefont {Davoudi}}, \bibinfo {author} {\bibfnamefont
  {W.}~\bibnamefont {Detmold}}, \ and\ \bibinfo {author} {\bibfnamefont
  {K.}~\bibnamefont {Orginos}},\ }\href {\doibase
  10.1103/PhysRevLett.119.062002} {\bibfield  {journal} {\bibinfo  {journal}
  {Phys. Rev. Lett.}\ }\textbf {\bibinfo {volume} {119}},\ \bibinfo {pages}
  {062002} (\bibinfo {year} {2017})},\ \Eprint
  {http://arxiv.org/abs/1610.04545} {arXiv:1610.04545 [hep-lat]} \BibitemShut
  {NoStop}%
\bibitem [{\citenamefont {Bouchard}\ \emph {et~al.}(2017)\citenamefont
  {Bouchard}, \citenamefont {Chang}, \citenamefont {Kurth}, \citenamefont
  {Orginos},\ and\ \citenamefont {Walker-Loud}}]{Bouchard:2016heu}%
  \BibitemOpen
  \bibfield  {author} {\bibinfo {author} {\bibfnamefont {C.}~\bibnamefont
  {Bouchard}}, \bibinfo {author} {\bibfnamefont {C.~C.}\ \bibnamefont {Chang}},
  \bibinfo {author} {\bibfnamefont {T.}~\bibnamefont {Kurth}}, \bibinfo
  {author} {\bibfnamefont {K.}~\bibnamefont {Orginos}}, \ and\ \bibinfo
  {author} {\bibfnamefont {A.}~\bibnamefont {Walker-Loud}},\ }\href {\doibase
  10.1103/PhysRevD.96.014504} {\bibfield  {journal} {\bibinfo  {journal} {Phys.
  Rev. D}\ }\textbf {\bibinfo {volume} {96}},\ \bibinfo {pages} {014504}
  (\bibinfo {year} {2017})},\ \Eprint {http://arxiv.org/abs/1612.06963}
  {arXiv:1612.06963 [hep-lat]} \BibitemShut {NoStop}%
\bibitem [{\citenamefont {Shanahan}\ \emph {et~al.}(2017)\citenamefont
  {Shanahan}, \citenamefont {Tiburzi}, \citenamefont {Wagman}, \citenamefont
  {Winter}, \citenamefont {Chang}, \citenamefont {Davoudi}, \citenamefont
  {Detmold}, \citenamefont {Orginos},\ and\ \citenamefont
  {Savage}}]{Shanahan:2017bgi}%
  \BibitemOpen
  \bibfield  {author} {\bibinfo {author} {\bibfnamefont {P.~E.}\ \bibnamefont
  {Shanahan}}, \bibinfo {author} {\bibfnamefont {B.~C.}\ \bibnamefont
  {Tiburzi}}, \bibinfo {author} {\bibfnamefont {M.~L.}\ \bibnamefont {Wagman}},
  \bibinfo {author} {\bibfnamefont {F.}~\bibnamefont {Winter}}, \bibinfo
  {author} {\bibfnamefont {E.}~\bibnamefont {Chang}}, \bibinfo {author}
  {\bibfnamefont {Z.}~\bibnamefont {Davoudi}}, \bibinfo {author} {\bibfnamefont
  {W.}~\bibnamefont {Detmold}}, \bibinfo {author} {\bibfnamefont
  {K.}~\bibnamefont {Orginos}}, \ and\ \bibinfo {author} {\bibfnamefont
  {M.~J.}\ \bibnamefont {Savage}},\ }\href {\doibase
  10.1103/PhysRevLett.119.062003} {\bibfield  {journal} {\bibinfo  {journal}
  {Phys. Rev. Lett.}\ }\textbf {\bibinfo {volume} {119}},\ \bibinfo {pages}
  {062003} (\bibinfo {year} {2017})},\ \Eprint
  {http://arxiv.org/abs/1701.03456} {arXiv:1701.03456 [hep-lat]} \BibitemShut
  {NoStop}%
\bibitem [{\citenamefont {Tiburzi}\ \emph {et~al.}(2017)\citenamefont
  {Tiburzi}, \citenamefont {Wagman}, \citenamefont {Winter}, \citenamefont
  {Chang}, \citenamefont {Davoudi}, \citenamefont {Detmold}, \citenamefont
  {Orginos}, \citenamefont {Savage},\ and\ \citenamefont
  {Shanahan}}]{Tiburzi:2017iux}%
  \BibitemOpen
  \bibfield  {author} {\bibinfo {author} {\bibfnamefont {B.~C.}\ \bibnamefont
  {Tiburzi}}, \bibinfo {author} {\bibfnamefont {M.~L.}\ \bibnamefont {Wagman}},
  \bibinfo {author} {\bibfnamefont {F.}~\bibnamefont {Winter}}, \bibinfo
  {author} {\bibfnamefont {E.}~\bibnamefont {Chang}}, \bibinfo {author}
  {\bibfnamefont {Z.}~\bibnamefont {Davoudi}}, \bibinfo {author} {\bibfnamefont
  {W.}~\bibnamefont {Detmold}}, \bibinfo {author} {\bibfnamefont
  {K.}~\bibnamefont {Orginos}}, \bibinfo {author} {\bibfnamefont {M.~J.}\
  \bibnamefont {Savage}}, \ and\ \bibinfo {author} {\bibfnamefont {P.~E.}\
  \bibnamefont {Shanahan}},\ }\href {\doibase 10.1103/PhysRevD.96.054505}
  {\bibfield  {journal} {\bibinfo  {journal} {Phys. Rev. D}\ }\textbf {\bibinfo
  {volume} {96}},\ \bibinfo {pages} {054505} (\bibinfo {year} {2017})},\
  \Eprint {http://arxiv.org/abs/1702.02929} {arXiv:1702.02929 [hep-lat]}
  \BibitemShut {NoStop}%
\bibitem [{\citenamefont {Agadjanov}\ \emph {et~al.}(2017)\citenamefont
  {Agadjanov}, \citenamefont {Mei{\ss}ner},\ and\ \citenamefont
  {Rusetsky}}]{Agadjanov:2016cjc}%
  \BibitemOpen
  \bibfield  {author} {\bibinfo {author} {\bibfnamefont {A.}~\bibnamefont
  {Agadjanov}}, \bibinfo {author} {\bibfnamefont {U.-G.}\ \bibnamefont
  {Mei{\ss}ner}}, \ and\ \bibinfo {author} {\bibfnamefont {A.}~\bibnamefont
  {Rusetsky}},\ }\href {\doibase 10.1103/PhysRevD.95.031502} {\bibfield
  {journal} {\bibinfo  {journal} {Phys. Rev. D}\ }\textbf {\bibinfo {volume}
  {95}},\ \bibinfo {pages} {031502(R)} (\bibinfo {year} {2017})},\ \Eprint
  {http://arxiv.org/abs/1610.05545} {arXiv:1610.05545 [hep-lat]} \BibitemShut
  {NoStop}%
\bibitem [{\citenamefont {Devenish}\ and\ \citenamefont
  {Cooper-Sarkar}(2004)}]{DevenishRobin2004Dis}%
  \BibitemOpen
  \bibfield  {author} {\bibinfo {author} {\bibfnamefont {R.}~\bibnamefont
  {Devenish}}\ and\ \bibinfo {author} {\bibfnamefont {A.}~\bibnamefont
  {Cooper-Sarkar}},\ }\href@noop {} {\emph {\bibinfo {title} {Deep inelastic
  scattering}}}\ (\bibinfo  {publisher} {Oxford University Press},\ \bibinfo
  {address} {Oxford},\ \bibinfo {year} {2004})\BibitemShut {NoStop}%
\bibitem [{\citenamefont {Manohar}(1992)}]{Manohar:1992tz}%
  \BibitemOpen
  \bibfield  {author} {\bibinfo {author} {\bibfnamefont {A.~V.}\ \bibnamefont
  {Manohar}},\ }in\ \href@noop {} {\emph {\bibinfo {booktitle} {{Lake Louise
  Winter Institute: Symmetry and Spin in the Standard Model Lake Louise,
  Alberta, Canada, February 23-29, 1992}}}}\ (\bibinfo {year} {1992})\ pp.\
  \bibinfo {pages} {1--46},\ \Eprint {http://arxiv.org/abs/hep-ph/9204208}
  {arXiv:hep-ph/9204208 [hep-ph]} \BibitemShut {NoStop}%
\bibitem [{\citenamefont {Liu}(2000)}]{Liu:1999ak}%
  \BibitemOpen
  \bibfield  {author} {\bibinfo {author} {\bibfnamefont {K.-F.}\ \bibnamefont
  {Liu}},\ }\href {\doibase 10.1103/PhysRevD.62.074501} {\bibfield  {journal}
  {\bibinfo  {journal} {Phys. Rev.}\ }\textbf {\bibinfo {volume} {D62}},\
  \bibinfo {pages} {074501} (\bibinfo {year} {2000})},\ \Eprint
  {http://arxiv.org/abs/hep-ph/9910306} {arXiv:hep-ph/9910306 [hep-ph]}
  \BibitemShut {NoStop}%
\bibitem [{\citenamefont {Drechsel}\ \emph {et~al.}(2003)\citenamefont
  {Drechsel}, \citenamefont {Pasquini},\ and\ \citenamefont
  {Vanderhaeghen}}]{Drechsel:2002ar}%
  \BibitemOpen
  \bibfield  {author} {\bibinfo {author} {\bibfnamefont {D.}~\bibnamefont
  {Drechsel}}, \bibinfo {author} {\bibfnamefont {B.}~\bibnamefont {Pasquini}},
  \ and\ \bibinfo {author} {\bibfnamefont {M.}~\bibnamefont {Vanderhaeghen}},\
  }\href {\doibase 10.1016/S0370-1573(02)00636-1} {\bibfield  {journal}
  {\bibinfo  {journal} {Phys. Rept.}\ }\textbf {\bibinfo {volume} {378}},\
  \bibinfo {pages} {99} (\bibinfo {year} {2003})},\ \Eprint
  {http://arxiv.org/abs/hep-ph/0212124} {arXiv:hep-ph/0212124} \BibitemShut
  {NoStop}%
\bibitem [{\citenamefont {Brice{\~n}o}\ \emph {et~al.}(2017)\citenamefont
  {Brice{\~n}o}, \citenamefont {Hansen},\ and\ \citenamefont
  {Monahan}}]{Briceno:2017cpo}%
  \BibitemOpen
  \bibfield  {author} {\bibinfo {author} {\bibfnamefont {R.~A.}\ \bibnamefont
  {Brice{\~n}o}}, \bibinfo {author} {\bibfnamefont {M.~T.}\ \bibnamefont
  {Hansen}}, \ and\ \bibinfo {author} {\bibfnamefont {C.~J.}\ \bibnamefont
  {Monahan}},\ }\href {\doibase 10.1103/PhysRevD.96.014502} {\bibfield
  {journal} {\bibinfo  {journal} {Phys. Rev. D}\ }\textbf {\bibinfo {volume}
  {96}},\ \bibinfo {pages} {014502} (\bibinfo {year} {2017})},\ \Eprint
  {http://arxiv.org/abs/1703.06072} {arXiv:1703.06072 [hep-lat]} \BibitemShut
  {NoStop}%
\bibitem [{\citenamefont {Gunn}\ \emph {et~al.}(2020)\citenamefont {Gunn} \emph
  {et~al.}}]{Alec:2020fp}%
  \BibitemOpen
  \bibfield  {author} {\bibinfo {author} {\bibfnamefont {A.~H.}\ \bibnamefont
  {Gunn}} \emph {et~al.},\ }\href@noop {} {\bibfield  {journal} {\bibinfo
  {journal} {in preperation}\ } (\bibinfo {year} {2020})}\BibitemShut {NoStop}%
\bibitem [{\citenamefont {Cundy}\ \emph {et~al.}(2009)\citenamefont {Cundy},
  \citenamefont {G{\"o}ckeler}, \citenamefont {Horsley}, \citenamefont
  {Kaltenbrunner}, \citenamefont {Kennedy}, \citenamefont {Nakamura},
  \citenamefont {Perlt}, \citenamefont {Pleiter}, \citenamefont {Rakow},
  \citenamefont {Sch{\"a}fer}, \citenamefont {Schierholz}, \citenamefont
  {Schiller}, \citenamefont {St{\"u}ben},\ and\ \citenamefont
  {Zanotti}}]{Cundy:2009yy}%
  \BibitemOpen
  \bibfield  {author} {\bibinfo {author} {\bibfnamefont {N.}~\bibnamefont
  {Cundy}}, \bibinfo {author} {\bibfnamefont {M.}~\bibnamefont {G{\"o}ckeler}},
  \bibinfo {author} {\bibfnamefont {R.}~\bibnamefont {Horsley}}, \bibinfo
  {author} {\bibfnamefont {T.}~\bibnamefont {Kaltenbrunner}}, \bibinfo {author}
  {\bibfnamefont {A.~D.}\ \bibnamefont {Kennedy}}, \bibinfo {author}
  {\bibfnamefont {Y.}~\bibnamefont {Nakamura}}, \bibinfo {author}
  {\bibfnamefont {H.}~\bibnamefont {Perlt}}, \bibinfo {author} {\bibfnamefont
  {D.}~\bibnamefont {Pleiter}}, \bibinfo {author} {\bibfnamefont {P.~E.~L.}\
  \bibnamefont {Rakow}}, \bibinfo {author} {\bibfnamefont {A.}~\bibnamefont
  {Sch{\"a}fer}}, \bibinfo {author} {\bibfnamefont {G.}~\bibnamefont
  {Schierholz}}, \bibinfo {author} {\bibfnamefont {A.}~\bibnamefont
  {Schiller}}, \bibinfo {author} {\bibfnamefont {H.}~\bibnamefont
  {St{\"u}ben}}, \ and\ \bibinfo {author} {\bibfnamefont {J.~M.}\ \bibnamefont
  {Zanotti}},\ }\href {\doibase 10.1103/PhysRevD.79.094507} {\bibfield
  {journal} {\bibinfo  {journal} {Phys. Rev. D}\ }\textbf {\bibinfo {volume}
  {79}},\ \bibinfo {pages} {094507} (\bibinfo {year} {2009})},\ \Eprint
  {http://arxiv.org/abs/0901.3302} {arXiv:0901.3302 [hep-lat]} \BibitemShut
  {NoStop}%
\bibitem [{\citenamefont {Bietenholz}\ \emph
  {et~al.}(2011{\natexlab{b}})\citenamefont {Bietenholz}, \citenamefont
  {Bornyakov}, \citenamefont {G{\"o}ckeler}, \citenamefont {Horsley},
  \citenamefont {Lockhart}, \citenamefont {Nakamura}, \citenamefont {Perlt},
  \citenamefont {Pleiter}, \citenamefont {Rakow}, \citenamefont {Schierholz},
  \citenamefont {Schiller}, \citenamefont {Streuer}, \citenamefont
  {St{\"u}ben}, \citenamefont {Winter},\ and\ \citenamefont
  {Zanotti}}]{PhysRevD.84.054509}%
  \BibitemOpen
  \bibfield  {author} {\bibinfo {author} {\bibfnamefont {W.}~\bibnamefont
  {Bietenholz}}, \bibinfo {author} {\bibfnamefont {V.}~\bibnamefont
  {Bornyakov}}, \bibinfo {author} {\bibfnamefont {M.}~\bibnamefont
  {G{\"o}ckeler}}, \bibinfo {author} {\bibfnamefont {R.}~\bibnamefont
  {Horsley}}, \bibinfo {author} {\bibfnamefont {W.~G.}\ \bibnamefont
  {Lockhart}}, \bibinfo {author} {\bibfnamefont {Y.}~\bibnamefont {Nakamura}},
  \bibinfo {author} {\bibfnamefont {H.}~\bibnamefont {Perlt}}, \bibinfo
  {author} {\bibfnamefont {D.}~\bibnamefont {Pleiter}}, \bibinfo {author}
  {\bibfnamefont {P.~E.~L.}\ \bibnamefont {Rakow}}, \bibinfo {author}
  {\bibfnamefont {G.}~\bibnamefont {Schierholz}}, \bibinfo {author}
  {\bibfnamefont {A.}~\bibnamefont {Schiller}}, \bibinfo {author}
  {\bibfnamefont {T.}~\bibnamefont {Streuer}}, \bibinfo {author} {\bibfnamefont
  {H.}~\bibnamefont {St{\"u}ben}}, \bibinfo {author} {\bibfnamefont
  {F.}~\bibnamefont {Winter}}, \ and\ \bibinfo {author} {\bibfnamefont {J.~M.}\
  \bibnamefont {Zanotti}} (\bibinfo {collaboration} {QCDSF Collaboration, UKQCD
  Collaboration}),\ }\href {\doibase 10.1103/PhysRevD.84.054509} {\bibfield
  {journal} {\bibinfo  {journal} {Phys. Rev. D}\ }\textbf {\bibinfo {volume}
  {84}},\ \bibinfo {pages} {054509} (\bibinfo {year}
  {2011}{\natexlab{b}})}\BibitemShut {NoStop}%
\bibitem [{\citenamefont {Bietenholz}\ \emph
  {et~al.}(2010{\natexlab{b}})\citenamefont {Bietenholz}, \citenamefont
  {Bornyakov}, \citenamefont {Cundy}, \citenamefont {G{\"o}ckeler},
  \citenamefont {Horsley}, \citenamefont {Kennedy}, \citenamefont {Lockhart},
  \citenamefont {Nakamura}, \citenamefont {Perlt}, \citenamefont {Pleiter},
  \citenamefont {Rakow}, \citenamefont {Sch{\"a}fer}, \citenamefont
  {Schierholz}, \citenamefont {Schiller}, \citenamefont {St{\"u}ben},\ and\
  \citenamefont {Zanotti}}]{Bietenholz:2010jr}%
  \BibitemOpen
  \bibfield  {author} {\bibinfo {author} {\bibfnamefont {W.}~\bibnamefont
  {Bietenholz}}, \bibinfo {author} {\bibfnamefont {V.}~\bibnamefont
  {Bornyakov}}, \bibinfo {author} {\bibfnamefont {N.}~\bibnamefont {Cundy}},
  \bibinfo {author} {\bibfnamefont {M.}~\bibnamefont {G{\"o}ckeler}}, \bibinfo
  {author} {\bibfnamefont {R.}~\bibnamefont {Horsley}}, \bibinfo {author}
  {\bibfnamefont {A.~D.}\ \bibnamefont {Kennedy}}, \bibinfo {author}
  {\bibfnamefont {W.~G.}\ \bibnamefont {Lockhart}}, \bibinfo {author}
  {\bibfnamefont {Y.}~\bibnamefont {Nakamura}}, \bibinfo {author}
  {\bibfnamefont {H.}~\bibnamefont {Perlt}}, \bibinfo {author} {\bibfnamefont
  {D.}~\bibnamefont {Pleiter}}, \bibinfo {author} {\bibfnamefont {P.~E.~L.}\
  \bibnamefont {Rakow}}, \bibinfo {author} {\bibfnamefont {A.}~\bibnamefont
  {Sch{\"a}fer}}, \bibinfo {author} {\bibfnamefont {G.}~\bibnamefont
  {Schierholz}}, \bibinfo {author} {\bibfnamefont {A.}~\bibnamefont
  {Schiller}}, \bibinfo {author} {\bibfnamefont {H.}~\bibnamefont
  {St{\"u}ben}}, \ and\ \bibinfo {author} {\bibfnamefont {J.~M.}\ \bibnamefont
  {Zanotti}},\ }\href {\doibase 10.1016/j.physletb.2010.05.067} {\bibfield
  {journal} {\bibinfo  {journal} {Phys. Lett.}\ }\textbf {\bibinfo {volume}
  {B690}},\ \bibinfo {pages} {436} (\bibinfo {year} {2010}{\natexlab{b}})},\
  \Eprint {http://arxiv.org/abs/1003.1114} {arXiv:1003.1114 [hep-lat]}
  \BibitemShut {NoStop}%
\bibitem [{\citenamefont {Horsley}\ \emph {et~al.}(2014)\citenamefont
  {Horsley}, \citenamefont {Najjar}, \citenamefont {Nakamura}, \citenamefont
  {Perlt}, \citenamefont {Pleiter}, \citenamefont {Rakow}, \citenamefont
  {Schierholz}, \citenamefont {Schiller}, \citenamefont {St{\"u}ben},\ and\
  \citenamefont {Zanotti}}]{Horsley:2013wqa}%
  \BibitemOpen
  \bibfield  {author} {\bibinfo {author} {\bibfnamefont {R.}~\bibnamefont
  {Horsley}}, \bibinfo {author} {\bibfnamefont {J.}~\bibnamefont {Najjar}},
  \bibinfo {author} {\bibfnamefont {Y.}~\bibnamefont {Nakamura}}, \bibinfo
  {author} {\bibfnamefont {H.}~\bibnamefont {Perlt}}, \bibinfo {author}
  {\bibfnamefont {D.}~\bibnamefont {Pleiter}}, \bibinfo {author} {\bibfnamefont
  {P.~E.~L.}\ \bibnamefont {Rakow}}, \bibinfo {author} {\bibfnamefont
  {G.}~\bibnamefont {Schierholz}}, \bibinfo {author} {\bibfnamefont
  {A.}~\bibnamefont {Schiller}}, \bibinfo {author} {\bibfnamefont
  {H.}~\bibnamefont {St{\"u}ben}}, \ and\ \bibinfo {author} {\bibfnamefont
  {J.~M.}\ \bibnamefont {Zanotti}} (\bibinfo {collaboration} {QCDSF
  Collaboration, UKQCD Collaboration}),\ }\bibfield  {booktitle} {\emph
  {\bibinfo {booktitle} {{Proceedings, 31st International Symposium on Lattice
  Field Theory (Lattice 2013): Mainz, Germany, July 29-August 3, 2013}}},\
  }\href {\doibase 10.22323/1.187.0249} {\bibfield  {journal} {\bibinfo
  {journal} {PoS}\ }\textbf {\bibinfo {volume} {LATTICE2013}},\ \bibinfo
  {pages} {249} (\bibinfo {year} {2014})},\ \Eprint
  {http://arxiv.org/abs/1311.5010} {arXiv:1311.5010 [hep-lat]} \BibitemShut
  {NoStop}%
\bibitem [{\citenamefont {Bornyakov}\ \emph {et~al.}(2015)\citenamefont
  {Bornyakov}, \citenamefont {Horsley}, \citenamefont {Hudspith}, \citenamefont
  {Nakamura}, \citenamefont {Perlt}, \citenamefont {Pleiter}, \citenamefont
  {Rakow}, \citenamefont {Schierholz}, \citenamefont {Schiller}, \citenamefont
  {St{\"u}ben},\ and\ \citenamefont {Zanotti}}]{Bornyakov:2015eaa}%
  \BibitemOpen
  \bibfield  {author} {\bibinfo {author} {\bibfnamefont {V.~G.}\ \bibnamefont
  {Bornyakov}}, \bibinfo {author} {\bibfnamefont {R.}~\bibnamefont {Horsley}},
  \bibinfo {author} {\bibfnamefont {R.}~\bibnamefont {Hudspith}}, \bibinfo
  {author} {\bibfnamefont {Y.}~\bibnamefont {Nakamura}}, \bibinfo {author}
  {\bibfnamefont {H.}~\bibnamefont {Perlt}}, \bibinfo {author} {\bibfnamefont
  {D.}~\bibnamefont {Pleiter}}, \bibinfo {author} {\bibfnamefont {P.~E.~L.}\
  \bibnamefont {Rakow}}, \bibinfo {author} {\bibfnamefont {G.}~\bibnamefont
  {Schierholz}}, \bibinfo {author} {\bibfnamefont {A.}~\bibnamefont
  {Schiller}}, \bibinfo {author} {\bibfnamefont {H.}~\bibnamefont
  {St{\"u}ben}}, \ and\ \bibinfo {author} {\bibfnamefont {J.~M.}\ \bibnamefont
  {Zanotti}},\ }\href@noop {} {\  (\bibinfo {year} {2015})},\ \Eprint
  {http://arxiv.org/abs/1508.05916} {arXiv:1508.05916 [hep-lat]} \BibitemShut
  {NoStop}%
\bibitem [{\citenamefont {Constantinou}\ \emph {et~al.}(2015)\citenamefont
  {Constantinou}, \citenamefont {Horsley}, \citenamefont {Panagopoulos},
  \citenamefont {Perlt}, \citenamefont {Rakow}, \citenamefont {Schierholz},
  \citenamefont {Schiller},\ and\ \citenamefont
  {Zanotti}}]{Constantinou:2014fka}%
  \BibitemOpen
  \bibfield  {author} {\bibinfo {author} {\bibfnamefont {M.}~\bibnamefont
  {Constantinou}}, \bibinfo {author} {\bibfnamefont {R.}~\bibnamefont
  {Horsley}}, \bibinfo {author} {\bibfnamefont {H.}~\bibnamefont
  {Panagopoulos}}, \bibinfo {author} {\bibfnamefont {H.}~\bibnamefont {Perlt}},
  \bibinfo {author} {\bibfnamefont {P.~E.~L.}\ \bibnamefont {Rakow}}, \bibinfo
  {author} {\bibfnamefont {G.}~\bibnamefont {Schierholz}}, \bibinfo {author}
  {\bibfnamefont {A.}~\bibnamefont {Schiller}}, \ and\ \bibinfo {author}
  {\bibfnamefont {J.~M.}\ \bibnamefont {Zanotti}},\ }\href {\doibase
  10.1103/PhysRevD.91.014502} {\bibfield  {journal} {\bibinfo  {journal} {Phys.
  Rev. D}\ }\textbf {\bibinfo {volume} {91}},\ \bibinfo {pages} {014502}
  (\bibinfo {year} {2015})},\ \Eprint {http://arxiv.org/abs/1408.6047}
  {arXiv:1408.6047 [hep-lat]} \BibitemShut {NoStop}%
\bibitem [{\citenamefont {Hannaford-Gunn}\ \emph {et~al.}(2020)\citenamefont
  {Hannaford-Gunn}, \citenamefont {Horsley}, \citenamefont {Nakamura},
  \citenamefont {Perlt}, \citenamefont {Rakow}, \citenamefont {Schierholz},
  \citenamefont {Somfleth}, \citenamefont {St{\"u}ben}, \citenamefont {Young},\
  and\ \citenamefont {Zanotti}}]{Hannaford-Gunn:2020pvu}%
  \BibitemOpen
  \bibfield  {author} {\bibinfo {author} {\bibfnamefont {A.}~\bibnamefont
  {Hannaford-Gunn}}, \bibinfo {author} {\bibfnamefont {R.}~\bibnamefont
  {Horsley}}, \bibinfo {author} {\bibfnamefont {Y.}~\bibnamefont {Nakamura}},
  \bibinfo {author} {\bibfnamefont {H.}~\bibnamefont {Perlt}}, \bibinfo
  {author} {\bibfnamefont {P.~E.~L.}\ \bibnamefont {Rakow}}, \bibinfo {author}
  {\bibfnamefont {G.}~\bibnamefont {Schierholz}}, \bibinfo {author}
  {\bibfnamefont {K.}~\bibnamefont {Somfleth}}, \bibinfo {author}
  {\bibfnamefont {H.}~\bibnamefont {St{\"u}ben}}, \bibinfo {author}
  {\bibfnamefont {R.~D.}\ \bibnamefont {Young}}, \ and\ \bibinfo {author}
  {\bibfnamefont {J.~M.}\ \bibnamefont {Zanotti}},\ }in\ \href@noop {} {\emph
  {\bibinfo {booktitle} {{37th International Symposium on Lattice Field Theory
  (Lattice 2019) Wuhan, Hubei, China, June 16-22, 2019}}}}\ (\bibinfo {year}
  {2020})\ \Eprint {http://arxiv.org/abs/2001.05090} {arXiv:2001.05090
  [hep-lat]} \BibitemShut {NoStop}%
\bibitem [{\citenamefont {Allton}\ \emph {et~al.}(1993)\citenamefont {Allton},
  \citenamefont {Sachrajda}, \citenamefont {Baxter}, \citenamefont {Booth},
  \citenamefont {Bowler}, \citenamefont {Collins}, \citenamefont {Henty},
  \citenamefont {Kenway}, \citenamefont {Pendleton}, \citenamefont {Richards},
  \citenamefont {Simone}, \citenamefont {Simpson}, \citenamefont {Wilkes},\
  and\ \citenamefont {Michael}}]{Allton:1993wc}%
  \BibitemOpen
  \bibfield  {author} {\bibinfo {author} {\bibfnamefont {C.~R.}\ \bibnamefont
  {Allton}}, \bibinfo {author} {\bibfnamefont {C.~T.}\ \bibnamefont
  {Sachrajda}}, \bibinfo {author} {\bibfnamefont {R.~M.}\ \bibnamefont
  {Baxter}}, \bibinfo {author} {\bibfnamefont {S.~P.}\ \bibnamefont {Booth}},
  \bibinfo {author} {\bibfnamefont {K.~C.}\ \bibnamefont {Bowler}}, \bibinfo
  {author} {\bibfnamefont {S.}~\bibnamefont {Collins}}, \bibinfo {author}
  {\bibfnamefont {D.~S.}\ \bibnamefont {Henty}}, \bibinfo {author}
  {\bibfnamefont {R.~D.}\ \bibnamefont {Kenway}}, \bibinfo {author}
  {\bibfnamefont {B.~J.}\ \bibnamefont {Pendleton}}, \bibinfo {author}
  {\bibfnamefont {D.~G.}\ \bibnamefont {Richards}}, \bibinfo {author}
  {\bibfnamefont {J.~N.}\ \bibnamefont {Simone}}, \bibinfo {author}
  {\bibfnamefont {A.~D.}\ \bibnamefont {Simpson}}, \bibinfo {author}
  {\bibfnamefont {B.~E.}\ \bibnamefont {Wilkes}}, \ and\ \bibinfo {author}
  {\bibfnamefont {C.}~\bibnamefont {Michael}} (\bibinfo {collaboration} {UKQCD
  Collaboration}),\ }\href {\doibase 10.1103/PhysRevD.47.5128} {\bibfield
  {journal} {\bibinfo  {journal} {Phys. Rev. D}\ }\textbf {\bibinfo {volume}
  {47}},\ \bibinfo {pages} {5128} (\bibinfo {year} {1993})},\ \Eprint
  {http://arxiv.org/abs/hep-lat/9303009} {arXiv:hep-lat/9303009} \BibitemShut
  {NoStop}%
\bibitem [{\citenamefont {Hausdorff}(1921)}]{Hausdorff}%
  \BibitemOpen
  \bibfield  {author} {\bibinfo {author} {\bibfnamefont {F.}~\bibnamefont
  {Hausdorff}},\ }\href@noop {} {\bibfield  {journal} {\bibinfo  {journal}
  {Mathematische Zeitschrift}\ }\textbf {\bibinfo {volume} {9}},\ \bibinfo
  {pages} {74} (\bibinfo {year} {1921})}\BibitemShut {NoStop}%
\bibitem [{\citenamefont {Salvatier}\ \emph {et~al.}(2016)\citenamefont
  {Salvatier}, \citenamefont {Wiecki},\ and\ \citenamefont
  {Fonnesbeck}}]{Salvatier:2016swf}%
  \BibitemOpen
  \bibfield  {author} {\bibinfo {author} {\bibfnamefont {J.}~\bibnamefont
  {Salvatier}}, \bibinfo {author} {\bibfnamefont {T.~V.}\ \bibnamefont
  {Wiecki}}, \ and\ \bibinfo {author} {\bibfnamefont {C.}~\bibnamefont
  {Fonnesbeck}},\ }\href@noop {} {\bibfield  {journal} {\bibinfo  {journal}
  {PeerJ Computer Science}\ }\textbf {\bibinfo {volume} {2:e55}} (\bibinfo
  {year} {2016})}\BibitemShut {NoStop}%
\bibitem [{\citenamefont {Hoffman}\ and\ \citenamefont
  {Gelman}(2014)}]{JMLR:v15:hoffman14a}%
  \BibitemOpen
  \bibfield  {author} {\bibinfo {author} {\bibfnamefont {M.~D.}\ \bibnamefont
  {Hoffman}}\ and\ \bibinfo {author} {\bibfnamefont {A.}~\bibnamefont
  {Gelman}},\ }\href {http://jmlr.org/papers/v15/hoffman14a.html} {\bibfield
  {journal} {\bibinfo  {journal} {Journal of Machine Learning Research}\
  }\textbf {\bibinfo {volume} {15}},\ \bibinfo {pages} {1593} (\bibinfo {year}
  {2014})}\BibitemShut {NoStop}%
\bibitem [{\citenamefont {Bloom}\ and\ \citenamefont
  {Gilman}(1970)}]{Bloom:1970xb}%
  \BibitemOpen
  \bibfield  {author} {\bibinfo {author} {\bibfnamefont {E.~D.}\ \bibnamefont
  {Bloom}}\ and\ \bibinfo {author} {\bibfnamefont {F.~J.}\ \bibnamefont
  {Gilman}},\ }\href {\doibase 10.1103/PhysRevLett.25.1140} {\bibfield
  {journal} {\bibinfo  {journal} {Phys. Rev. Lett.}\ }\textbf {\bibinfo
  {volume} {25}},\ \bibinfo {pages} {1140} (\bibinfo {year}
  {1970})}\BibitemShut {NoStop}%
\bibitem [{\citenamefont {Nachtmann}(1973)}]{Nachtmann:1973mr}%
  \BibitemOpen
  \bibfield  {author} {\bibinfo {author} {\bibfnamefont {O.}~\bibnamefont
  {Nachtmann}},\ }\href {\doibase 10.1016/0550-3213(73)90144-2} {\bibfield
  {journal} {\bibinfo  {journal} {Nucl. Phys. B}\ }\textbf {\bibinfo {volume}
  {63}},\ \bibinfo {pages} {237} (\bibinfo {year} {1973})}\BibitemShut
  {NoStop}%
\bibitem [{\citenamefont {Kuroda}\ and\ \citenamefont
  {Schierholz}(1979)}]{Kuroda:1978jx}%
  \BibitemOpen
  \bibfield  {author} {\bibinfo {author} {\bibfnamefont {M.}~\bibnamefont
  {Kuroda}}\ and\ \bibinfo {author} {\bibfnamefont {G.}~\bibnamefont
  {Schierholz}},\ }\href {\doibase 10.1016/0550-3213(79)90020-8} {\bibfield
  {journal} {\bibinfo  {journal} {Nucl. Phys. B}\ }\textbf {\bibinfo {volume}
  {148}},\ \bibinfo {pages} {148} (\bibinfo {year} {1979})}\BibitemShut
  {NoStop}%
\bibitem [{\citenamefont {Cornwall}\ and\ \citenamefont
  {Norton}(1969)}]{Cornwall:1968cx}%
  \BibitemOpen
  \bibfield  {author} {\bibinfo {author} {\bibfnamefont {J.~M.}\ \bibnamefont
  {Cornwall}}\ and\ \bibinfo {author} {\bibfnamefont {R.~E.}\ \bibnamefont
  {Norton}},\ }\href {\doibase 10.1103/PhysRev.177.2584} {\bibfield  {journal}
  {\bibinfo  {journal} {Phys. Rev.}\ }\textbf {\bibinfo {volume} {177}},\
  \bibinfo {pages} {2584} (\bibinfo {year} {1969})}\BibitemShut {NoStop}%
\bibitem [{\citenamefont {Capitani}\ \emph
  {et~al.}(1999{\natexlab{b}})\citenamefont {Capitani}, \citenamefont
  {G{\"o}ckeler}, \citenamefont {Horsley}, \citenamefont {Petters},
  \citenamefont {Pleiter}, \citenamefont {Rakow},\ and\ \citenamefont
  {Schierholz}}]{CAPITANI1999173}%
  \BibitemOpen
  \bibfield  {author} {\bibinfo {author} {\bibfnamefont {S.}~\bibnamefont
  {Capitani}}, \bibinfo {author} {\bibfnamefont {M.}~\bibnamefont
  {G{\"o}ckeler}}, \bibinfo {author} {\bibfnamefont {R.}~\bibnamefont
  {Horsley}}, \bibinfo {author} {\bibfnamefont {D.}~\bibnamefont {Petters}},
  \bibinfo {author} {\bibfnamefont {D.}~\bibnamefont {Pleiter}}, \bibinfo
  {author} {\bibfnamefont {P.}~\bibnamefont {Rakow}}, \ and\ \bibinfo {author}
  {\bibfnamefont {G.}~\bibnamefont {Schierholz}},\ }\href {\doibase
  https://doi.org/10.1016/S0920-5632(99)00666-0} {\bibfield  {journal}
  {\bibinfo  {journal} {Nuclear Physics B - Proceedings Supplements}\ }\textbf
  {\bibinfo {volume} {79}},\ \bibinfo {pages} {173 } (\bibinfo {year}
  {1999}{\natexlab{b}})},\ \bibinfo {note} {proceedings of the 7th
  International Workshop on Deep Inelastic Scattering and QCD}\BibitemShut
  {NoStop}%
\bibitem [{\citenamefont {Bietenholz}\ \emph
  {et~al.}(2009{\natexlab{b}})\citenamefont {Bietenholz}, \citenamefont
  {Cundy}, \citenamefont {G{\"o}ckeler}, \citenamefont {Horsley}, \citenamefont
  {Perlt}, \citenamefont {Pleiter}, \citenamefont {Rakow}, \citenamefont
  {Schierholz}, \citenamefont {Schiller}, \citenamefont {Streuer},\ and\
  \citenamefont {Zanotti}}]{Bietenholz:2009rb}%
  \BibitemOpen
  \bibfield  {author} {\bibinfo {author} {\bibfnamefont {W.}~\bibnamefont
  {Bietenholz}}, \bibinfo {author} {\bibfnamefont {N.}~\bibnamefont {Cundy}},
  \bibinfo {author} {\bibfnamefont {M.}~\bibnamefont {G{\"o}ckeler}}, \bibinfo
  {author} {\bibfnamefont {R.}~\bibnamefont {Horsley}}, \bibinfo {author}
  {\bibfnamefont {H.}~\bibnamefont {Perlt}}, \bibinfo {author} {\bibfnamefont
  {D.}~\bibnamefont {Pleiter}}, \bibinfo {author} {\bibfnamefont {P.~E.~L.}\
  \bibnamefont {Rakow}}, \bibinfo {author} {\bibfnamefont {G.}~\bibnamefont
  {Schierholz}}, \bibinfo {author} {\bibfnamefont {A.}~\bibnamefont
  {Schiller}}, \bibinfo {author} {\bibfnamefont {T.}~\bibnamefont {Streuer}}, \
  and\ \bibinfo {author} {\bibfnamefont {J.~M.}\ \bibnamefont {Zanotti}},\
  }\href {\doibase 10.22323/1.091.0138} {\bibfield  {journal} {\bibinfo
  {journal} {PoS}\ }\textbf {\bibinfo {volume} {LAT2009}},\ \bibinfo {pages}
  {138} (\bibinfo {year} {2009}{\natexlab{b}})},\ \Eprint
  {http://arxiv.org/abs/0910.2437} {arXiv:0910.2437 [hep-lat]} \BibitemShut
  {NoStop}%
\bibitem [{\citenamefont {Melnitchouk}\ \emph {et~al.}(2005)\citenamefont
  {Melnitchouk}, \citenamefont {Ent},\ and\ \citenamefont
  {Keppel}}]{Melnitchouk:2005zr}%
  \BibitemOpen
  \bibfield  {author} {\bibinfo {author} {\bibfnamefont {W.}~\bibnamefont
  {Melnitchouk}}, \bibinfo {author} {\bibfnamefont {R.}~\bibnamefont {Ent}}, \
  and\ \bibinfo {author} {\bibfnamefont {C.}~\bibnamefont {Keppel}},\ }\href
  {\doibase 10.1016/j.physrep.2004.10.004} {\bibfield  {journal} {\bibinfo
  {journal} {Phys. Rept.}\ }\textbf {\bibinfo {volume} {406}},\ \bibinfo
  {pages} {127} (\bibinfo {year} {2005})},\ \Eprint
  {http://arxiv.org/abs/hep-ph/0501217} {arXiv:hep-ph/0501217} \BibitemShut
  {NoStop}%
\bibitem [{\citenamefont {L{\"u}scher}(1986)}]{Luscher:1985dn}%
  \BibitemOpen
  \bibfield  {author} {\bibinfo {author} {\bibfnamefont {M.}~\bibnamefont
  {L{\"u}scher}},\ }\href {\doibase 10.1007/BF01211589} {\bibfield  {journal}
  {\bibinfo  {journal} {Commun. Math. Phys.}\ }\textbf {\bibinfo {volume}
  {104}},\ \bibinfo {pages} {177} (\bibinfo {year} {1986})}\BibitemShut
  {NoStop}%
\bibitem [{\citenamefont {Brice{\~n}o}\ \emph {et~al.}(2018)\citenamefont
  {Brice{\~n}o}, \citenamefont {Guerrero}, \citenamefont {Hansen},\ and\
  \citenamefont {Monahan}}]{Briceno:2018lfj}%
  \BibitemOpen
  \bibfield  {author} {\bibinfo {author} {\bibfnamefont {R.~A.}\ \bibnamefont
  {Brice{\~n}o}}, \bibinfo {author} {\bibfnamefont {J.~V.}\ \bibnamefont
  {Guerrero}}, \bibinfo {author} {\bibfnamefont {M.~T.}\ \bibnamefont
  {Hansen}}, \ and\ \bibinfo {author} {\bibfnamefont {C.~J.}\ \bibnamefont
  {Monahan}},\ }\href {\doibase 10.1103/PhysRevD.98.014511} {\bibfield
  {journal} {\bibinfo  {journal} {Phys. Rev. D}\ }\textbf {\bibinfo {volume}
  {98}},\ \bibinfo {pages} {014511} (\bibinfo {year} {2018})},\ \Eprint
  {http://arxiv.org/abs/1805.01034} {arXiv:1805.01034 [hep-lat]} \BibitemShut
  {NoStop}%
\bibitem [{\citenamefont {Lensky}\ \emph {et~al.}(2018)\citenamefont {Lensky},
  \citenamefont {Hagelstein}, \citenamefont {Pascalutsa},\ and\ \citenamefont
  {Vanderhaeghen}}]{Lensky:2017bwi}%
  \BibitemOpen
  \bibfield  {author} {\bibinfo {author} {\bibfnamefont {V.}~\bibnamefont
  {Lensky}}, \bibinfo {author} {\bibfnamefont {F.}~\bibnamefont {Hagelstein}},
  \bibinfo {author} {\bibfnamefont {V.}~\bibnamefont {Pascalutsa}}, \ and\
  \bibinfo {author} {\bibfnamefont {M.}~\bibnamefont {Vanderhaeghen}},\ }\href
  {\doibase 10.1103/PhysRevD.97.074012} {\bibfield  {journal} {\bibinfo
  {journal} {Phys. Rev. D}\ }\textbf {\bibinfo {volume} {97}},\ \bibinfo
  {pages} {074012} (\bibinfo {year} {2018})},\ \Eprint
  {http://arxiv.org/abs/1712.03886} {arXiv:1712.03886 [hep-ph]} \BibitemShut
  {NoStop}%
\bibitem [{\citenamefont {Seng}\ and\ \citenamefont
  {Mei{\ss}ner}(2019)}]{Seng:2019plg}%
  \BibitemOpen
  \bibfield  {author} {\bibinfo {author} {\bibfnamefont {C.-Y.}\ \bibnamefont
  {Seng}}\ and\ \bibinfo {author} {\bibfnamefont {U.-G.}\ \bibnamefont
  {Mei{\ss}ner}},\ }\href {\doibase 10.1103/PhysRevLett.122.211802} {\bibfield
  {journal} {\bibinfo  {journal} {Phys. Rev. Lett.}\ }\textbf {\bibinfo
  {volume} {122}},\ \bibinfo {pages} {211802} (\bibinfo {year} {2019})},\
  \Eprint {http://arxiv.org/abs/1903.07969} {arXiv:1903.07969 [hep-ph]}
  \BibitemShut {NoStop}%
\bibitem [{\citenamefont {Horsley}\ \emph {et~al.}(2020)\citenamefont
  {Horsley}, \citenamefont {Nakamura}, \citenamefont {Perlt}, \citenamefont
  {Rakow}, \citenamefont {Schierholz}, \citenamefont {Somfleth}, \citenamefont
  {Young},\ and\ \citenamefont {Zanotti}}]{Horsley:2020ltc}%
  \BibitemOpen
  \bibfield  {author} {\bibinfo {author} {\bibfnamefont {R.}~\bibnamefont
  {Horsley}}, \bibinfo {author} {\bibfnamefont {Y.}~\bibnamefont {Nakamura}},
  \bibinfo {author} {\bibfnamefont {H.}~\bibnamefont {Perlt}}, \bibinfo
  {author} {\bibfnamefont {P.~E.~L.}\ \bibnamefont {Rakow}}, \bibinfo {author}
  {\bibfnamefont {G.}~\bibnamefont {Schierholz}}, \bibinfo {author}
  {\bibfnamefont {K.}~\bibnamefont {Somfleth}}, \bibinfo {author}
  {\bibfnamefont {R.~D.}\ \bibnamefont {Young}}, \ and\ \bibinfo {author}
  {\bibfnamefont {J.~M.}\ \bibnamefont {Zanotti}} (\bibinfo {collaboration}
  {CSSM Collaboration, QCDSF Collaboration, UKQCD Collaboration}),\ }in\
  \href@noop {} {\emph {\bibinfo {booktitle} {{37th International Symposium on
  Lattice Field Theory (Lattice 2019) Wuhan, Hubei, China, June 16-22,
  2019}}}}\ (\bibinfo {year} {2020})\ \Eprint {http://arxiv.org/abs/2001.05366}
  {arXiv:2001.05366 [hep-lat]} \BibitemShut {NoStop}%
\bibitem [{\citenamefont {Haar}\ \emph {et~al.}(2018)\citenamefont {Haar},
  \citenamefont {Nakamura},\ and\ \citenamefont {St{\"u}ben}}]{Haar:2017ubh}%
  \BibitemOpen
  \bibfield  {author} {\bibinfo {author} {\bibfnamefont {T.~R.}\ \bibnamefont
  {Haar}}, \bibinfo {author} {\bibfnamefont {Y.}~\bibnamefont {Nakamura}}, \
  and\ \bibinfo {author} {\bibfnamefont {H.}~\bibnamefont {St{\"u}ben}},\
  }\href {\doibase 10.1051/epjconf/201817514011} {\bibfield  {journal}
  {\bibinfo  {journal} {EPJ Web Conf.}\ }\textbf {\bibinfo {volume} {175}},\
  \bibinfo {pages} {14011} (\bibinfo {year} {2018})},\ \Eprint
  {http://arxiv.org/abs/1711.03836} {arXiv:1711.03836 [hep-lat]} \BibitemShut
  {NoStop}%
\bibitem [{\citenamefont {Edwards}\ and\ \citenamefont
  {Joo}(2005)}]{Edwards:2004sx}%
  \BibitemOpen
  \bibfield  {author} {\bibinfo {author} {\bibfnamefont {R.~G.}\ \bibnamefont
  {Edwards}}\ and\ \bibinfo {author} {\bibfnamefont {B.}~\bibnamefont {Joo}}
  (\bibinfo {collaboration} {SciDAC Collaboration, LHPC Collaboration, UKQCD
  Collaboration}),\ }\href {\doibase 10.1016/j.nuclphysbps.2004.11.254}
  {\bibfield  {journal} {\bibinfo  {journal} {Nucl.Phys.Proc.Suppl.}\ }\textbf
  {\bibinfo {volume} {140}},\ \bibinfo {pages} {832} (\bibinfo {year}
  {2005})},\ \Eprint {http://arxiv.org/abs/hep-lat/0409003}
  {arXiv:hep-lat/0409003 [hep-lat]} \BibitemShut {NoStop}%
\end{thebibliography}
\end{document}